\documentclass[aps,prb,reprint]{revtex4-2}
\usepackage{physics} 
\usepackage{siunitx} 
\usepackage{enumerate} 
\usepackage{pgfplotstable}
\usepackage{tikz,pgfplots}
\usepackage{pifont}
\usepackage{braket}
\usepackage{bm}
\usepackage{amssymb}
\usepackage{bbm}
\usepackage{amsmath}
\usepackage{booktabs}  
\usepackage{amsthm}
\usepackage{geometry}
\usepackage{tabularx}
\usepackage{yhmath}
\usepackage{float}
\usepackage[graphicx]{realboxes}
\usepackage{dcolumn}
\usepackage{hyperref}
\usepackage{my_symbols}
\usepackage{natbib}
\begin{document}

\title{Point-group symmetry enriched topological orders}

\author{Zhaoyang Ding}
\affiliation{State Key Laboratory of Surface Physics and Department of Physics, Fudan University, Shanghai 200433, China}

\author{Yang Qi}
\thanks{qiyang@fudan.edu.cn}
\affiliation{State Key Laboratory of Surface Physics and Department of Physics, Fudan University, Shanghai 200433, China}

\date{\today}
\begin{abstract}
	We study the classification of two-dimensional (2D) topological orders enriched by point-group symmetries, by generalizing the folding appraoch which was previously developed for mirror-symmetry-enriched topological orders.
	We fold the 2D plane hosting the topological order into the foundamental domain of the group group, which is a sector with an angle $2\pi/n$ for the cyclic point group $C_n$ and a sector with an angle $\pi/n$ for the dihedral point group $D_{2n}$, and the point-group symmetries becomes onsite unitary symmetries on the sector.
	The enrichment of the point-group symmetries is then fully encoded at the boundary of the sector and the apex of the section, which forms a junction between the two boundaries.
	The mirror-symmetry enrichment encoded on the boundaries is analyzed by the classification theory of symmetric gapped boundaries,
	and the point-group-symmetry enrichment encoded on the junction is analyzed by a framework for classifying symmetric gapped junctions between boundaries which we develop in this work.
	We show that at the junction, there are two potential obstructions, which we refer to as an $H^1$ obstruction and an $H^2$ obstruction, respectively.
	When the obstruction vanishes, the junction, and therefore the point-group-symmetry-enriched topological orders, are classified by an $H^0$ cohomology class and an $H^1$ cohomology class, which can be understood as an additional Abelian anyon and a symmetry charge attached to the rotation center, respectively.
	These results are consistent with the classification of onsite-symmetry-enriched topological orders, where the $H^1$ and $H^2$ obstructions and the junction corresponds to the $H^3$ and $H^4$ obstructions for onsite symmetries, respectively.
\end{abstract}

\maketitle
\section{Introduction}
\label{sec:intro}

The interplay between symmetry and topology has become an important theme in modern condensed-matter physics.
In particular, states with intrinsic topological orders host fractional quasiparticle excitations known as the anyons~\cite{Wen1990,wilczek1990fractional,Wen2015nsr}.
Intrinsic topological orders can be further enriched by global symmetries, forming symmetry-enriched topological (SET) orders~\cite{XGWenPSG2002,Kitaev2006,XieChen2010LUT,Mesaros2013}.
Such symmetry enrichment can be used as signatures for detecting fractionalization~\cite{Essin2013,YQi2015,Tarantino_2016,Heinrich2016,MengCheng2017} and nontrivial topological excitation in numerical simulations and experiments~\cite{sun2018dynamical, barkeshli_coherent_2014},
and symmetry defects in SET phases may host degrees of freedom that can be used to store and manipulate quantum information~\cite{barkeshli2013twist}.
These motivates us to study SET orders, especially the classification of SET orders for a given underlying intrinsic topological order and a global symmetry group.

For an onsite symmetry group $G$, the classification of SET orders is described by the $G$-crossed braided tensor category theory~\cite{barkeshli2019symmetry}.
However, when $G$ contains crystalline symmetries, the problem of classifying crystalline-SET orders is still not fully solved.
The crystalline equivalence principle proposed by \citet{thorngren2018gauging} can be used to compute the classification, using an one-to-one correspondence between crystalline-SETs and onsite-SETs.
However, this equivalence principle does not directly offer a transparent understanding of the crystalline-SET states.
The idea of real-space construction has been a fruitful approach to study topological phases protected or enriched by crystalline symmetries~\cite{Song2020,Else2019}.
Dimensional reduction~\cite{isobe2015theory,song2017topological,huang2017building}, or more generally a real-space recipe, can be used to study crystalline-symmetry-protected topological states, or the topological crystalline states, by breaking them down to lower-dimensional symmetry-protected topological (SPT) states decorated on lower-dimensional cells in the crystal, such as mirror planes and rotation axis.
For two-dimensional (2D) intrinsic topological orders, the folding approach has been applied to study mirror-SET orders~\cite{qi2019folding}.
Similar to the dimensional-reduction scheme for crystalline-SPT states, one can argue that the information of mirror-enrichment in a mirror-SET is encoded on the mirror axis.

In this work, we generalize the folding approach to all 2D point groups, especially ones with rotational symmetries.
Similar to the mirror-SETs, we show that information of rotation-enrichment is encoded at the rotation center.
In the folding approach, a rotational-SET is transformed into a wedge-shape multilayer system, where the rotation center is a junction between two gapped boundaries.
We can then convert the problem of classifying rotational-SET states into classifying junctions between two gapped boundaries.
We study the latter problem by describing the junction with a bimodule category, and require that the rotational symmetry acts in a consistent way on this bimodule category.
In this way, we find two obstruction classes, dubbed the $H^1$ and $H^2$ obstruction respectively, which indicates that such consistent symmetry action does not exist.
Using the crystalline equivalence principle, the $H^1$ and $H^2$ obstructions are mapped to $H^3$ and $H^4$ obstructions for onsite-SETs, respectively.

The rest of the paper is oganized as the following:
In Sec.~\ref{sec:overview}, we give intuitive pictures and a brief overview of our main results.
In Sec.~\ref{sec:junction}, we provide a general description of junctions between two gapped boundaries of a 2D topological order.
In Sec.~\ref{sec:symmetric}, we add symmetry to the whole system and discuss the classification of symmetric junctions between two symmetric gapped boundaries of a 2D SET order.
Next, we apply these theories to study the folded multilayer system of a rotational-SET state.
In Sec.~\ref{sec:Cn} and Sec.~\ref{sec:D2n}, we study topological orders enriched by $C_n$ and $D_{2n}$ point-group symmetries, respectively.
Next, we discuss some concrete examples of obstruction functions.
In Sec.~\ref{sec:H1} and Sec.~\ref{sec:H2}, we discuss examples of nontrivial $H^1$ and $H^2$ obstruction functions, respectively.
Finally, conclusions and discussion on future studies are given in Sec.~\ref{sec:conclusion}.

\section{Overview of main results}
\label{sec:overview}

In this section, we give an intuitive argument that we can deform a 2D crystalline-SET such that how point-group symmetries enrich a 2D topological order is encoded near the regions, which are invariant under corresponding nontrivial symmetry operations.
The regions may involve the mirror axes and the rotation center.
To study the information of such enrichment, we use the folding approach to convert the 2D crystalline-SET into a multilayer system, where the point-group symmetries are transformed into onsite symmetries that interchange the layers~\footnote{In this paper, we use blackboard bold letters to denote the onsite symmetry groups in order to distinguish them from point-groups. For example, an interlayer exchange symmetry group in multilayer systems is denoted as $\mathbb{Z}_n$, which originates from the rotation symmetry group $C_n$.}.
In this case, the mirror axes can be regarded as 1D boundaries of associated bilayer systems and the rotation center become a junction between two boundaries of the multilayer system.
And then, the symmetry data of mirror axes can be encoded by the theory of symmetric boundaries~\cite{cheng2020relative, ding2024anomalies} in bilayer geometry.
The rotation center, carrying the information of rotation-symmetry-enrichment, can be studied by using the theory of symmetric junctions between symmetric boundaries in multilayer geometry.
We briefly summarize the classification results of the mirror axes and the rotation center, which give the classification of 2D crystalline-SET orders, focusing on intuitive understandings, while leaving full details to following sections.

In this work, we consider an intrinsic topological order, described by a unitary modular fusion category (UMTC) $\mathcal C$, enriched by a symmetry group $G$.
For clarity, in most part of this work, we consider the simple case where $G$ is a 2D point group and contains no additional onsite symmetries.
Our approach can be easily generalized to cases where $G$ also contains onsite symmetries, which will be demonstrated in an example in Sec.~\ref{sec:semion-c2-z2}.
2D point groups including the rotation groups $C_n$ and the dihedral groups $D_{2n}$~\footnote{In this work, we adapt the notation that $D_{2n}$ is the order-$2n$ dihedral group.
This notation is often used in mathematical literatures, while the dihedral groups $D_{2n}$ is denoted as $D_n$ in crystallography.}.
In a 2D crystal, $n$ can only be $1, 2, 3, 4, 6$~\footnote{$C_1$ is the trivial point group containing no nontrivial symmetry operations; $D_2$ is the mirror symmetry group.}.

\begin{figure}[H]
	\centering
	\includegraphics[width=0.48\textwidth, height=3.6cm]{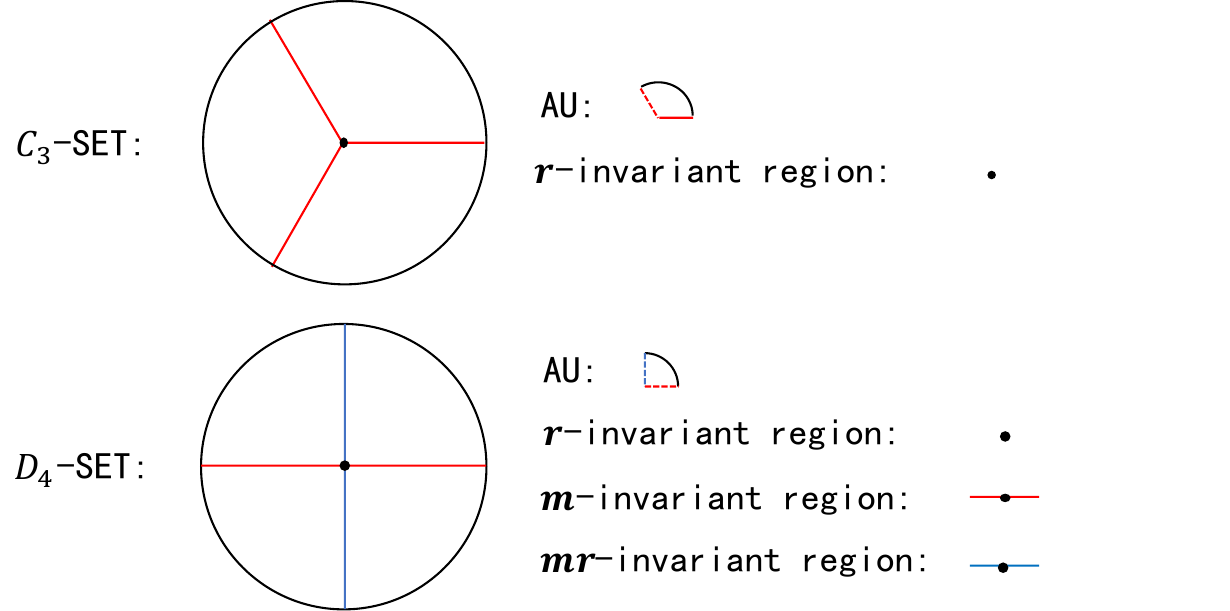}
	\caption{AU and invariant regions}
	\label{fig:AU}
\end{figure} 
A 2D system with point-group symmetry $G$ can be divided in several regions known as the ``asymmetric units (AU)'' or the fundamental domain of $G$.
As depicted in Fig.~\ref{fig:AU}, for rotation groups $C_n= \{\br^k| \br^n=\be \}$, an AU can be chosen to be a sector of the infinite 2d plane with an angle $2\pi/n$, whose apex locates at the rotation center.
The rotation center is invariant under the rotation action.
In the case of dihedral groups $D_{2n}= \{ {\mb}^l {\br}^k|{\br}^n = {\mb}^2 = \be, \mb \br \mb = {\br}^{-1} \}$, an AU is the sector between two adjacent mirror axes, with the sector angle being $\pi/n$, and similarly, its apex locates at the rotation center.
The mirror axes is invariant under the corresponding mirror actions, and the rotation center is invariant under all symmetry actions.
Since the AU has the property that any two points in the interior of the AU are not related by point-group symmetries,
the point-group symmetry posts no constraints on the wave function within the interior of the AU.
Therefore, we can use local unitary transformations (LUTs)~\cite{XieChen2010LUT,chen2013symmetry} to deform the wave function on an AU.
Such LUTs cannot change the underlying topological order, but they can deform arbitrary states on the AU to a standard reference state of the same topological order.
If we extend such LUTs to the whole 2D plane in a symmetric way, they will deform the states in all sectors to the standard reference state, except at the regions invariant under nontrivial symmetry operations.
Therefore, all information of $G$-enrichment is encoded at the invariant regions.
More specifically, the mirror axes and the rotation center respectively encode the information of the corresponding mirror symmetries and rotation symmetry.

\begin{figure}[H]
	\centering
	\includegraphics[width=0.4\textwidth, height=3.6cm]{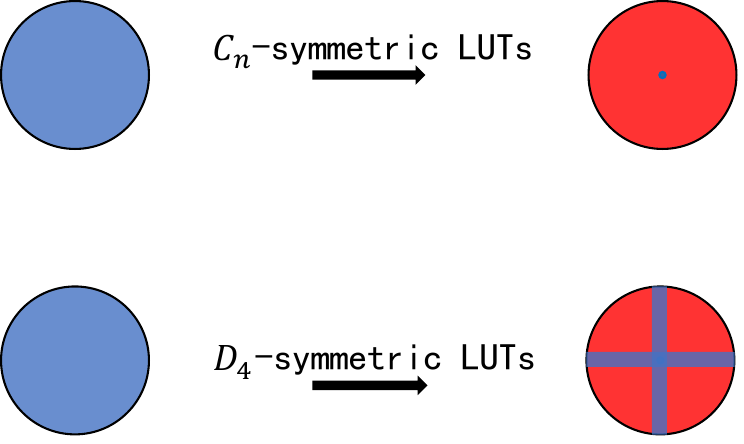}
	\caption{Symmetric LUTs}
	\label{fig:LUT}
	\end{figure}
\noindent

\begin{figure}[H]
	\centering
	\includegraphics[width=0.4\textwidth, height=2.9cm]{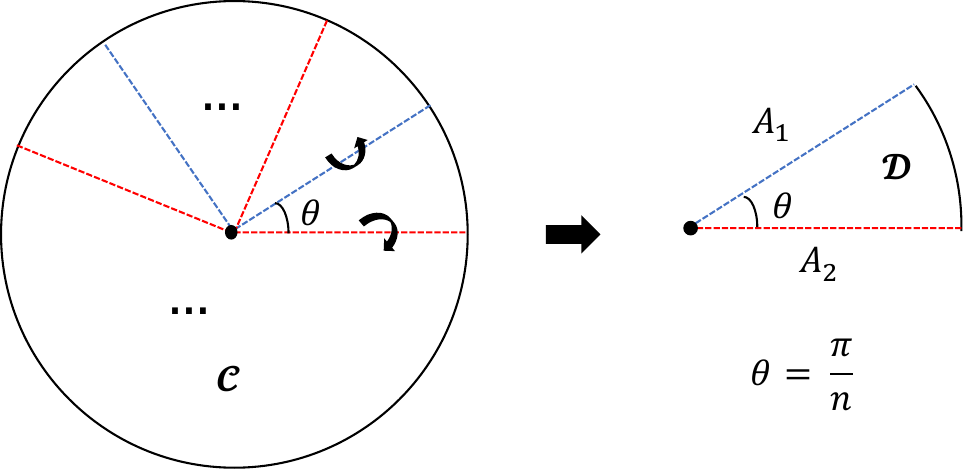}
	\caption{Folding approach}
	\label{fig:fold}
\end{figure}

To access such symmetry-enrichment information, we apply the folding approach to fold the 2D plane into one sector containing multiple layers.
For $G=C_n$, in order to encode symmetry data at the rotation center, we fold along two arbitrary lines that are separated by an angle of $\pi/n$, as shown in Fig.~\ref{fig:fold}.
\noindent
The even layers and odd layers host the intrinsic topological orders $\mathcal C$ and $\mathcal C^{\text{op}}$, respectively.
Then, the $2\pi/n$ rotation symmetry actions transform into interlayer exchanges, which are onsite $\mathbb{Z}_n$ symmetry actions.
The rotation center can be regarded as $\mathbb{Z}_n$-symmetric junction between two $\mathbb{Z}_n$-symmetric gapped boundaries in the folded system.
In particular, the symmetry data of the two $\mathbb{Z}_n$-symmetric gapped boundaries are trivial.
Applying the theory of symmetric junction, introduced in Sec.~\ref{sec:symmetric}, we find that there is a potential $H^1$ obstruction which is characterized by a cohomology class in $H^1_{\rho}[C_n, \cA_{\cC}]$.
Here, $\cA_\cC$ denotes the fusion group of the Abelian anyons in the topological order $\cC$.
When the $H^1$ obstruction vanishes~\footnote{Generally, there might be another $H^2$ obstruction class in $H^2[G, \mathrm U(1)]$, but the cohomology group $H^2[C_n, \mathrm U(1)]$ is trivial.}, the symmetric junction is classified by $[\cA^{C_n}_{\cC}]_{n\cA_{\cC}}$ and $H^1[C_n, \mathrm U(1)]$, which classify the $C_n$ SET order.
Intuitively, they can be viewed as decorating the rotation center with an Abelian anyon and a charge of $C_n$ symmetry, respectively.
The former decoration classifies rotation-symmetry fractionalization:
In simple cases where rotation does not change anyon types, rotation-symmetry fractionalization is manifested as an anyon $a$ carries a projective representation with $\br^n=e^{i\theta}$,
where $e^{i\theta}$ can be understood as the Berry phase the anyon acquires as it moves around the rotation center when it is acted by the rotation symmetry $\br$ $n$ times.
Therefore, decorating the rotation center by an anyon $b$ adds to this Berry phase a phase factor $M_{ba}$, which denotes the mutual braiding between the two anyons.
On the other hand, the latter decoration corresponds to stacking a bosonic SPT state to the SET order, because rotation-SPT states can be constructed by decorating the rotation center with a symmetry charge according to the real-space construction recipe~\cite{Song2020,Else2019}.

For $G=D_{2n}$, in addition to rotation symmetries, it also contains mirror-reflection symmetries.
We fold the system along the mirror axes to make it a $2n$-layers system, as shown in Fig.~\ref{fig:fold}.
The $D_{2n}$ symmetry transforms into an onsite $\mathbb{D}_{2n}$ symmetry in the folded system, which permutes the layers.
As shown in Refs.~\cite{qi2019folding,ding2024anomalies}, the region of mirror axes, which become $\mathbb{Z}_2$-symmetric boundaries of the bilayer system, encodes the information of mirror-symmetry enrichment.
Furthermore, the rotation center becomes a $\mathbb{D}_{2n}$-symmetric junction between the two boundaries.
At the junction, there are potential $H^1$ and $H^2$ obstructions which are characterized by cohomology classes in $H^1_{\rho}[D_{2n}, \cA_{\cC}]$ and $H^2[D_{2n}, \mathrm U(1)]$, respectively.
When the obstruction vanishes, the symmetric junction is classified by $[\cA^{D_{2n}}_{\cC}]_{2n\cA_{\cC}}$ and $H^1[D_{2n}, \mathrm U(1)]/\mathbb{Z}_{(n+1, 2)}$~\footnote{(m, n) is the greatest divisor of m and n.}, which classify the $D_{2n}$ SET order.

Similar to the case of $C_n$ symmetry, they can be viewed as decorating the rotation center with an Abelian anyon and a charge of $D_n$ symmetry, respectively.
The former decoration describes symmetry fractionalization, and the latter decoration describes an additional bosonic SPT layer.

\begin{figure}[H]
	\centering
	\includegraphics[width=0.4\textwidth, height=2.9cm]{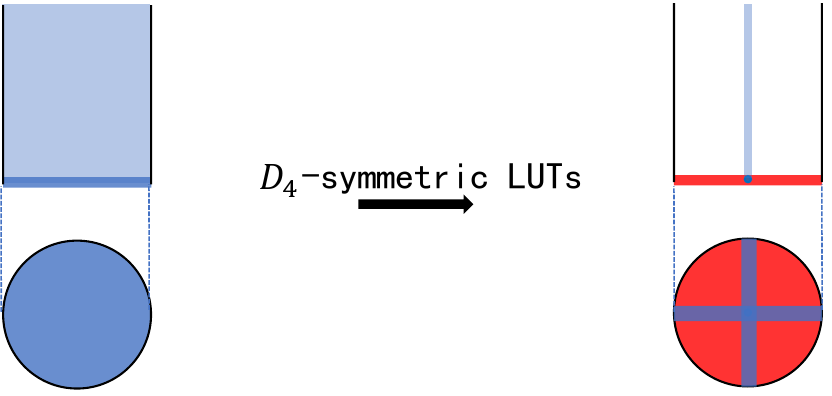}
	\caption{Anomalous $D_4$-SET order}
	\label{fig:D4SET}
\end{figure}

As discussed above, considering anomalous $D_{2n}$-SET orders, there are potential anomalies localized at the mirror axes and the rotation center.
In particular, the anomalies of mirror axes can be detected in corresponding bilayer systems, which have been understood in Ref.~\cite{qi2015anomalous, barkeshli_reflection_2020, barkeshli2018time, ding2024anomalies}.
And then, we focus on the the rotation center of the anomalous $D_{2n}$-SET orders and assume that the mirror axes are anomaly free.
When the $H^1$ obstruction is absent, we can easily understand the $H^2$ obstruction by the dimensional reduction procedure for 3D SPT phases~\cite{cheng2022rotation}.
It is known that an anomalous 2D SET order should localize at the surface of a proper 3D topological phase~\cite{chen2015anomalous,fidkowski2017realizing,barkeshli2018time}.
As depicted in Fig.~\ref{fig:D4SET}, the anomalous $D_{2n}$-SET order is localized at the bottom of a 3D $D_{2n}$-SPT phase, represented as an infinite blue disk.
\noindent
Since the wave function of the 3D $D_{2n}$-SPT phase is short-range entanglement, we can construct proper $D_{2n}$-symmetric LUTs to transform the wave function into a product state, except for the rotation axis and mirror planes.
Furthermore, due to the absence of anomalies at the mirror axes, we can construct another proper $D_{2n}$-symmetric LUTs, which only acts on the mirror planes, to transform the the wave function into a product state, except for the rotation axis.
The remaining rotation axis can be viewed as a 1D $D_{2n}$-SPT phase, characterized by an element of $H^2[D_{2n}, \mathrm U(1)]$, which cancels out the $H^2$ obstruction at the rotation center.

\section{The junction between two gapped boundaries}
\label{sec:junction}
In this section, based on the works~\cite{eliens2014diagrammatics, kong2014anyon, cong2016topological}, we briefly review descriptions of a gapped boundary of an intrinsic topological order. 
We then consider a junction between two such gapped boundaries. 
It is known that a defect, localized at the junction, corresponds to an irreducible bimodule~\cite{kong2014anyon, lou2021dummy}.
Thus, we argue that the junction is described by a bimodule category, which involve these irreducible bimodules.
Furthermore, we find that the bimodule category can be characterized by a fundamental bimodule $(X, \rho^L, \rho^R)$, which also simultaneously characterizes the junction.
To encode the data of the fundamental bimodule, we define $L$-symbols that serve to construct its operation tensors.
More specifically, these $L$-symbols originate from $L$-moves, which should be compatible with the fusion of boundary excitations on the two gapped boundaries.
This compatibility leads to the pentagon consistency equations.
Thus, the $L$-symbols obtained from these consistency equations can  determine operation tensors of the fundamental bimodule.
In addition, we examine the \textit{topological equivalent relations} of the $L$-symbols.
These relations suggest that junctions, characterized by distinct fundamental bimodules constructed from topologically equivalent $L$-symbols, are topologically equivalent.
It should be noted that when the bulk is a Levin-Wen model, the junction between two gapped boundaries has been studied in Ref.~\cite{wang_extend_2022}.
In our formulation, the bulk is described by a unitary modular tensor category(UMTC), which enriches the previous results.

\subsection{The description of a gapped boundary}
\begin{figure}[H]
	\centering
	\includegraphics[width=0.22\textwidth, height=2.5cm]{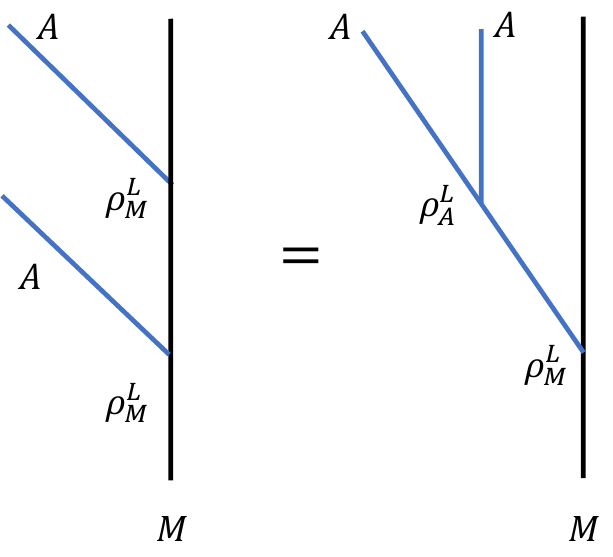}
	\caption{ Consistency equations of the left operation}
	\label{fig:Lm}
\end{figure}

\begin{figure}[H]
	\centering
	\includegraphics[width=0.28\textwidth, height=2.4cm]{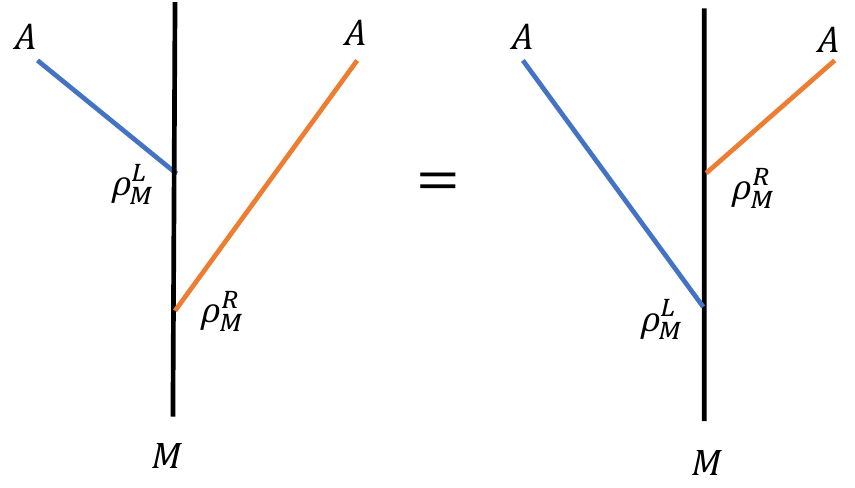}
	\caption{The consistency equations of $A$-$A$-bimodules}
	\label{fig:AA}
\end{figure}

\begin{figure}[H]
	\centering
	\includegraphics[width=0.28\textwidth, height=2.8cm]{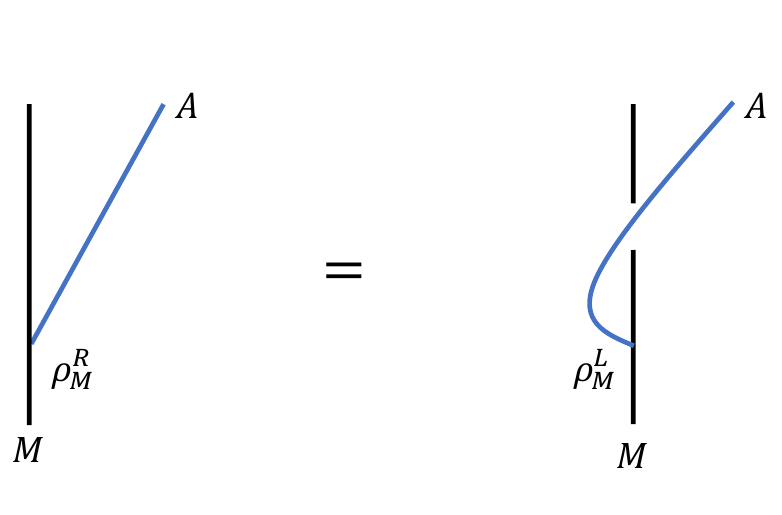}
	\caption{The $\rho^L_M$-$\rho^R_M$ relation}
	\label{fig:R-L}
\end{figure}

We start from an intrinsic topological order described by a UMTC $\cC$.
The types of anyons (or the simple objects) of $\cC$ are labeled as letters $a, b, c,\dots$, and we abuse the notation $a$ to represent an anyon.
It is known that a gapped boundary of this topological order can be characterized by a Lagrangian algebra $(A, \phi)$ over $\cC$~\cite{kong2014anyon}.
In this case, a boundary excitation corresponds to an irreducible $A$-$A$-bimodules.
In general, an $A$-$A$-bimodule is a triplet $(M, \rho^L_M, \rho^R_M)$, where $M$ is an object of $\cC$ and $\rho^{L(R)}_M$ is a left(right) operation.
The left operation $\rho^L_M$ is defined as
\begin{gather}
    \rho^L_M : A \otimes M \rightarrow M	,
\end{gather}
which satisfies consistency equations, as illsutrated in Fig.~\ref{fig:Lm}.
\noindent
For simplicity of notation, we omit certain indices and present them in an abbreviated form as follows:
\begin{equation}
      [{\rho}^L_M(a_1)][{\rho}^L_M(a_2)] = [{\rho}^L_A(a_1)][{\rho}^L_M(a_3)]	[F^{a_1 a_2 m_1}_{m_3}].
\end{equation}
Here, $[F^{a_1 a_2 m_1}_{m_3}]$ denote the $F$-symbols of $\cC$, $a_i$ represent condensed anyons on the $A$-type boundary and $m_i$ are anyons involved in the $A$-$A$-bimodule $(M, \rho^L_M, \rho^R_M)$.
The definition and consistency equations for the right operation are similar.
In addition, the compatibility of these two operations leads to consistency equations of $A$-$A$-bimodules, as illustrated in Fig.~\ref{fig:AA}.
\noindent
Since there are braiding structures of $\cC$, the compatible right operation $\rho^R_M$ can be related to the left operation $\rho^L_M$, as illustrated in Fig.~\ref{fig:R-L}.
\noindent
That is, 
\begin{gather}
     \rho^R_M = \rho^L_M \circ c_{MA}	
     \label{RL},
\end{gather}
where $c_{MA}$ is a braiding map $M \otimes A \xrightarrow{c_{MA}} A \otimes M$, involved in $\cC$.
Without ambiguity, we employ $M$ to represent both the object and the $A$-$A$-bimodule itself.
In particular, the algebra $A$ is a special irreducible $A$-$A$-bimodule, corresponding to the trivial boundary excitation.
The $A$-$A$-bimodules constitute a $A$-$A$-bimodule category, denoted as $\cC_{A|A}$, which describes the $A$-type boundary.
Next, we introduce the vertex lifting coefficients(VLCs), which are determined by data of the Lagrangian algebra $(A, \phi)$~\cite{eliens2014diagrammatics}.
The operation tensors of irreducible $A$-$A$-bimodules can be constructed from the VLCs, which explains why the Lagrangian algebra $(A, \phi)$ characterizes the $A$-type boundary.

\begin{figure}[H]
	\centering
	\includegraphics[width=0.4\textwidth, height=4cm]{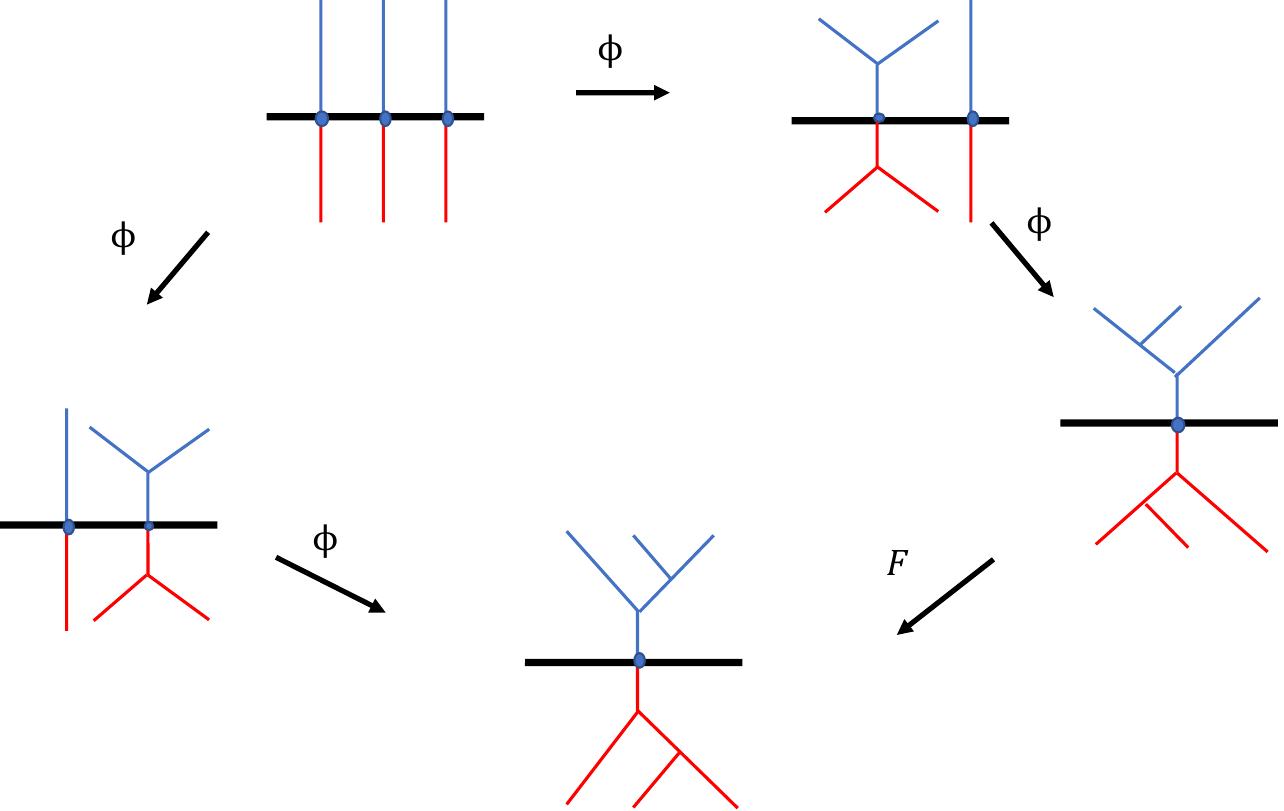}
	\caption{ The pentagon consistency equations of VLCs}
	\label{fig:Eq(VLCs)}
\end{figure}

\begin{figure}[H]
	\centering
	\includegraphics[width=0.4\textwidth, height=2.2cm]{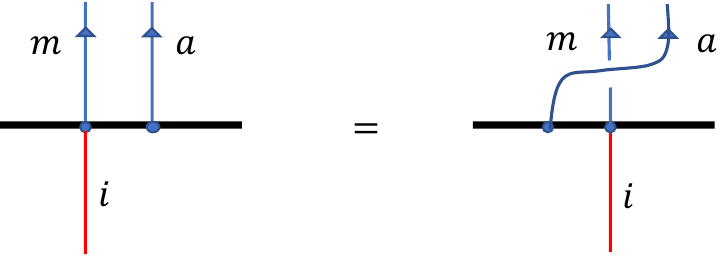}
	\caption{Additional constraints}
	\label{fig:Eq(braiding)}
\end{figure}

In the following, we adopt the indices $i, j, k\dots$ to represent boundary excitations, with the associated $A$-$A$-bimodules denoted as $M_i, M_j, M_k \dots$.   
Based on the physical fact that bulk anyons can transform into boundary excitations  upon reaching the $A$-type boundary, we define topological spaces $V^a_{i} = \text{hom}_{\cC}(a, M_i)$, with bases denoted as $\ket{a; i, \mu}_{A}$.
Particularly, the bases of condensed spaces $V^a_A$~\cite{ding2024anomalies} are $\ket{a; A, \mu}_A$.
In this case, vertex lifting coefficients(VLCs)~\cite{eliens2014diagrammatics} can be defined as projection coefficients of bases of spaces $V^a_{i} \otimes V^b_{j}$ onto bases of spaces $V^{ab}_c \otimes V^c_{k} \otimes V^{k}_{ij}$ 
\begin{widetext}
\begin{align}
	\ket{a; i, \mu}_A \otimes \ket{b; j, \nu}_A = \sum_{c, k, \alpha, \beta, \lambda}[\phi^{ab; k,\beta}_{c, \alpha; ij}]^{\mu \nu}_{\lambda}\ket{a, b; c, \alpha}\otimes \ket{c; k,\lambda}_A\otimes \ket{k, \beta; i, j},
\end{align}
where $V^{ab}_c$ are fusion spaces with bases $\ket{a, b; c, \alpha}$, originating from the fusion of anyons in the bulk, and $V^{k}_{ij}$ are fusion spaces with bases $\ket{k, \beta; i, j}$, corresponding to the fusion of boundary excitations on the $A$-type boundary.
Diagrammatically, the above definition of VLCs can be illustrated in Fig.~\ref{fig:VLCs}.
\begin{figure}[H]
	\centering
	\includegraphics[width=0.6\textwidth, height=3.5cm]{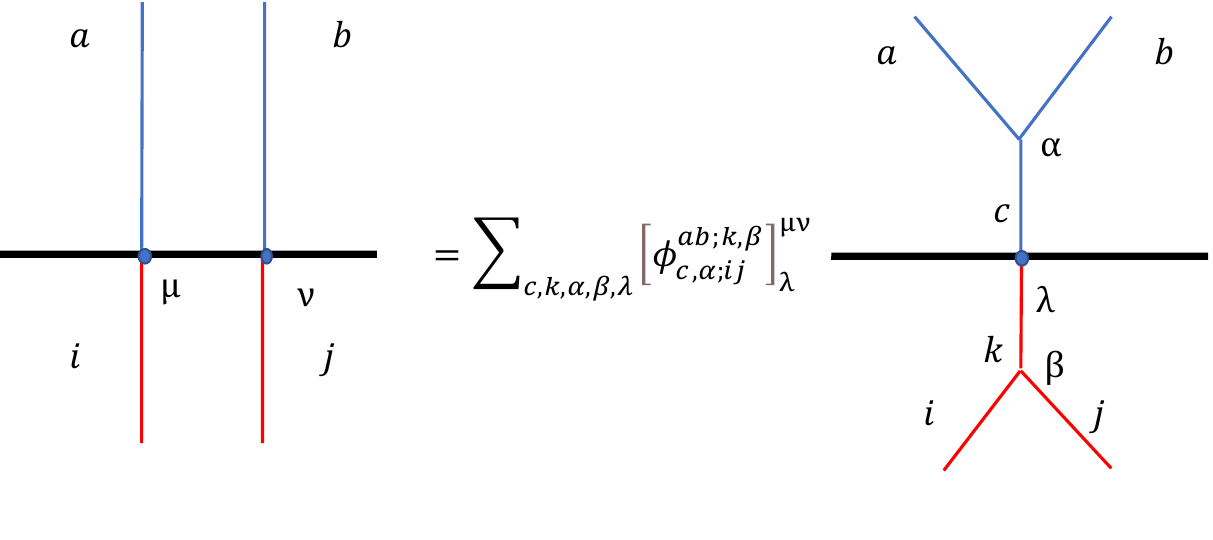}
	\caption{Definition of VLCs}
	\label{fig:VLCs}
\end{figure}
\noindent
Furthermore, the VLCs should be compatible with the fusions of anyons and boundary excitations, leading to pentagon consistency equations, as depicted in Fig.~\ref{fig:Eq(VLCs)}.
That is~\cite{eliens2014diagrammatics} 
\begin{align}
	\sum_{e,\alpha,\beta, m,\sigma,\rho} \phi^{ab; m,\sigma}_{e,\alpha; ij}\phi^{e c; l,\rho}_{d,\beta; m k}[F^{abc}_d]_{(f,\mu,\nu)(e,\alpha,\beta)}[F^{ijk}_l]^{-1}_{(n,\lambda,\tau)(m,\sigma,\rho)}=\phi^{b c; n,\lambda}_{f,\mu; j k}\phi^{a f; l,\tau}_{d,\nu; i n},
	\label{2}
\end{align}
\end{widetext}
where $[F^{abc}_d]$ and $[F^{ijk}_l]$ represent the bulk $F$-symbols and boundary $F$-symbols, respectively.
\noindent
Considering that a bulk anyon $a$ condenses and becomes the trivial boundary excitation on the $A$-type boundary, there are additional constraints on the process, as illustrated in Fig.~\ref{fig:Eq(braiding)}.
More specifically, for a boundary excitation $i$, transformed by a bulk anyon $m$, the additional constraints require that the associated VLCs should satisfy~\cite{eliens2014diagrammatics} 
\begin{gather}
	[\phi^{m a; i}_{m',\alpha;i A}]^{\mu_i \mu}_{\mu'_i}=\sum_{\beta}[\phi^{a m; i}_{m',\beta; A i}]^{\mu \mu_i}_{\mu'_i} [R^{m a}_{m'}]_{\beta \alpha}
	\label{3},
\end{gather} 
where $[R^{m a}_{m'}]_{\beta \alpha}$ is the $R$-symbol of $\cC$.
\noindent
Then, we can use the VLCs obtained from \Eqs{2}{3} to construct operation tensors of $A$-$A$-bimodules, shown as
\begin{equation}
\begin{gathered}
    [\rho_{\alpha}^L(a, \mu)]^{m, i, \mu_i}_{m', i, \mu'_{i}} = [\phi^{a m; i}_{m', \alpha; A i}]^{\mu \mu_{i}}_{\mu'_{i}}, \\
    [\rho_{\alpha}^R(a, \mu)]^{m, i, \mu_i}_{m', i, \mu'_{i}} = [\phi^{m a; i}_{m',\alpha; i A}]^{\mu_i \mu}_{\mu'_i}.
    \label{6}
\end{gathered}
\end{equation}
Clearly, the constructed operation tensors satisfy the consistency equations of the left operation and the $\rho^L_M$-$\rho^R_M$ relation.
In other words, the data of the $A$-$A$-bimodules can be constructed from the VLCs.

\subsection{The description of a junction}
Next, we examine a junction between an $A_1$-type boundary and an $A_2$-type boundary~\footnote{In the subsequent sections of this paper, we focus exclusively on boundaries where the multiplicity of anyon condensing is single.}. 
We argue that the junction is described by a $A_1$-$A_2$-bimodule category, denoted as $\cC_{A_1|A_2}$. 
A defect, localized at the junction, corresponds to an irreducible $A_1$-$A_2$-bimodule of $\cC_{A_1|A_2}$~\cite{kong2014anyon, lou2021dummy}.  
Considering a special irreducible $A_1$-$A_2$-bimodule $(X, \rho^{L}, \rho^R)$, referred to as the fundamental $A_1$-$A_2$-bimodule, the object $X$ is constructed by objects of the two algebras.
To be precise, the objects of the two algebras are written as
\begin{equation}
\begin{gathered}
	A_1 = \oplus_a W^{A_1}_{a}a ,\\
	A_2 = \oplus_a W^{A_2}_{a}a ,
\end{gathered}
\end{equation}
where $W^{A_{1(2)}}_{a}$ is the dimension of $V^a_{A_{1(2)}}$, representing the multiplicity of condensation channels for $a$.
The object $X$ can be constructed as:
\begin{align}
	A_1 \otimes A_2 = N_{12}X \oplus X^{\prime},
\end{align} 
where $N_{12} = \Sigma_{a} W^{A_1}_{a} W^{A_2}_{\bar{a}} $ counts types of boundary defects localized at the junction~\cite{lan2015gapped} and we denote the dual anyon of $a$ as $\bar{a}$.
The object $X$, which involves the trivial anyon $\bm{1}$, is formed by collecting anyons on the right side of the equation whose coefficients are multiple of the $N_{12}$, while the object $X'$ are formed by the remaining anyons.
Furthermore, given an object $C \in \cC$, an associated induced $A_1$-$A_2$-bimodule $(C \otimes X, \rho^L_{C\otimes X}, \rho^R_{C\otimes X})$ can be determined by the data of the fundamental $A_1$-$A_2$-bimodule:
\begin{equation}
\begin{aligned}
	\rho^L_{C\otimes X} =& (\textbf{id}_{C} \otimes \rho^L)\circ (c^{-1}_{CA_1} \otimes \textbf{id}_X),\\
	\rho^R_{C\otimes X} =& \textbf{id}_C \otimes \rho^R.
\end{aligned}
\end{equation}
The induced $A_1$-$A_2$-bimodules correspond to defects produced by absorbing bulk anyons at the junction. 
In addition, it is known that every bimodule in $\cC_{A_1|A_2}$ can be regarded as a submodule of an induced bimodule\cite{fuchs2002tft}.
Consequently, the data of the bimodule category $\cC_{A_1|A_2}$ can be entirely characterized by the fundamental bimodule $(X, \rho^L, \rho^R)$.

\begin{figure}[H]
	\centering
	\includegraphics[width=0.4\textwidth,height=2.7cm]{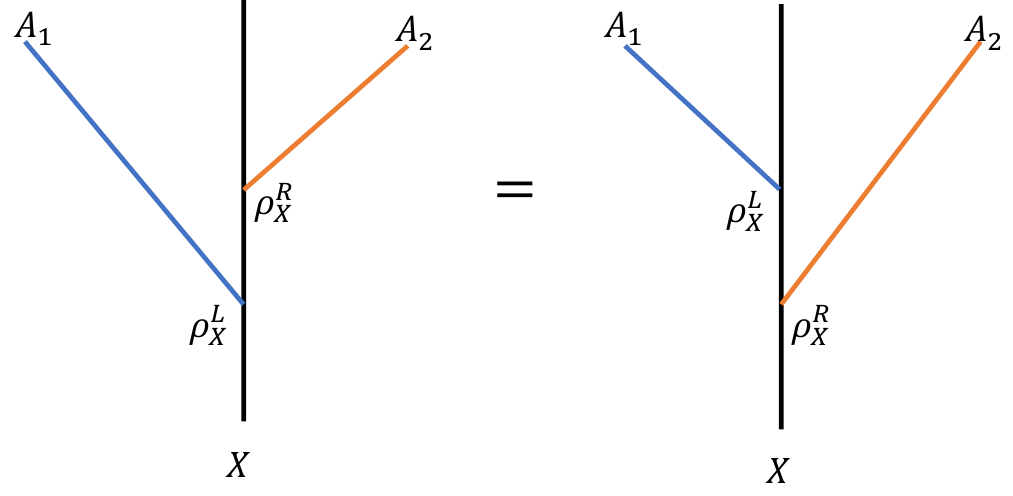}
	\caption{Consistency equations of fundamental $A_1$-$A_2$-bimodule}
	\label{fig:Bimodule}
\end{figure}
\begin{figure}[H]
	\centering
	\includegraphics[width=0.4\textwidth,height=1.6cm]{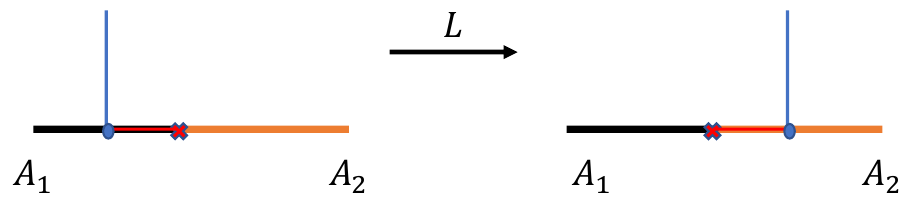}
	\caption{$L$-moves}
	\label{fig:Lmove}
\end{figure}
\begin{figure}[H]
	\centering
	\includegraphics[width=0.5\textwidth,height=4.5cm]{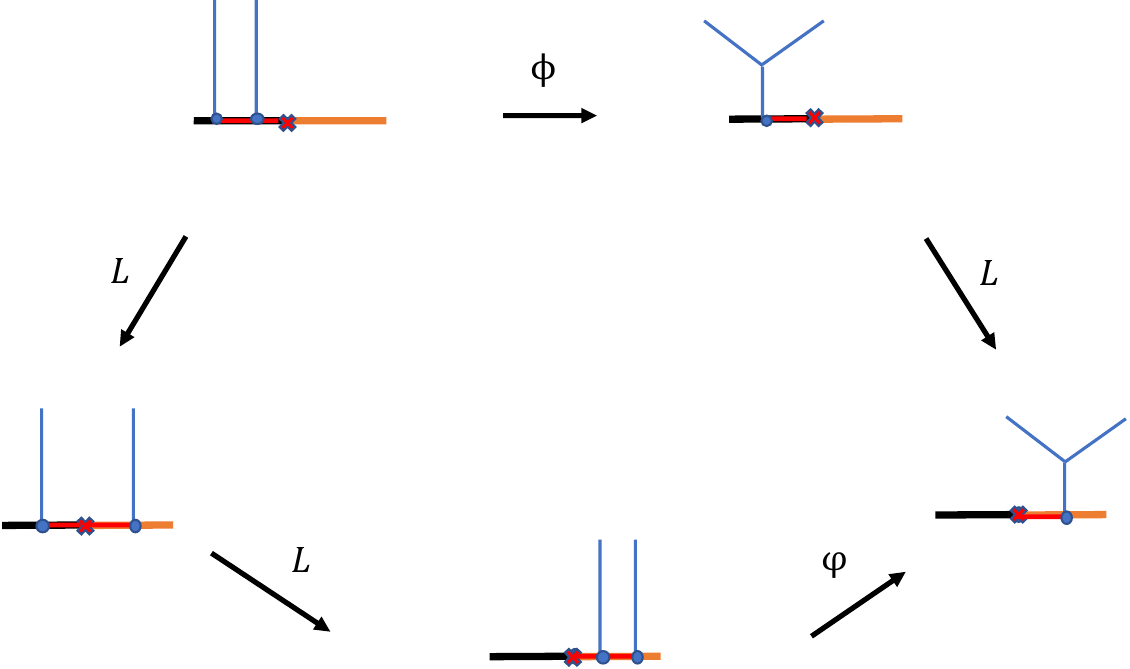}
	\caption{Consistency equations of $L$-symbols}
	\label{fig:Eq(L)}
\end{figure}

Different from $A_i$-$A_i$-bimodules, the fundamental $A_1$-$A_2$-bimodule has distinct consistency equations, as illustrated in Fig.~\ref{fig:Bimodule}.
As a result, the right operation cannot be directly related to the left operation as \Eq{RL}.
It implies that the VLCs of two boundaries alone are insufficient to construct the operation tensors of the fundamental $A_1$-$A_2$-bimodules.
Therefore, we aim to seek additional data to determine the fundamental bimodule.
Since the fundamental bimodule can be regarded both as a left $A_1$-module and as a right $A_2$-module, the object $X$ can be decomposed into objects of irreducible $A_i$-$A_i$-bimodules in two distinct ways, shown as
\begin{align}
	X = \oplus_{i}W^{(1)}_{i}M^{(1)}_{i} = \oplus_{j}W^{(2)}_{j}M^{(2)}_{j},
\end{align}
where each $M^{(k)}_{i}$ is the object of an irreducible $A_{k}$-$A_{k}$ bimodule and each $W^{(k)}_{i}$ is a non-negative integer.
Particularly, we have
\begin{align}
	A_1 \otimes A_2 &= \sum_{a, b, c}W^{A_1}_{a} W^{A_2}_{b}N^c_{ab}c \notag \\
	                &= \sum_{a, c}W^{A_1}_{a}c \sum_{b} W^{A_2}_{b} N^b_{\bar{a}c} \notag \\
	                &= N_{12}A_2 \oplus \sum_{a, c, i'_2, j'_2} W^{A_1}_{a} W^{i'_2}_{\bar{a}}W^{j'_2}_c N^{A_2}_{i'_2 j'_2} c,
\end{align}
where $i'_2$ and $j'_2$ represent nontrivial boundary excitations on the $A_2$-type boundary.  
Here, we apply the constraints of anyon condensation in the third line, expressed as~\cite{qi2019folding}:
\begin{align}
	\sum_{c} W^{A_k}_{c} N^c_{a b} = \sum_{i_{k}, j_{k}} W^{i_{k}}_a W^{j_{k}}_b N^{A_k}_{i_k j_k},
\end{align} 
where $N^c_{a b}$, $N^{A_k}_{i_k j_k}$, $W^{i_{k}}_a$ and $W^{j_{k}}_b$ represent the dimensions of $V^{ab}_c$, $V^{A_k}_{i_k j_k}$ $V^a_{i_{k}}$ and $V^a_{j_{k}}$, respectively.
Thus, we have proven that $W^{(2)}_{A_2} = 1$ and the proof for $W^{(2)}_{A_2} = 1$ is similar.
The two decompositions above lead to two induced decompositions of topological space $\text{hom}_{\cC}(a, X)$, shown as
\begin{equation}
\begin{aligned}
	\text{hom}_{\cC}(a, X) =& \oplus _{i, n^{(1)}_{i}} \text{hom}_{\cC}(a, M^{(1)}_{i})_{n^{(1)}_{i}}  \\
	=& \oplus _{j, n^{(2)}_{j}} \text{hom}_{\cC}(a, M^{(2)}_{j})_{n^{(2)}_{j}},
\end{aligned}
\end{equation}
where $n^{(k)}_{i} = 1,\dots,W^{(k)}_{i}$.
Physically, the two induced decompositions of topological spaces correspond to two distinct paths along which the junction absorbs the anyon $a$.
These two paths can be related by $L$-moves, as illustrated in Fig.~\ref{fig:Lmove}.
\noindent
Therefore, we can define $L$-symbols by basis transformations of the topological space $\text{hom}_{\cC}(a, X)$ as
\begin{align}
	\ket{a; i, \mu_{i}, n^{(1)}_{i}}_{X} &= \notag \\
	&\sum_{j, \mu_{j}, n^{(2)}_{j}}  [L^{a}_{X}]_{(i, \mu_{i})(j, \mu_{j})} \ket{a; j, \mu_{j}, n^{(2)}_{j}}_{X},
\end{align}
where $\ket{a; i, \mu_{i}, n^{(k)}_{i}}_{X}$ is a basis of topological subspace $\text{hom}_{\cC}(a, M^{(k)}_{i})_{n^{(k)}_{i}}$ and $\mu_i$ label the multiplicity originating from the dimension of those topological subspaces.
Furthermore, the $L$-moves should be compatible with the fusions of boundary excitations on the two gapped boundaries.
This leads to pentagon consistency equations, as illustrated in Fig.~\ref{fig:Eq(L)}.
\noindent
That is
\begin{widetext}
\begin{align}
	\sum_{i_3, \mu_{i_3}} W^{(1)}_{i_3} [\phi^{a_1 a_2; i_3, \beta}_{a_3, \alpha; i_1 i_2}]^{\mu_{i_1} \mu_{i_2}}_{\mu_{i_3}} & [L^{a_3}_X]_{(i_3, \mu_{i_3})(j_3, \mu_{j_3})} =\notag \\ & \sum_{\mu_{j_1}, \mu_{j_2}}  W^{(2)}_{j_1}  W^{(2)}_{j_2}
        [L^{a_1}_X]_{(i_1, \mu_{i_1})(j_1, \mu_{j_1})} [L^{a_2}_X]_{(i_2, \mu_{i_2})(j_2, \mu_{j_2})}[\varphi^{a_1 a_2; j_3, \beta}_{a_3, \alpha; j_1 j_2}]^{\mu_{j_1}\mu_{j_2}}_{\mu_{j_3}}
        \label{9},   
\end{align}
where $\phi$ represent VLCs of the $A_1$-type boundary and $\varphi$ represent VLCs of the $A_2$-type boundary.
Solving \Eq{9}, we obtain $L$-symbols, which can be used to construct operation tensors of the fundamental bimodule.
Investigating a basis $\ket{m; i, \mu_{i}, n^{(1)}_{i}}_X$ of $\text{hom}_{\cC}(m, X)$, we select
\begin{equation}
\begin{gathered}
	[\rho_{\alpha}^L(a_1, \mu)]^{m, i, \mu_i}_{m', i', \mu'_{i'}} = \delta_{i i'}[\phi^{a_1 m; i}_{m', \alpha; A_1 i}]^{\mu \mu_{i}}_{\mu'_{i}}, \\
	[\rho_{\alpha}^R(a_2, \mu)]^{m, i, \mu_i}_{m', i', \mu'_{i'}}=\sum_{j, \mu_{j}, \mu'_{j}} W^{(2)}_{j} [L^{m'}_{X}]^{-1}_{(i', \mu'_{i'})(j, \mu'_{j})}[\varphi^{m a_2; j}_{m', \alpha; jA_2}]^{\mu_{j} \mu}_{\mu'_{j}}[L^{m}_{X}]_{(i, \mu_{i})(j, \mu_{j})} \label{15},
\end{gathered}
\end{equation}
\end{widetext}
where $a_i$ is a condensed anyon on the $A_i$-type boundary.
According to \Eq{9}, we can easily verify that the constructed operation tensors of the fundamental bimodule satisfy consistency equations of bimodules, illustrated in Fig.~\ref{fig:Bimodule}. 

Next, we consider the gauge redundancy of $L$-symbols.
When studying the junction, the VLCs of two boundaries are considered as already determined, so the allowable gauge transformations are expressed as:
\begin{equation}	   
\begin{gathered}
   \widetilde{\ket{a; i, \mu_{i}, n^{(1)}_{i}}}_{X} = \zeta^a_{i} \ket{a; i, \mu_{i}, n^{(1)}_{i}}_{X}, \\
   \widetilde{\ket{a; j, \mu_{j}, n^{(2)}_{j}}}_{X} = \zeta^a_{j} \ket{a; j, \mu_{j}, n^{(2)}_{j}}_{X},
   \label{gt} 	
\end{gathered}
\end{equation}
where the condition $\zeta^a_{i_1} \zeta^b_{i_2} = \zeta^c_{i_3}$ and $\zeta^a_{j_1} \zeta^b_{j_2} = \zeta^c_{j_3}$ must be satisfied to keep $[\phi^{ab; i_3,\beta}_{c, \alpha; i_1 i_2}]^{\mu \nu}_{\lambda}$ and $[\varphi^{ab; j_3,\beta}_{c, \alpha; j_1 j_2}]^{\mu \nu}_{\lambda}$ invariant.
In addition, since both $A_1$ and $A_2$ represent vacuum after anyon condensation~\cite{kong2014anyon}, we treat $A_1$ and $A_2$ as identical on the other side.
Thus, the phase factors, arising from the gauge transformations~(\ref{gt}), must satisfy the additional constraints $\zeta^a_{A_1} = \zeta^a_{A_2}$.
The additional constraints ensure that the special $L$-symbols $[L^a_{X}]_{(A_1, 1)(A_2, 1)}$ are gauge invariant under the gauge transformations~(\ref{gt}).
In contrast, the other $L$-symbols are gauge dependent.
To ease notation, we denote the special gauge invariant $L$-symbols $[L^a_{X}]_{(A_1, 1)(A_2, 1)}$ as $\hL^a$, which is another version of half-linkings discussed in other literatures~\cite{lou2021dummy, shen2019defect}.
As shown in Appendix~\ref{App:gauge}, considering two set of $L$-symbols, obtained from \Eq{9}, when their special $L$-symbols are identical, the two set of  $L$-symbols can be related by a gauge transformation.
In this case, we consider the two set of $L$-symbols to be gauge equivalent.
As a result, a gauge equivalence class of $L$-symbols can be represented by their special $L$-symbols $\hL^a$. 
Since the operation tensors of the fundamental bimodule $X$ are constructed from a set of $L$-symbols, we argue that the special $L$-symbols $\hL^a$ can determine a gauge equivalence class of operation tensors for the fundamental bimodule, which characterize the junction.

\begin{figure}[H]
	\centering
	\includegraphics[width=0.4\textwidth,height=4cm]{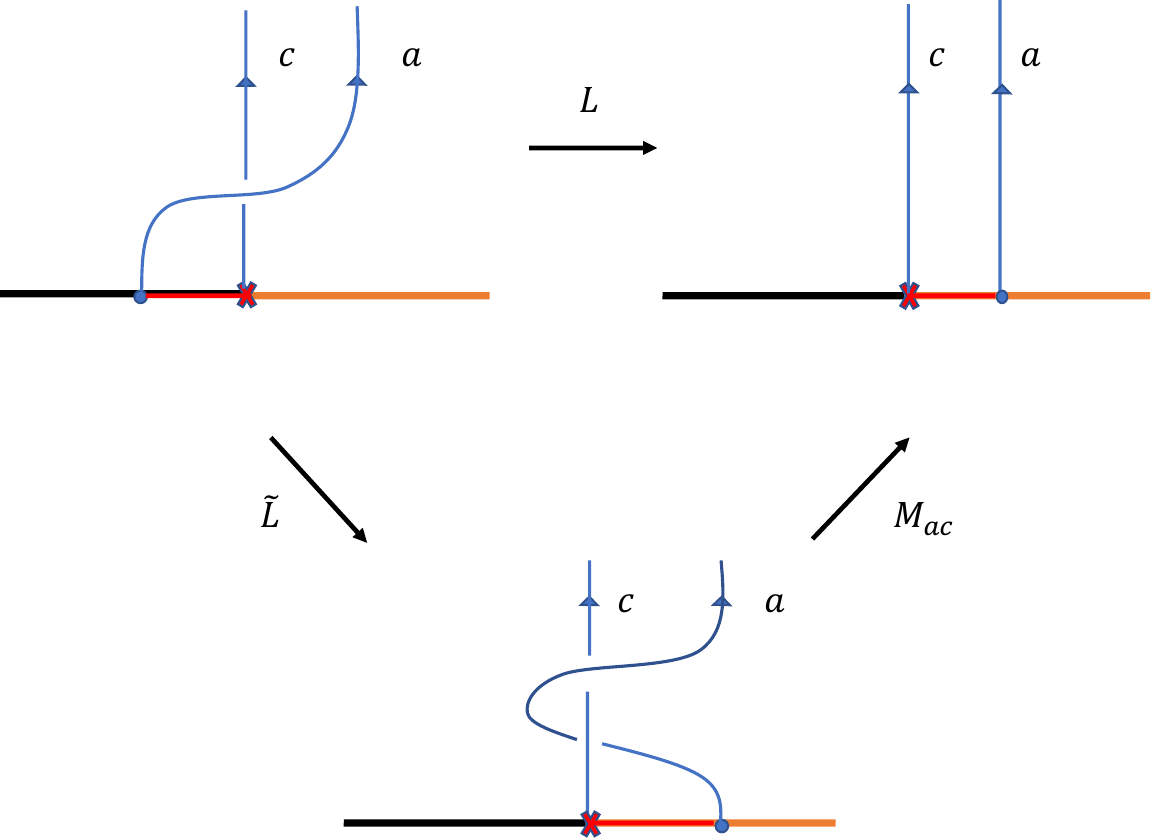}
	\caption{The junction modified by a string operator}
	\label{fig:Twisted L}
\end{figure}

However, junctions, characterized by distinct gauge inequivalent fundamental bimodules, may be topologically equivalent.
Here, since the junctions are always associated with both bulk and boundary, we can consider a series of LUTs acting on the whole system.
These may modify the special $L$-symbols of the junctions, thereby leading to the topological equivalence of two junctions characterized by gauge-inequivalent fundamental bimodules.
We consider an open string operator $\hW_{c}$ that produce a pair of abelian anyons, $c$ and $\bar{c}$.
One endpoint of the open string, responsible for producing anyon $c$, is localized at the junction, as depicted in Fig.~\ref{fig:Twisted L}.
Clearly, this string operator can be viewed as a LUT on the system, suggesting that the junction modified by the string operator is topologically equivalent to the original one.
In this case, attaching of the abelian anyon $c$ at the junction results in a modification of the special $L$-symbols 
\begin{gather}
    \tilde{\hL}^a = \hL^aM^*_{ac}
    \label{TER},
\end{gather}
where $a$ is a common condensed anyon and $M_{ac}$ is mutual braiding statistics between anyons $a$ and $c$.
Consequently, \Eq{TER} are referred to as the \textit{topologically equivalent relations} of junctions, which relate two sets of topologically equivalent special $L$-symbols.
\noindent

\section{Symmetric junction between symmetric gapped boundaries}
\label{sec:symmetric}
In this section, we begin by reviewing the introduction of an intrinsic topological order $\cC$ enriched by an onsite $G$-symmetry~\cite{barkeshli2019symmetry}.
Then, we examine an $A$-type gapped boundary of $\cC$ enriched by the introduced onsite $G$-symmetry.
Referring to the previous works~\cite{bischoff2019spontaneous, ding2024anomalies}, we review the consistency equations of boundary $U$-symbols $U_{\bg}(a)$.
Furthermore, we introduce the additional VLCs $[\phi^{a_{\bg} b_{\bh}}_{c_{\bg \bh}, \alpha}]$ and derive their corresponding consistency equations.
These equations potentially lead to an $H^3$ obstruction, which characterizes the inconsistency of them.
To facilitate calculations, we define the relative $H^3$ obstruction.
When the $G$-symmetric gapped boundary is anomaly-free, we argue that it is characterized by a grand algebra $(\textbf{A}, \tilde{\phi})$, which can be constructed from a triplet $(A, U_{\bg}(a), [\phi^{a_{\bg} b_{\bh}}_{c_{\bg \bh}, \alpha}])$.
In this case, the boundary $U$-symbols $U_{\bg}(a)$ and additional VLCs $[\phi^{a_{\bg} b_{\bh}}_{c_{\bg \bh}, \alpha}]$ are classified by $H^1_{\rho}[G, [\cA_{\cC}]_{\cA_A}]$ and $H^2[G, \mathrm U(1)]$, respectively.
Concurrently, these two groups classify the $G$-symmetric gapped boundary.

Based on the results for $G$-symmetric boundaries, we examine a $G$-symmetric junction between two $G$-symmetric gapped boundaries.
Similar to the $G$-symmetric boundary case, the $G$-symmetric junction can be obtained by enriching a junction between two boundaries of $\cC$ with $G$-symmetry.
Then, we derive the consistency equations for the $L$-symbols of the intrinsic junction.
According to these consistency equations, we define an obstruction function to diagnose the compatibility between the intrinsic junction and the $G$-symmetry.
This obstruction is characterized by an element of an $H^1$ cohomology group, which we refer to as $H^1$ obstruction. 
When the $H^1$ obstruction vanishes, we introduce common $\bg$-local defects and define associated additional $L$-symbols.
In this case, we derive the consistency equations for the additional $L$-symbols.
These consistency equations lead to potential $H^2$ obstruction at the junction.
In general, calculating the $H^2$ obstruction is challenging.
Therefore, we derive a specific expression for the relative $H^2$ obstruction, which is easier to compute.
When the $H^1$ and $H^2$ obstructions vanish, we argue that the $G$-symmetric junction is characterized by a grand fundamental bimodule $(\textbf{X}, \rho^L_{\textbf{X}}, \rho^R_{\textbf{X}})$, which can be constructed from a $L$-symbols pair $(\hL^a, \hL^{a_{\bg}}_{\mu \nu})$.
These $L$-symbols are classified by $ [\cA_{\cC}]^G_{\cA_{A_1} \vee \cA_{A_2}}/([[\cA_{\cC}]^G_{\cA_{A_1}}]_{\cA_{A_2}} \vee [[\cA_{\cC}]^G_{\cA_{A_2}}]_{\cA_{A_1}})$ and $H^1[G, \mathrm U(1)]$, respectively.
Similarly, these two groups simultaneously classify the $G$-symmetric junction. 

\subsection{The enrichment of a gapped boundary by a onsite $G$-symmetry}
Considering an intrinsic topological order $\cC$ enriched by on-site symmetry $G$, we assume that the bulk SET order is anomaly free.
The SET order is described by a $G$-crossed braided tensor category, denoted as $\cC^{\times}_{G}$, which involves three levels data, $[\rho]$, $[\mathfrak{w}]$ and $[\alpha]$~\cite{barkeshli2019symmetry}.
More specifically, $\rho: G \rightarrow \text{Aut}(\cC)$ assigns to each element $\bg \in G$ a symmetry action $\rho_{\bg}$ on anyons of $\cC$, where $\rho$ is a representative of equivalence class $[\rho]$.
The torsor $[\mathfrak{w}] \in H^2_{\rho}[G, \cA_{\cC}]$ is responsible for modifying the symmetry fractionalization, and the torsor $[\alpha] \in H^3[G, \mathrm U(1)]$ specifies a way of modifying the $F$-symbols of defects in $\cC^{\times}_{G}$.
Here, $\cA_{\cC}$ is a fusion subcategory of $\cC$, whose simple objects consist of abelian anyons in $\cC$.
Without ambiguity, we also use $\cA_{\cC}$ to denote the associated fusion ring of the fusion subcategory itself.

Once the symmetry data of the bulk SET are determined, we then examine one of its boundaries, which can be regarded as an $A$-type boundary of $\cC$ enriched by $G$ symmetry in a compatible manner.
We stress that there are two equivalent ways to understand the enrichment of the $A$-type boundary.
First, the enriched $A$-type boundary can be regarded as a $G$-symmetric gapped of the bulk SET.
Second, it can be viewed as an interface between the bulk SET order and a $G$-SPT phase.
The $G$-SPT phase represents a $H^3$ obstruction(we will discuss later in this section), which potential appear at the enrichment of the $A$-type boundary.
For the anomaly-free cases, these two understandings are equivalent.
Since the second approach is more convenient for introducing the potential $H^3$ obstruction, we adopt this perspective throughout this article and refer to the enriched $A$-type boundary as the $G$-symmetric $A$-type boundary of the bulk SET.
In this case, the $A$-type boundary should be preserved under the $G$-symmetry action, implying that the Lagrangian algebra $A$ is $G$-invariant~\cite{bischoff2019spontaneous}.
Thus, the induced symmetry actions on the condensed spaces $V^a_A$ can be defined as
\begin{align}
	\rho_{\bg}\ket{a; A} = U_{\bg}({\ubg a})\ket{{\ubg a}; A},
\end{align}       
where the anyon permutation $\rho_{\bg}(a)$ is denoted as $\ubg a$ and  $U_{\bg}(a)$ is called as boundary $U$-symbols~\cite{ding2024anomalies}.
Since the symmetry operators of the system are unitary(anti-unitary), it leads to the consistency equations of the boundary $U$-symbols, shown as
\begin{align}
	\phi^{ab}_{c, \alpha} = \sum_{\beta} \phi^{{\ubbg a} {\ubbg b}}_{{\ubbg c}, \beta}  [U_{\bg}(a, b; c)]_{\beta \alpha}\frac{U_{\bg}(c)}{U_{\bg}(a)U_{\bg}(b)} 
	\label{13},
\end{align}
where the bulk $U$-symbols $[U_{\bg}(a, b;c)]_{\beta \alpha}$ originate from symmetry actions on fusion spaces $V^{a b}_c$~\footnote{In this paper, for the convenience of exposition, we only present the consistency equations for unitary cases. For anti-unitary cases, the extension is straightforward.}.
Furthermore, there are additional constraints on the boundary $U$-symbols:
\begin{align}
	\eta_{a}(\bg, \bh) = \frac{U_{\bg \bh}(a)}{U_{\bg}(a)U_{\bh}({\ubbg a})}
	\label{14},
\end{align}
where $\eta_{a}(\bg, \bh)$ is the symmetry data of the bulk SET, characterizing the symmetry fractionalization.
Physically, \Eq{14} originates from the fact that the condensed anyons must also be  consistent with the description of the symmetry fractionalization in the bulk, which ensures the group homomorphism of symmetry operators~\cite{ding2024anomalies}.
When the $H^2$ obstruction vanishes on the $A$-type boundary~\cite{meir2012module, bischoff2019spontaneous, ding2024anomalies}, there exist solutions for the boundary $U$-symbols in \Eqs{13}{14}.
Considering the gauge transformations for the bases of condensed spaces $V^a_A$, which are expressed as $\widetilde{\ket{a, A}} = \gamma_a \ket{a, A}$, we note that only the boundary $U$-symbols $U_{\bg}(a)$ are physical, where $\bg$ is an element in the stabilizer subgroup of $a$, defined as $G_a = \{\bg \in G| {^{\bg}a} = a  \}$. 
These boundary $U$-symbols characterize an irreducible $\eta_a$-projective representation of $G_a $.
Thus, after excluding the gauge redundancy, the nonequivalent boundary $U$-symbols can be modified by a torsor $\mathfrak{v} \in  H^1_{\rho}[G, [\cA_{\cC}]_{\cA_A}]$ as 
\begin{align}
	\tilde{U}_{\bg}(a) = U_{\bg}(a) M^*_{a \mathfrak{v}(\bg)},
\end{align}
where $\cA_{A}$ is a fusion subcategory of $\cC$, whose simple objects consist of condensed abelian anyons on the $A$-type boundary.
The notation $[\cA_{\cC}]_{\cA_A}$ denotes a set of equivalence classes in $\cA_{\cC}$, which can be also regarded as a fusion ring of a fusion category.
Additionally, $\mathfrak{v}(\bg)$ is a representative element of the equivalence class $[\mathfrak{v}(\bg)]_{\cA_A}$.
The equivalence relation is defined with respect to $\cA_A$ as follows: for $a, b \in \cA_{\cC}$, $a \sim b$ if and only if there exists $c \in \cA_A$ such that $a = b \otimes c$.
Consequently, the boundary $U$-symbols are classified by the cohomology group, $H^1_{\rho}[G, [\cA_{\cC}]_{\cA_A}]$.

\begin{figure}[H]
	\centering
	\includegraphics[width=0.45\textwidth, height=2.5cm]{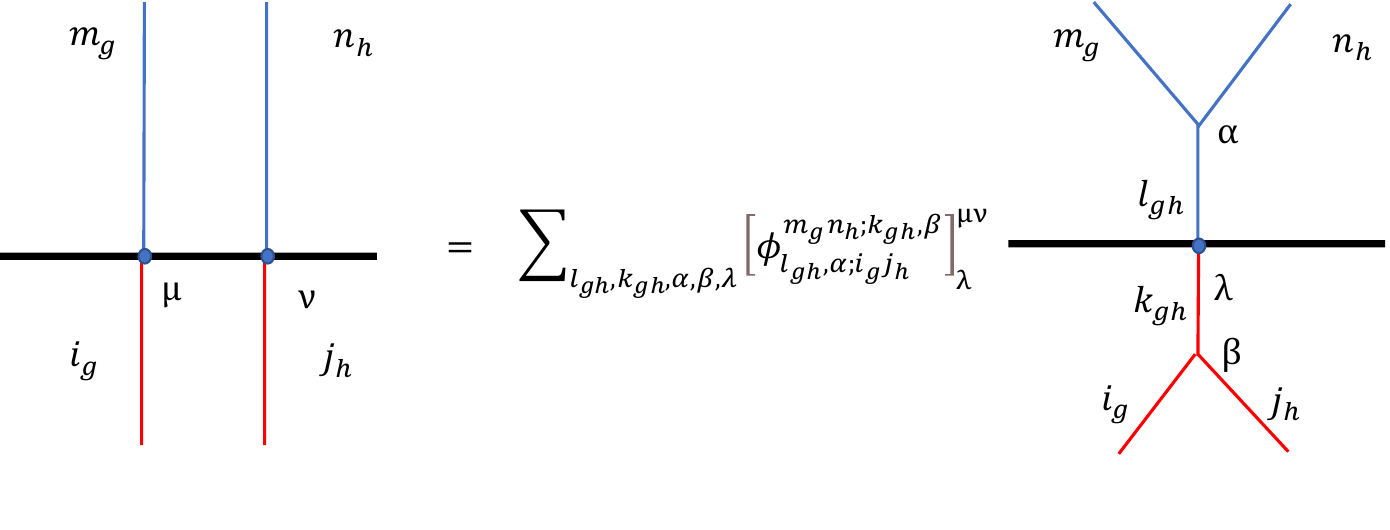}
	\caption{Definition of additional VLCs}
	\label{fig:VLCs(defect)}
\end{figure}

\begin{figure}[H]
	\centering
	\includegraphics[width=0.4\textwidth, height=2.8cm]{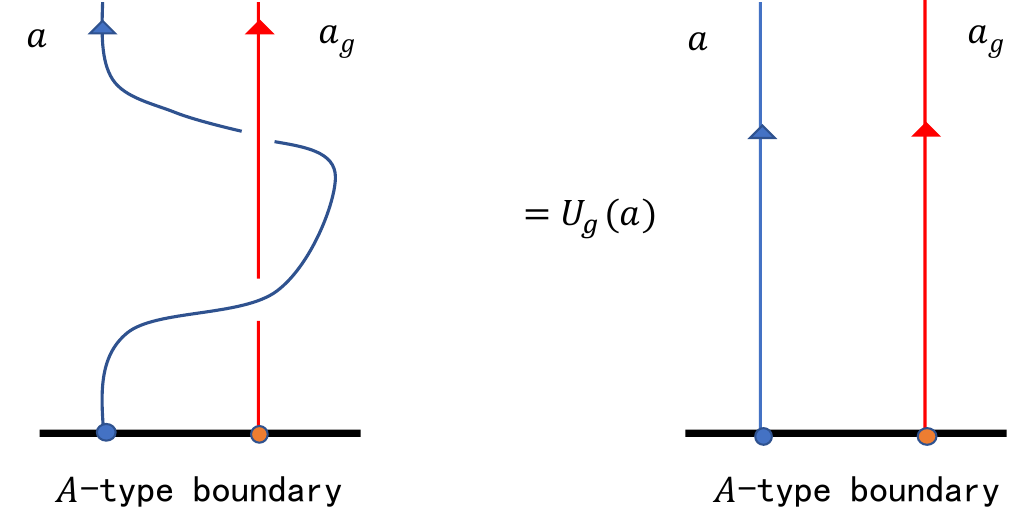}
	\caption{$\bg$-local condition}
	\label{fig:local}
\end{figure}

In addition to anyons, the bulk SET order also involve symmetry defects.
Mathematically, since $\cC^{\times}_{G}$ is an extension of $\cC$, the algebra $A$ can also be regraded as a condensable algebra over $\cC^{\times}_{G}$.
Thus, there emerge additional $A$-$A$-bimodules, whose objects are constituted by the symmetry defects.
Similar to the consistency equations of the $A$-$A$-bimodules, discussed in Sec.~\ref{sec:junction}, there exists a defect version of these consistency equations for the newly introduced additional $A$-$A$-bimodules. 
In addition, the presence of braiding structures in $\cC^{\times}_{G}$ allows for the compatibility of right operations with left operations, as presented in \Eq{RL}, for the additional $A$-$A$-bimodules. 
To construct the data of the additional $A$-$A$-bimodules, we define additional VLCs as detailed below.
Considering an irreducible additional $A$-$A$-bimodules $(M_{i_{\bg}}, \rho^L_{i_{\bg}}, \rho^R_{i_{\bg}})$, it corresponds to a boundary defect $i_{\bg}$.
Given that the boundary defect $i_{\bg}$ can originate from a bulk symmetry defect $m_{\bg}$, involved in $M_{i_{\bg}}$, we define additional topological spaces $V^{m_{\bg}}_{M_{i_{\bg}}} = \text{hom}_{\cC^{\times}_{G}}(m_{\bg}, M_{i_{\bg}})$. 
The additional topological spaces induce the definition of additional VLCs, as illustrated in Fig.~\ref{fig:VLCs(defect)},
\noindent
which is a defect version of VLCs.
Considering the fusion of boundary defects and their braiding with condensed anyons, we encounter a defect-version of the consistency equations for the additional VLCs, akin to \Eqs{2}{3}.
Consequently, for an irreducible additional $A$-$A$-bimodules $(M_{i_{\bg}}, \rho^L_{i_{\bg}}, \rho^R_{i_{\bg}})$, we can use the additional VLCs to construct operation tensors, as shown
\begin{equation}
\begin{gathered}
    [\rho_{\alpha}^L(a, \mu)]^{m_{\bg}, i_{\bg}, \mu_{i_{\bg}}}_{m'_{\bg}, i_{\bg}, \mu'_{i_{\bg}}} = [\phi^{a m_{\bg}; i_{\bg}}_{m'_{\bg}, \alpha; A i_{\bg}}]^{\mu \mu_{i_{\bg}}}_{\mu'_{i_{\bg}}}, \\
    [\rho_{\alpha}^R(a, \mu)]^{m_{\bg}, i_{\bg}, \mu_{i_{\bg}}}_{m'_{\bg}, i_{\bg}, \mu'_{i_{\bg}}} = [\phi^{m_{\bg} a; i_{\bg}}_{m'_{\bg}, \alpha; i_{\bg} A}]^{\mu_{i_{\bg}} \mu}_{\mu'_{i_{\bg}}}, \label{Operation(defect)}
\end{gathered}
\end{equation}
where $a$ is a condensed anyon.
In particular, among the additional $A$-$A$-bimodules, for a given element $\bg$, there exists a special irreducible $\bg$-local $A$-$A$-bimodules~\cite{bischoff2019spontaneous}, denoted as $(A_{\bg}, \rho^L_{A_{\bg}}, \rho^R_{A_{\bg}})$.
The object $A_{\bg}$ consists of defects capable of traversing the gapped boundary to manifest as a $\bg$-defect on the opposite side, referred to as $\bg$-local defects.
Since the operation tensors of additional $A$-$A$-bimodules are constructed from additional VLCs as \Eq{Operation(defect)}, the $\bg$-local conditions can be formulated as consistency equations for the VLCs.
These equations serve to filtrate the $\bg$-local defects $a_{\bg}$, as illustrated in Fig.~\ref{fig:local}.
For identity element $\be$, due to \Eq{2}, we note that associated $\be$-local $A$-$A$-bimodule $A$ satisfies the $\be$-local condition automatically.
\noindent 

\begin{figure}[H]
	\centering
	\includegraphics[width=0.42\textwidth, height=3.2cm]{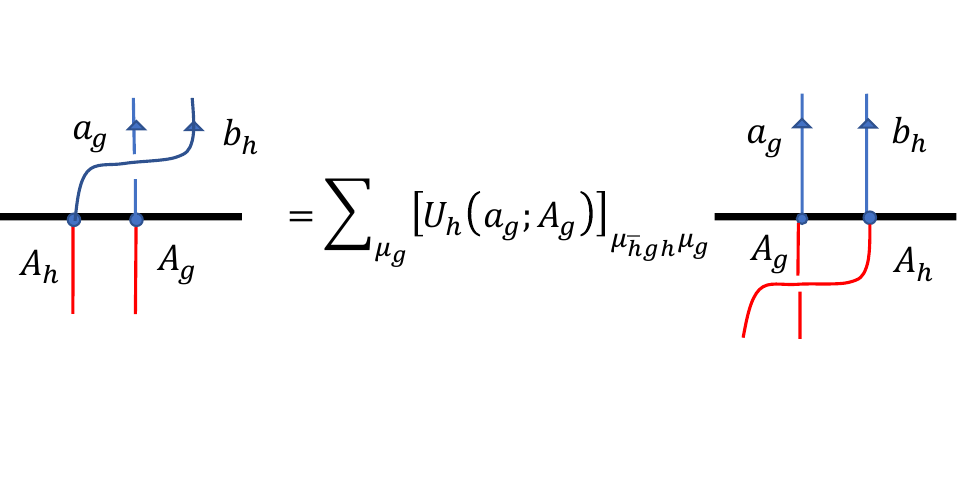}
	\caption{Additional constraints for $\bg$-local defects}
	\label{fig:DEq(braiding)}
\end{figure}

\begin{figure}[H]
	\centering
	\includegraphics[width=0.4\textwidth, height=0.2\textwidth]{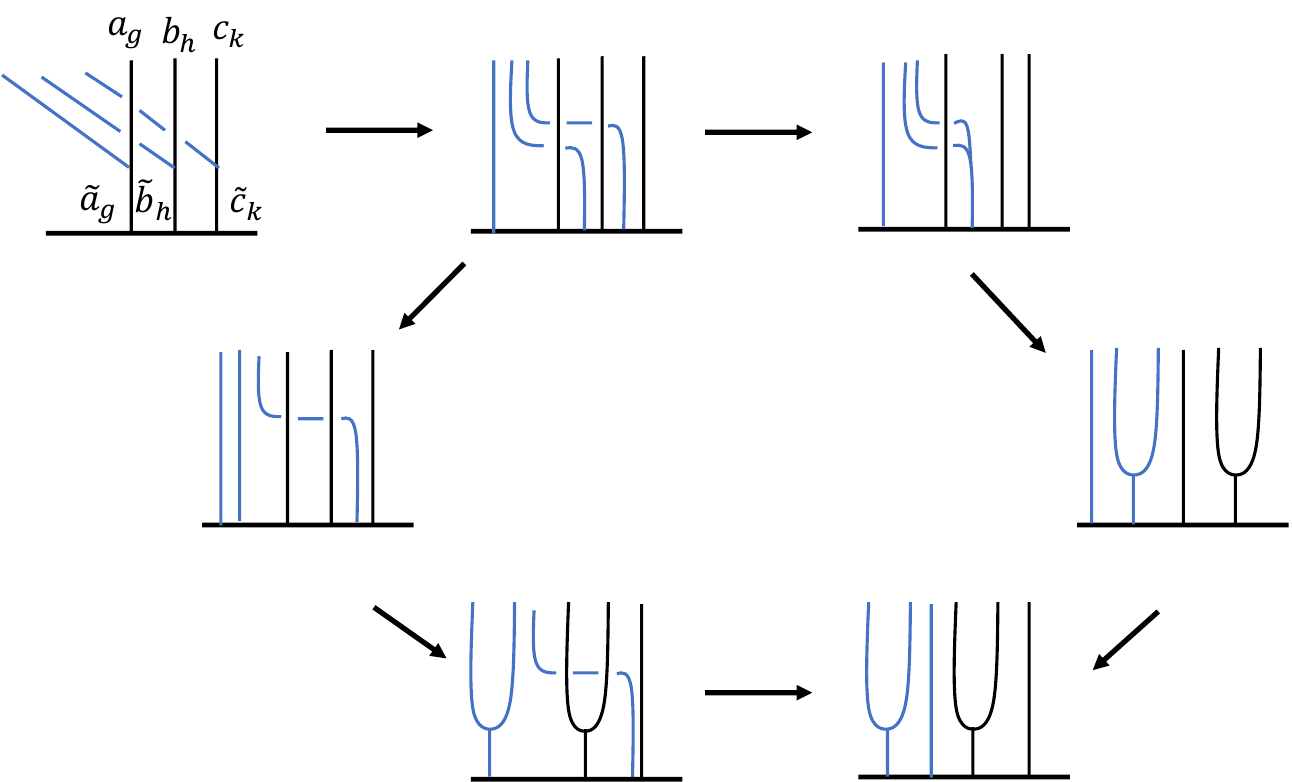}
	\caption{The appearance of the relative $H^3$ obstruction}
	\label{fig:H3Anomaly}
\end{figure}

In analogy to the condensed spaces, for $\bg$-local bimodules, we introduce symmetry actions on the associated topological spaces $V^{a_{\bg}}_{A_{\bg}}$, defined as
\begin{widetext}
\begin{align}
	\rho_{\bk}\ket{a_{\bg}; A_{\bg}, \mu_{\bg}} = \sum_{\mu_{\bk \bg \bar{\bk}}}[U_{\bk}({^{\bk} a}_{\bg}; A_{\bk \bg \bar{\bk}})]_{\mu_{\bg} \mu_{\bk \bg \bar{\bk}}} \ket{{{^{\bk}a}_{\bg}}; A_{\bk \bg \bar{\bk}}, \mu_{\bk \bg \bar{\bk}}},
\end{align}
where $[U_{\bk}(a_{\bg}; A_{ \bg })]_{\mu_{\bg} \mu_{\bk \bg \bar{\bk}}}$ represent boundary $U$-symbols associated with $\bg$-local defects.
For the boundary $U$-symbols of $\bg$-local defects, considering the unitarity and the group homomorphism of symmetry operators, we derive consistency equations
\begin{equation}
\begin{gathered}
	[\phi^{a_{\bg} b_{\bh}}_{c_{\bg \bh}, \alpha}]^{\mu_{\bg} \mu_{\bh}}_{\mu_{\bg \bh}} = \sum_{\substack{\mu_{\bar{\bk} \bg \bk}, \mu_{\bar{\bk} \bh \bk} \\ \beta, \mu_{\bar{\bk} \bg \bh \bk} } } \frac{[U_{\bk}(c_{\bg \bh}; A_{ \bg \bh})]_{ \mu_{\bar{\bk} \bg \bh \bk} \mu_{\bg \bh} } }{ [U_{\bk}(a_{\bg}; A_{ \bg })]_{ \mu_{\bar{\bk} \bg \bk } \mu_{\bg} } [U_{\bk}(b_{\bh}; A_{ \bh })]_{\mu_{\bar{\bk} \bh \bk} \mu_{\bh} } } \frac{[U_{\bk}(a_{\bg}, b_{\bh}; c_{\bg \bh} )]_{\beta \alpha}}{U_{\bk}(A_{\bg}, A_{\bh}; A_{\bg \bh})} [\phi^{{^{\bar{\bk}}a}_{\bg} {^{\bar{\bk}}b}_{\bh}}_{{^{\bar{\bk}}c}_{\bg \bh}, \beta}]^{\mu_{\bar{\bk} \bg \bk} \mu_{\bar{\bk} \bh \bk}}_{\mu_{\bar{\bk} \bg \bh \bk}}, \\
	\frac{\eta_{a_{\bk}}(\bg, \bh) }{\eta_{A_{\bk}}(\bg, \bh)} = \frac{[U_{\bg \bh}(a_{\bk}; A_{ \bk })]_{\mu_{ \bar{\bh} \bar{\bg} \bk \bg \bh} \mu_{\bk}  } }{ \sum_{\mu_{\bar{\bg}\bk \bg}} [U_{\bg}(a_{\bk}; A_{ \bk })]_{ \mu_{\bar{\bg}\bk \bg } \mu_{\bk} } [U_{\bh}({^{\bar{\bg}} a}_{\bk}; A_{\bar{\bh} \bk \bh})]_{\mu_{\bar{\bg}\bk \bg} \mu_{ \bar{\bh} \bar{\bg} \bk \bg \bh}} } \label{17}, 
\end{gathered}
\end{equation}
where the notation $[\phi^{a_{\bg} b_{\bh}}_{c_{\bg \bh}, \alpha}]^{\mu_{\bg} \mu_{\bh}}_{\mu_{\bg \bh}}$ represents a simplified form of $[\phi^{a_{\bg} b_{\bh}; A_{\bg \bh}}_{c_{\bg \bh}, \alpha; A_{\bg} A_{\bh}}]^{\mu_{\bg} \mu_{\bh}}_{\mu_{\bg \bh}}$.
Additionally, $[U_{\bk}(a_{\bg}, b_{\bh}; c_{\bg \bh} )]_{\beta \alpha}$ and  $\eta_{a_{\bk}}(\bg, \bh)$ represent the $U$-symbols and $\eta$-symbols of the bulk SET, respectively.
Similarly, $U_{\bk}(A_{\bg}, A_{\bh}; A_{\bg \bh})$  and $\eta_{A_{\bk}}(\bg, \bh)$ denote the $U$-symbols and $\eta$-symbols on the opposite side of the boundary. 
It is known that the opposite side, which only includes defects, is treated as a $G$-SPT phase.
This $G$-SPT phase is characterized by $F$-symbols of the defects, denoted as $F^{A_{\bg} A_{\bh} A_{\bk}}_{A_{\bg \bh \bk}}$, which are related to $F$-symbols of the bulk SET, shown as
\begin{align}
	\sum_{e_{\bg \bh}, \mu_{\bg \bh}, \alpha, \beta} [\phi^{a_{\bg} b_{\bh}}_{e_{\bg \bh}, \alpha}]^{\mu_{\bg} \mu_{\bh}}_{\mu_{\bg \bh}} [\phi^{e_{\bg \bh} c_{\bk}}_{d_{\bg \bh \bk}, \beta}]^{\mu_{\bg \bh} \mu_{\bk}}_{\mu_{\bg \bh \bk}} [F^{a_{\bg} b_{\bh} c_{\bk}}_{d_{\bg \bh \bk}}]_{(e_{\bg \bh}, \alpha, \beta)(f_{\bh \bk}, \rho, \sigma)} & [F^{A_{\bg} A_{\bh} A_{\bk}}_{A_{\bg \bh \bk}}]^{-1} = \notag \\
	  &\sum_{\mu_{\bh \bk}} [\phi^{b_{\bh} c_{\bk}}_{f_{\bh \bk}, \rho}]^{\mu_{\bh} \mu_{\bk}}_{\mu_{\bh \bk}} [\phi^{a_{\bg} f_{\bh \bk}}_{d_{\bg \bh \bk}, \sigma}]^{\mu_{\bg} \mu_{\bh \bk}}_{\mu_{\bg \bh \bk}},
	  \label{18}
\end{align}
\end{widetext}
We note that the $F$-symbols of defects $F^{A_{\bg} A_{\bh} A_{\bk}}_{A_{\bg \bh \bk}}$, can be interpreted as a cocycle $\alpha \in Z^3[G, \mathrm U(1)]$.
When $\alpha(\bg, \bh, \bk)$ can be expressed as a coboundary $d\omega(\bg, \bh, \bk)$, we can adjust the VLCs as follows: $[\tilde{\phi}^{a_{\bg} b_{\bh}}_{c_{\bk}}] = \omega(\bg, \bh)[\tilde{\phi}^{a_{\bg} b_{\bh}}_{c_{\bk}}]$.
This adjustment leads to the disappearance of the $F$-symbols $F^{A_{\bg} A_{\bh} A_{\bk}}_{A_{\bg \bh \bk}}$, as indicated in \Eq{17}.
Thus, when the cocycle $\alpha$ is viewed as an obstruction to \Eq{17}, it should belong to $H^3[G, \mathrm U(1)]$, referred to as the $H^3$ obstruction.
On the opposite side, when we appropriately choose the gauge to fix the $R$-symbols as $R^{A_{\bg} A_{\bh}}_{A_{\bg \bh}} = 1$, the $U$-symbols and $\eta$-symbols are determined by the cocycle $\alpha$~\cite{barkeshli2019symmetry}.
Additionally, considering the braiding between $\bg$-local defects, there are other constraints on the associated additional VLCs, as illustrated in Fig.~\ref{fig:DEq(braiding)}.
\noindent
That is
\begin{gather}
      \frac{ \sum_{\beta} [\phi^{b_{\bh} {^{\bar{\bh}}a}_{\bg}}_{e_{\bg \bh}, \beta}]^{\mu_{\bh} \mu_{\bar{\bh} \bg \bh }}_{\mu_{\bg \bh}} [R^{a_{\bg} b_{\bh}}_{e_{\bg \bh}}]_{\beta \alpha}	} {\sum_{\mu_{\bg}} [U_{\bh}(a_{\bg}; A_{ \bg })]_{\mu_{\bar{\bh} \bg \bh} \mu_{\bg}} [\phi^{a_{\bg} b_{\bh}}_{e_{\bg \bh}, \alpha}]^{\mu_{\bg} \mu_{\bh}}_{\mu_{\bg \bh}}} = 1 
     \label{24}.
\end{gather}
For a special case $\bh = \be$, we note that \Eq{24} turns into the defect version of \Eq{3}.
In general, determining the $H^3$ obstruction involves solving \Eqss{17}{24}, which is an extremely challenging task.
Fortunately, drawing on the heuristic definition of the relative $H^4$ anomaly in Ref.~\cite{barkeshli2020relative}, we can, in this case, define a relative $H^3$ obstruction $\alpha_r$, which is more computationally feasible.
Beginning with an initial system where the $H^3$ obstruction vanishes on the $G$-symmetric boundary, we examine the modification of the original $G$-symmetric boundary by a torsor $\mathfrak{v} \in H^1[G, [\cA_{\cC}]_{\cA_{A}}]$.
For the modified boundary, the torsor $\mathfrak{v}$ not only modify the boundary $U$-symbols, but also further modify the $\bg$-local defects as follows:
\begin{align}
	\tilde{a}_{\bg} = a_{\bg} \otimes \mathfrak{v}(\bg)
\end{align}
Considering the moves, corresponding to \Eq{18} for the modified $\bg$-local defects $\tilde{a}_{\bg}$, as illustrated in Fig.~\ref{fig:H3Anomaly}, these lead to  emergence of the relative $H^3$ obstruction.
\noindent
Here, the blue lines represent abelian anyon $\mathfrak{v}(\bg)$, while the black lines represent defects $a_{\bg}$ and $\tilde{a}_{\bg}$.
We emphasize that the $F$-symbols and the additional VLCs for modified $\bg$-local defects can be related to the $F$-symbols and the additional VLCs for the original $\bg$-local defects by a series of moves, as shown in Appendix~\ref{app:modified}.
After disregarding the redundant moves, the modified consistency equations can be simplified and illustrated in Fig.~\ref{fig:H3Anomaly}.
Thus, we define the relative $H^3$ obstruction as:
\begin{widetext}
\begin{align}
	\alpha_r(\bg, \bh, \bk) = \frac{\eta_{{^{\bg \bh}\mathfrak{v}(\bk)}}(\bg, \bh)}{U_{\bg}({^{\bg}\mathfrak{v}(\bh)}, {^{\bg \bh}\mathfrak{v}(\bk)})}\frac{\epsilon_{\bg}(\bh) \epsilon_{\bg \bh}(\bk)}{\epsilon_{\bh}(\bk) \epsilon_{\bg}(\bh \bk)} \frac{\phi^{\mathfrak{v}(\bg) {^{\bg}\mathfrak{v}(\bh)} } \phi^{\mathfrak{v}(\bg \bh) {^{\bg \bh}\mathfrak{v}(\bk)} } }{\phi^{\mathfrak{v}(\bh) {^{\bh}\mathfrak{v}(\bk)} }  \phi^{\mathfrak{v}(\bg) {^{\bg}\mathfrak{v}(\bh \bk)} } } F^{\mathfrak{v}(\bg) {^{\bg}\mathfrak{v}(\bh)}  {^{\bg \bh}\mathfrak{v}(\bk)} } ;
	\label{ReH3}
\end{align}
\end{widetext}
where $\epsilon_{\bg}(\bh) = \frac{\phi^{a_{\bg} \mathfrak{v}(\bh)} }{\phi^{{^{\bg}\mathfrak{v}(\bh) a_{\bg}} }} R^{{^{\bg}\mathfrak{v}(\bh) } a_{\bg}}$ and $\mathfrak{v}(\bg \bh) = \mathfrak{v}(\bg) {^{\bg}\mathfrak{v}(\bh) }$  .
Here, since the fusion channels and the channels transforming into boundary excitations associated with abelian anyons are unique, we omit some indices in \Eq{ReH3}.
For instance, $\phi^{a_{\bg} \mathfrak{v}(\bh) }$ and $U_{\bg}({^{\bg}\mathfrak{v}(\bh)}, {^{\bg \bh}\mathfrak{v}(\bk)})$ represent $[\phi^{a_{\bg} \mathfrak{v}(\bh); i_{\bg} }_{a'_{\bg}, A_{\bg} i}]^{\mu_{\bg} 1}_{\mu'_{\bg}} $ and bulk $U$-symbols $U_{\bg}({^{\bg}\mathfrak{v}(\bh)}, {^{\bg \bh}\mathfrak{v}(\bk)}; {^{\bg}\mathfrak{v}(\bh \bk)})$, respectively.
In \Eq{ReH3}, the unknown quantities are $\phi^{\mathfrak{v}(\bh) a_{\bg} }$ and $\phi^{a_{\bg}{^{\bg}\mathfrak{v}(\bh) } }$.
Both of them can be obtained by \Eq{2}, which involves only condensed anyons $a$, abelian anyons $\mathfrak{v}(\bg)$ and a single $\bg$-local defect $a_{\bg}$.
This procedure is much simpler than solving \Eqss{17}{24}.

At the $G$-symmetric boundary, when the $H^3$ obstruction vanishes, the $U$-symbols and $\eta$-symbols of the $G$-SPT phase are all trivial.
Thus, the additional VLCs of $\bg$-local defects can be obtained by \Eqss{17}{24}.
Observing these equations, the VLCs $[\phi^{a_{\bg} b_{\bh}}_{c_{\bg \bh}, \alpha}]$ can be modified by a torsor $\beta \in H^2[G, \mathrm U(1)]$ as 
\begin{align}
	[\tilde{\phi}^{a_{\bg} b_{\bh}}_{c_{\bg \bh}, \alpha}] = \beta(\bg, \bh)[\phi^{a_{\bg} b_{\bh}}_{c_{\bg \bh}, \alpha}].
\end{align} 
Meanwhile, the boundary $U$-symbols are modified as follows:
\begin{gather}
     [\tilde{U}_{\bh}(a_{\bg}; A_{ \bg })] = [U_{\bh}(a_{\bg}; A_{ \bg })] \frac{\beta(\bh, \bar{\bh} \bg \bh)}{\beta(\bg, \bh)}.\label{Twist BdryU}
\end{gather}
Thus, we argue that the additional VLCs $[\phi^{a_{\bg} b_{\bh}}_{c_{\bg \bh}, \alpha}]$ characterize partial data of the $G$-symmetric gapped boundary~\footnote{In the rest of paper, without ambiguity, we use additional VLCs to specifically refer to as the VLCs $[\phi^{a_{\bg} b_{\bh}}_{c_{\bg \bh}, \alpha}]$, associated with $\bg$-local defects.}.
Mathematically, by combining the data of the $\bg$-local $A$-$A$-bimodule, the grand algebra $(\textbf{A}, \tilde{\phi})$ over $\cC^{\times}_{G}$ can be constructed as follows:~\cite{meir2012module}
\begin{equation}
\begin{gathered}
	\textbf{A} = \oplus_{\bg} A_{\bg}, \\
	[\tilde{\phi}^{a_{\bg} b_{\bh}}_{c_{\bg \bh}, \alpha}] = [\phi^{a_{\bg} b_{\bh}}_{c_{\bg \bh}, \alpha}],
\end{gathered}
\end{equation}
which characterizes the $G$-symmetric gapped boundary.
Based on the discussion above, the grand algebra are constructed from a triplet $(A, U_{\bg}(a), [\phi^{a_{\bg} b_{\bh}}_{c_{\bg \bh}, \alpha}])$.
In other word, when the $G$-symmetric gapped boundary is anomaly-free,  it is determined by the triplet $(A, U_{\bg}(a), [\phi^{a_{\bg} b_{\bh}}_{c_{\bg \bh}, \alpha}])$ and classified by $H^1_{\rho}[G,[\cA_{\cC}]_{\cA_A}]$ and $H^2[G, \mathrm U(1)]$. 

\subsection{The enrichment of a junction by a onsite $G$-symmetry}
Then, we study a $G$-symmetric junction between two non-anomalous $G$-symmetric gapped boundaries, characterized by $(A_1, U^{(1)}_{\bg}(a), [\phi^{a_{\bg} b_{\bh}}_{c_{\bg \bh}, \alpha}])$ and  $(A_2, U^{(2)}_{\bg}(a), [\varphi^{a_{\bg} b_{\bh}}_{c_{\bg \bh}, \alpha}])$, respectively. 
When the $G$-symmetric junction is anomaly-free, we argue that it is characterized by a grand fundamental $\textbf{A}_1$-$\textbf{A}_2$-bimodule $(\textbf{X}, \rho^L_{\textbf{X}}, \rho^R_{\textbf{X}})$.
In order to fine the compatible data to construct the grand fundamental bimodule, we examine the enrichment of an intrinsic junction between the $A_1$-type boundary and the $A_2$-type boundary.

For the intrinsic junction, considering \Eq{9} for gauge invariant special $L$-symbols, we derive
\begin{align}
	\hL^c \phi^{ab}_{c, \alpha} = \hL^a \hL^b \varphi^{a b}_{c, \alpha}
	\label{19}.
\end{align}  
Because of the unitarity of symmetry operators, a constraint on the special $L$-symbols is derived
\begin{align}
	U^{(1)}_{\bg}(a)\hL^{a} = U^{(2)}_{\bg}(a)\hL^{\ubbg a},
	\label{20}
\end{align}
where $U^{(i)}_{\bg}(a)$ are $A_i$-type boundary $U$-symbols.
In principle, we can obtain $L$-symbols by solving \Eq{19} and then select proper solutions by \Eq{20}.
However, the proper solutions may not always exist. 
Thus, we define an obstruction function to diagnose the existence of the proper solutions
\begin{align}
	O_a(\bg) = \frac{U^{(1)}_{\bg}(a)}{U^{(2)}_{\bg}(a)}\frac{\hL^a}{\hL^{\ubbg a}}
	\label{21}.
\end{align}
Employing \Eqs{13}{14}, we derive
\begin{align}
	O_a(\bg)O_b(\bg) &= O_c(\bg), \\
	O_{\ubbg a}(\bh)O_a(\bg) &= O_a(\bg \bh),
\end{align}
whenever fusion coefficients of common condensed anyons satisfy $N^c_{a b} \neq 0 $.
Moreover, the factor $\frac{\hL^a}{\hL^{\ubbg a}}$ of $O_a(\bg)$ can be changed by a coboundary $\frac{M^*_{a \tau}}{M^*_{a \rho_{\bg}(\tau)}}$ where $\tau \in C^0_{\rho}[G, [\cA_{\cC}]_{\cA_{A_1} \vee \cA_{A_2}}]$.
Here, for the two fusion subcategories $\cA_{A_1}$ and $\cA_{A_1}$, the smallest fusion subcategory of $\cC$ that contains both $\cA_{A_1}$ and $\cA_{A_1}$ is denoted as $\cA_{A_1} \vee \cA_{A_2}$~\cite{etingof2015tensor}.
The equivalence relation of $[\cA_{\cC}]_{\cA_{A_1} \vee \cA_{A_2}}$ is determined by the fusion subcategory $\cA_{A_1} \vee \cA_{A_2}$.
Thus, the obstruction function $O_a(\bg)$ can be written as a phase $M^*_{a \cO(\bg)}$, where
\begin{gather}
    \cO \in H^1_{\rho}[G, [\cA_{\cC}]_{\cA_{A_1} \vee \cA_{A_2}}], 	
\end{gather}
so we refer to as $O_a(\bg)$ as the $H^1$ obstruction.
We will discuss an anomalous example with the $H^1$ obstruction in Sec.\ref{sec:H1}. 
Without loss of generality, we set $O_a(\bg) = 1$ for non-anomalous cases $\cO = [0]$.
Then, special $L$-symbols $\hL^a$ can be modified by a factor $\chi(a)$, which satisfies the following conditions:
\begin{equation}
\begin{gathered}
	\chi(a)\chi(b)=\chi(c),\\
	\chi(a)=\chi(\ubbg a),
\end{gathered}
\end{equation}
where $N^c_{a b} \neq 0$.
Thus, the factor $\chi(a)$ can be written as $M^*_{a\mathfrak{c}}$, where
\begin{gather}
    \mathfrak{c} \in  [\cA_{\cC}]^G_{\cA_{A_1} \vee \cA_{A_2}},
\end{gather}
where $[\cA_{\cC}]^G_{\cA_{A_1} \vee \cA_{A_2}}$ represents a fusion category, whose simple objects consist of the equivalence classes in $[\cA_{\cC}]_{\cA_{A_1} \vee \cA_{A_2}}$ that remain invariant under the actions of the $G$-symmetry.
In addition, as discussed in the Sec.\ref{sec:junction}, there are \textit{topologically equivalent relations} for junctions.
In this case, due to the presence of symmetry, only the string operators, that preserve the $G$-symmetry, are allowable to be used to relate topologically equivalent junctions.
Specifically, we examine string operators $\hW_{c_1}$ along the $A_1$-type boundary and string operators $\hW_{c_2}$ along the $A_2$-type boundary, each with one endpoint localized at the junction.
Here, $[c_1] \in [\cA_{\cC}]^G_{\cA_{A_1}}$ and $[c_2] \in [\cA_{\cC}]^G_{\cA_{A_2}}$. 
Since the special $L$-symbols only involve the common condensed anyons, the nonequivalent modifications resulting from the string operators $\hW_{c_1}$ and $\hW_{c_2}$ are classified by two fusion categories: $[[\cA_{\cC}]^G_{\cA_{A_1}}]_{\cA_{A_2}}$ and $[[\cA_{\cC}]^G_{\cA_{A_2}}]_{\cA_{A_1}}$.
Apparently, both $[[\cA_{\cC}]^G_{\cA_{A_1}}]_{\cA_{A_2}}$ and $[[\cA_{\cC}]^G_{\cA_{A_2}}]_{\cA_{A_1}}$ are fusion subcategories of $[\cA_{\cC}]^G_{\cA_{A_1} \vee \cA_{A_2}}$.
Consequently, the topologically inequivalent special $L$-symbols are classified by the quotient group: $ [\cA_{\cC}]^G_{\cA_{A_1} \vee \cA_{A_2}}/([[\cA_{\cC}]^G_{\cA_{A_1}}]_{\cA_{A_2}} \vee [[\cA_{\cC}]^G_{\cA_{A_2}}]_{\cA_{A_1}})$.

Next, we study $\bg$-local symmetry defects on the two $G$-symmetric boundaries.
At the junction, we assume that the $H^1$ obstruction vanishes.
Then, for each $\bg$-local symmetry defect $a_{\bg}$ on the $A_1$-type boundary, considering only the common condensed anyons, we can verify that the $\bg$-local condition also holds on $A_2$-type boundary using \Eq{20}.
Thus, we conjecture that common $\bg$-local symmetry defects must exist on the two boundaries, when the $H^1$ obstruction vanishes.
In fact, we can prove the conjecture for the cases that the anyons are all abelian, as shown in Appendix~\ref{app:com-def}.
Based on this, common $\bg$-local symmetry defect can move to the junction to produce additional defects, which are described by additional $A_1$-$A_2$-bimodules $(X_{\bg}, \rho^L_{X_{\bg}}, \rho^R_{X_{\bg}})$.
The object $X_{\bg}$ can be obtained by objects of additional irreducible $\bg$-local bimodules $(A^1_{\bg}, \rho^L_{A^1_{\bg}}, \rho^R_{A^1_{\bg}})$ and $(A^2_{\bg}, \rho^L_{A^2_{\bg}}, \rho^R_{A^2_{\bg}})$, shown as
\begin{align}
	X_{\bg} = A^1_{\bg} \otimes_{A_1} X 
	        = X \otimes_{A_2} A^2_{\bg},
\end{align}
where $\otimes_{A_i}$ is tensor product of $A_i$-$A_i$ bimodule category~\cite{kong2014anyon}.
It leads to two decompositions
\begin{equation}
	X_{\bg} = \oplus_{i_{\bg}} W^{(1)}_{i_{\bg}} M^{(1)}_{i_{\bg}} = \oplus_{j_{\bg}} W^{(2)}_{j_{\bg}} M^{(2)}_{j_{\bg}} ,	
\end{equation}
where $M^{(k)}_{i_{\bg}}$ is an irreducible additional $A_{k}$-$A_{k}$ bimodule and $W^{(k)}_{i_{\bg}}$ is a non-negative integer.
We note that $W^{(1)}_{A^1_{\bg}} = W^{(2)}_{A^2_{\bg}}=1$.
The object of the grand fundamental $\textbf{A}_1$-$\textbf{A}_2$-bimodule can be constructed as
\begin{align}
	\textbf{X} = \oplus_{\bg} X_{\bg},
\end{align}
where $X_{\be} = X$.
Furthermore, we can also define additional $L$-symbols by basis transformations of topological spaces $\text{hom}_{\cC^{\times}_{G}}(m_{\bg}, \textbf{X})$, shown as
\begin{align}
	&\ket{m_{\bg}; i_{\bg}, \mu_{i_{\bg}}, n^{(1)}_{i_{\bg}}}_{\textbf{X}} = \notag \\
	&\sum_{j_{\bg}, \mu_{j_{\bg}}, n^{(2)}_{j_{\bg}}}  [L^{m_{\bg}}_{\textbf{X}}]_{(i_{\bg}, \mu_{i_{\bg}})(j_{\bg}, \mu_{j_{\bg}})} \ket{m_{\bg}; j_{\bg}, \mu_{j_{\bg}}, n^{(2)}_{j_{\bg}}}_{\textbf{X}},
\end{align}
where $n^{(k)}_{i} = 1,\dots,W^{(k)}_{i}$.
When $\bg = \be$, the additional $L$-symbols $[L^{m_{\bg}}_{\textbf{X}}]_{(i_{\bg}, \mu_{i_{\bg}})(j_{\bg}, \mu_{j_{\bg}})}$ reduce to $L$-symbols $[L^{m}_{X}]_{(i, \mu_{i})(j, \mu_{j})}$.
Similar to the bases of $\text{hom}_{\cC}(a, X)$, the bases of $\text{hom}_{\cC^{\times}_{G}}(m_{\bg}, \textbf{X})$ are subject to gauge transformations in precisely the same manner as those listed in~(\ref{gt}).
Since both $A^1_{\bg}$ and $A^2_{\bg}$ represent same $\bg$-defect after anyon condensation, they are identical on the other side.
In this case, the phase factors of gauge transformations should satisfy $\zeta^{a_{\bg}}_{A^1_{\bg}} = \zeta^{a_{\bg}}_{A^2_{\bg}}$.
As a result, for common $\bg$-local symmetry defects $a_{\bg}$, the special additional $L$-symbols $[L^{a_{\bg}}_{\textbf{X}}]_{(A^1_{\bg}, \mu)(A^1_{\bg}, \nu)}$ remain gauge invariant.
Furthermore, the additional $L$-symbols should be compatible with boundary defects fusion on the two boundaries in same way that we have shown in Fig.~\ref{fig:Eq(L)}.
It results in consistency equations similar to \Eq{9}.
We list partial equations, which only involve special additional $L$-symbols $[L^{a_{\bg}}_{\textbf{X}}]_{(A^1_{\bg}, \mu)(A^2_{\bg}, \nu)}$.
That is
\begin{widetext}
\begin{gather}
  \sum_{\mu_{\bg \bh}} [\phi^{a_{\bg} b_{\bh}}_{c_{\bg \bh}, \alpha}]^{\mu_{\bg} \mu_{\bh}}_{\mu_{\bg \bh}} \hL^{c_{\bg \bh}}_{\mu_{\bg \bh} \mu'_{\bg \bh}} = \sum_{\mu'_{\bg}, \mu'_{\bh}}  \hL^{a_{\bg}}_{\mu_{\bg} \mu'_{\bg}} \hL^{b_{\bh}}_{\mu_{\bh} \mu'_{\bh}} [\varphi^{a_{\bg} b_{\bh}}_{c_{\bg \bh}, \alpha}]^{\mu'_{\bg} \mu'_{\bh}}_{\mu'_{\bg \bh}}
    \label{26}	.
\end{gather}
\end{widetext}
Here, we denote the additional special $L$-symbols $[L^{a_{\bg}}_{\textbf{X}}]_{(A^1_{\bg}, \mu)(A^2_{\bg}, \nu)}$ as $\hL^{a_{\bg}}_{\mu \nu}$.

\begin{figure}[H]
	\centering
	\includegraphics[width=0.4\textwidth, height=0.2\textwidth]{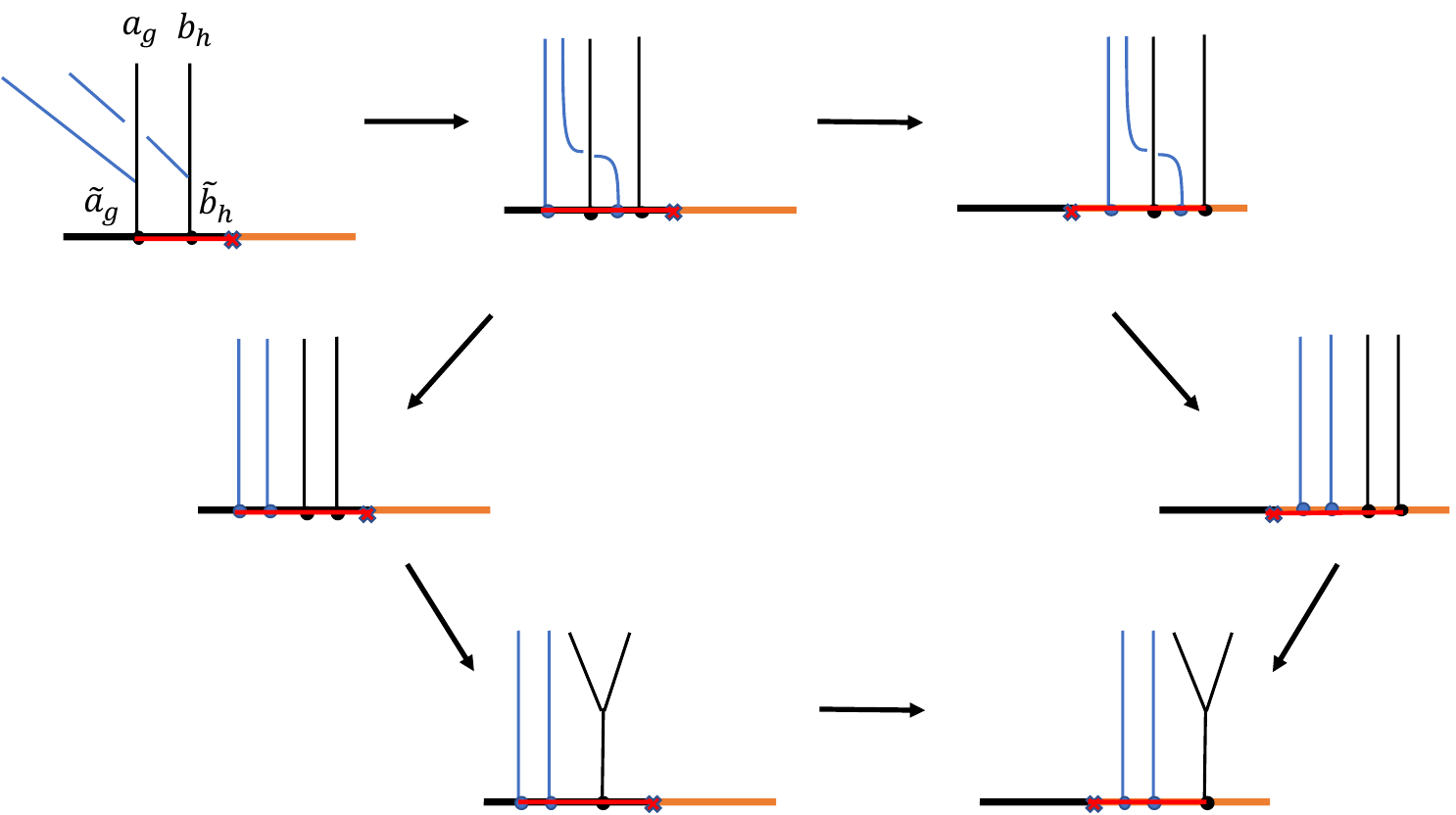}
	\caption{The appearance of the relative $H^2$ obstruction}
	\label{fig:H2Anomaly}
\end{figure}

However, for \Eq{26}, there may be no compatible solutions of the additional $L$-symbols.
In general, we do not yet know how to define the appropriate obstruction, which can characterizes the inconsistency of \Eq{26}.
Fortunately, similar to the $H^3$ obstructions on boundaries , in this case, we can define a relative $H^2$ obstruction $\gamma_r$ to characterize the inconsistency of \Eq{26}.
As previously discussed, there are five torsors $\beta_1$, $\beta_2$, $[\mathfrak{v}^{1}]$, $[\mathfrak{v}^{2}]$, and $[\mathfrak{c}]$ in the system.
Here, we require that the torsors $[\mathfrak{v}^{1}]$ and $[\mathfrak{v}^{2}]$ satisfy $M^*_{a\mathfrak{v}^{1}(\bg)} = M^*_{a\mathfrak{v}^{2}(\bg)}$ for every common condensed anyon $a$, which ensure that \Eq{21} remains invariant.
The contribution of the torsors $\beta_1$ and $\beta_2$ is obvious for $\gamma_r$ , which is a factor $\frac{\beta_1(\bg, \bh)}{\beta_2(\bg, \bh)}$.
Then, we focus on the last three torsors.
The torsor $\mathfrak{v}^{(i)}$ will alter the common $\bg$-local defects as
\begin{gather}
    \tilde{a}_{\bg} = a_{\bg} \otimes \mathfrak{v}(\bg).
\end{gather}
Here, since the $H^1$ obstruction vanishes at the junction, we can choose $\mathfrak{v}(\bg) \in [\mathfrak{v}^{1}(\bg)]_{\cA_{A_1}}$ and $\mathfrak{v}(\bg) \in [\mathfrak{v}^{2}(\bg)]_{\cA_{A_2}}$.
We emphasize that the attached abelian anyon $\mathfrak{v}(\bg)$ takes values in the equivalence class $[\mathfrak{v}(\bg)]_{\cA_{A_1} \cap \cA_{A_2}}$, where $\cA_{A_1} \cap \cA_{A_2}$ denotes the set of the common abelian condensed anyons on the two boundaries.
In addition, according to \Eq{9}, the torsor $\mathfrak{c}$ modify the $L$-symbols as
\begin{equation}
	[L^{m}_{X}]_{(i, \mu_{i})(j, \mu_{j})} = [\tilde{L}^{m}_{X}]_{(i, \mu_{i})(j, \mu_{j})} M_{m \mathfrak{c}}.
\end{equation}
In this case, for additional special $L$-symbols, the modification is
\begin{gather}
    	\hL^{a_{\bg}}_{\mu \nu} = \sum_{\nu', \nu''} \tilde{\hL}^{a_{\bg}}_{\mu \nu'} [\varphi^{{^{\bar{\bg}}\mathfrak{c} } a_{\bg} }_{c_{\bg}}]^{1 \nu'}_{\nu''}  [(\varphi^{\mathfrak{c} a_{\bg}}_{c_{\bg}})^{-1} ]^{1\nu}_{\nu''} R^{a_{\bg} {^{\bar{\bg}}\mathfrak{c} }} R^{\mathfrak{c} a_{\bg}},
\end{gather}
which ensure that the consistency equations \Eq{26}, which involve only a single defect and a common condensed anyon on the right side, are satisfied.
For the original model, we suppose that the associated \Eq{26} are consistency.
Then, we analyze the moves that correspond to the consistency equations \Eq{9} of a modified system, which involves the three torsors $\mathfrak{v}^{1}$, $\mathfrak{v}^{2}$, and $\mathfrak{c}$, illustrated in Fig.~\ref{fig:H2Anomaly}.
Here, the blue lines represent abelian anyon $\mathfrak{v}(\bg)$, while the black lines represent common $\bg$-local defects $a_{\bg}$ and $\tilde{a}_{\bg}$.
\noindent
Here, similar to the derivation of the relative $H^3$ obstruction on a symmetric boundary, the additional VLCs of the modified $\bg$-local defects can be related to the additional VLCs of the origial $\bg$-local defects, as shown in Appendix~\ref{app:modified}.
Therefore, after disregarding the redundant moves, the consistency equations for the additional special $L$-symbols of the modified system can be illustrated in Fig.~\ref{fig:H2Anomaly}.
It leads to the definition of the relative $H^2$ obstruction:
\begin{align}
	\gamma_{r}(\bg, \bh) = \eta_{\mathfrak{c}}(\bg, \bh) \frac{M^*_{{^{\bg} \mathfrak{v}(\bh)} \mathfrak{c}}}{M^*_{\mathfrak{v}(\bh) \mathfrak{c}}} \frac{\sigma_{\bh}({^{\bar{\bg}}\mathfrak{c} })}{\sigma_{\bh}(\mathfrak{c} )}\epsilon_{\bg \bh}(\bg) \epsilon_{\bg \bh}(\bh) ,
	\label{Def.H^2}
\end{align}
where $\sigma_{\bh} = \frac{\varphi^{b_{\bh} {^{\bar{\bh}}\mathfrak{c} }} }{\varphi^{\mathfrak{c} b_{\bh}} }R^{\mathfrak{c}b_{\bh} } $ and $\epsilon_{\bg \bh}(\bg) = \frac{\varphi^{\mathfrak{c}^{\bg}_{\bg \bh}a_{\bg}} }{\varphi^{a_{\bg} \mathfrak{c}^{\bg}_{\bg \bh}}} R^{a_{\bg} \mathfrak{c}^{\bg}_{\bg \bh}} $, with $\mathfrak{c}^{\bg}_{\bg \bh} = {^{\bar{\bg}}\bar{\mathfrak{c}}}\otimes {^{\bar{\bh}\bar{\bg}}\mathfrak{c} } $.
The derivation of $\gamma_r(\bg, \bh)$ above is independent of the transformation channels of symmetry defects on the two boundaries.
It is easy to verify that the relative obstruction $\gamma(\bg, \bh)$ is invariant under the gauge transformations~(\ref{gt}).
Applying \Eqs{9}{18}, we can derive
\begin{align}
    	\frac{\gamma_r(\bg, \bh)\gamma_r(\bg \bh, \bk)}{\gamma_r(\bh, \bk)\gamma_r(\bg, \bh \bk)} = 1.
\end{align}
It implies that the relative obstruction $\gamma_r$ is a 2-cocycle.
Besides, since $\gamma_r(\be, \bg) = \gamma_r(\bg, \be) =1 $, in cases where $\gamma_r(\bg, \bh)$ can be regraded as a coboundary $\frac{\mu(\bg)\mu(\bh)}{\mu(\bg \bh)}$, we have $\mu(\be) = 1$.
This factor $\frac{\mu(\bg)\mu(\bh)}{\mu(\bg \bh)}$ can always be eliminated by adjusting additional $L$-symbols such that $\tilde{\hL}^{a_{\bg}}_{\mu \nu} = \mu_{\bg} \hL^{a_{\bg}}_{\mu \nu}$ in the consistency equations \Eq{26}, resulting in the vanishing of the relative obstruction.
Thus, after quotienting out the coboundary cases, the relative obstruction $\gamma$  should be an element of $H^2[G, \mathrm U(1)]$, which is the reason that $\gamma_r$ is dubbed $H^2$ relative anomaly.
Physically, this $H^2$ obstruction means that the $G$-symmetric junction is assigned to an irreducible projective $\gamma$-representation of $G$, which can localize at a boundary of 1D $G$-SPT phase, characterized by $\gamma \in H^2[G, \mathrm U(1)]$.
We will discuss three simple examples where the the $H^2$ obstruction appear at the $G$-symmetric junction in Sec.\ref{sec:H2}.
In fact, we can always adjust additional VLCs $[\phi^{a_{\bg} b_{\bh}}_{c_{\bg \bh}, \alpha}]$ and $[\varphi^{a_{\bg} b_{\bh}}_{c_{\bg \bh}, \alpha}]$ by two torsors $\beta_1, \beta_2 \in H^2[G, \mathrm U(1)]$ respectively such that $\gamma(g, h) = 1$.
In the case that $H^2$ obstruction of the $G$-symmetric junction vanishes, using \Eqs{24}{26}, we can derive
\begin{gather}
    \sum_{\sigma}[U^{(1)}_{\bg}(a_{\bk}; A^1_{\bk})]_{\mu \sigma} \hL^{a_{\bk}}_{\sigma \nu} = \sum_{\rho} [U^{(2)}_{\bg}(a_{\bk}; A^2_{\bk})]_{\mu \rho} \hL^{{^{\bg}a}_{\bk}}_{\rho \nu}	.
\end{gather}
It implies that the additional $L$-symbols are compatible with the boundary $U$-symbols of common $\bg$-local defects.
Furthermore, we note that $L$-symbols can be modified by a torsor $\nu \in H^1[G, \mathrm U(1)]$, so it implies that the solutions of \Eq{26} are classified by the cohomology group $H^1[G, \mathrm U(1)]$.
Then, similar to \Eq{15}, the obtained additional $L$-symbols and additional VLCs can be used to construct operation tensors of the grand fundamental $\textbf{A}_1$-$\textbf{A}_2$-bimodules $(\textbf{X}, \rho^L_{\textbf{X}}, \rho^R_{\textbf{X}})$ as follows:
\begin{widetext}
\begin{equation}
\begin{gathered}
	[\rho_{\alpha}^L(a_{\bh}, \mu)]^{m_{\bg}, i_{\bg}, \mu_{i_{\bg}}}_{m_{\bh \bg}, i_{\bh \bg}, \mu_{i_{\bh \bg}}} =  [\phi^{a_{\bh} m_{\bg}; i_{\bh \bg}}_{m_{\bh \bg}, \alpha; A^1_{\bh} i_{\bg}}]^{\mu_{\bh} \mu_{i_{\bg}}}_{\mu_{i_{\bh \bg}}}, \\
    [\rho_{\alpha}^R(b_{\bk}, \mu)]^{m_{\bg}, i_{\bg}, \mu_{i_{\bg}}}_{m_{\bg \bk}, i_{\bg \bk}, \mu_{i_{\bg \bk}}} = \sum_{j_{\bg}, \mu_{j_{\bg}}, \mu_{j_{\bg \bk}}} W^{(2)}_{j_{\bg}} [L^{m_{\bg \bk}}_{\textbf{X}}]^{-1}_{(i_{\bg \bk}, \mu_{i_{\bg \bk}})(j_{\bg \bk}, \mu_{j_{\bg \bk}})}[\varphi^{m_{\bg} b_{\bk}; j_{\bg \bk}}_{m_{\bg \bk}, \alpha; j_{\bg}A^2_{\bk}}]^{\mu_{j_{\bg}} \mu_{\bk}}_{\mu_{j_{\bg \bk}}}[L^{m_{\bg}}_{\textbf{X}}]_{(i_{\bg}, \mu_{i_{\bg}})(j_{\bg}, \mu_{j_{\bg}})},
    \label{OpT}
\end{gathered}	
\end{equation}
\end{widetext}
where $a_{\bh}$ and $b_{\bk}$ are involved in grand algebras $\textbf{A}_1$ and $\textbf{A}_2$ respectively.
When $a_{\bh}$ and $b_{\bk}$ are respectively restricted in $A_1$-type and $A_2$-type condensed anyons, \Eq{OpT} shows the operation tensors of additional $A_1$-$A_2$-bimodules $(X_{\bg}, \rho^L_{X_{\bg}}, \rho^R_{X_{\bg}})$.
Consequently, the $G$-symmetric junction, characterized by the grand fundamental bimodule $(\textbf{X}, \rho^L_{\textbf{X}}, \rho^R_{\textbf{X}})$, is constructed from a $L$-symbols pair $(\hL^a, \hL^{a_{\bg}}_{\mu \nu})$ and classified by $ [\cA_{\cC}]^G_{\cA_{A_1} \vee \cA_{A_2}}/([[\cA_{\cC}]^G_{\cA_{A_1}}]_{\cA_{A_2}} \vee [[\cA_{\cC}]^G_{\cA_{A_2}}]_{\cA_{A_1}})$ and $H^1[G, \mathrm U(1)]$.

\section{$C_n$ symmetry enrich topological orders}
\label{sec:Cn}
In this section, we study $C_n$ rotation symmetry enriched topological orders.
As discussed in Sec.~\ref{sec:overview},
given an intrinsic topological order, the differences between different $C_n$ SET orders only appear at the rotation center.
After folding the system $2n$ times, the rotation center transforms into a $\mathbb{Z}^{\bR}_n$-symmetric junction in multi-layers geometry, which can be studied using the general theory of symmetric junctions in Sec.~\ref{sec:symmetric}.
At the junction, there is an $H^1$ obstruction and an $H^2$ obstruction, although the latter always vanishes because $H^2[C_n, \mathrm U(1)]$ is trivial.
When the obstruction vanishes, the topologically inequivalent rotation center, and thus the $C_n$-SET phase, is classified by $[\cA^{C_n}_{\cC}]_{n\cA_{\cC}}$ and $H^1[C_n, \mathrm U(1)]$.

As outlined in Sec.~\ref{sec:overview}, we fold the 2d system along $2n$ dashed lines which intersect at rotation center to produce a $2n$ layers system, denoted as $\cD$, as illustrated in Fig.~\ref{fig:fold}.
More specifically, the blue(red) ceases turn into an $A_1(A_2)$-type gapped boundary and rotation center becomes a junction between the two boundaries.
We note that $\cC^{\text{rev}}$ is reverse of $\cC$ resulting from reversing normal direction of $\cC$ by folding.
Anyons of $\cD$ are labeled by $(a_1, a^{\text{rev}}_2,\dots, a_{2n-1}, a^{\text{rev}}_{2n})$ where $a_i$ are anyons of $\cC$, and $a^{\text{rev}}_i$ are reverse counterparts of $a_i$ with the topological spins relation $\theta_{a^{\text{rev}}} = \theta^*_{a}$.
Due to the fold, $C_n = \{\br^k| \br^n = \be  \}$ rotation symmetry turns into $\mathbb{Z}^{\bR}_n = \{\bR^k| \bR^n = \be  \}$ onsite symmetry.

Next, we introduce $\mathbb{Z}^{\bR}_n$ symmetry into the topological order $\cD$.
Since the symmetry-enrichment is only encoded at the rotation center, we expect that the symmetry-enrichment in the 2d bulk and the one-dimensional boundaries are trivial, i.e. there is a unique choice of enrichment.
As discussed in Sec.~\ref{sec:symmetric}, the bulk $\mathbb{Z}^{\bR}_n$ SET order is characterized by three levels symmetry data, $[\rho]$, $\mathfrak{w}$ and $\alpha$.
Specifically, the $\mathbb{Z}^{\bR}_n$-actions on anyons of $\cD$ are induced by the following map
\begin{align}
	\rho : C_n \rightarrow \text{Aut}(\cC).
\end{align}
Similar to Ref.~\cite{qi2019folding}, we relabel the anyon $(a_1, a_2^{\text{rev}}, {\ubr a}_3, {\ubr a}_4^{\text{rev}},\dots, {{\ubr}^{n-1} a}_{2n-1}, {{\ubr}^{n-1} a}_{2n}^{\text{rev}})$ as $(a_1, a_2^{\text{rev}}, a_3, a_4^{\text{rev}},\dots, a_{2n-1}, a_{2n}^{\text{rev}})$.
Applying the new notation, the nontrivial symmetry action of $\mathbb{Z}^{\bR}_2$ can be regarded as layer exchange
\begin{align}
	\tilde{\rho}_{\bR} : &(a_1, a_2^{\text{rev}}, a_3, a_4^{\text{rev}},
	\dots, a_{2n-1}, a_{2n}^{\text{rev}})\notag \\ &\rightarrow (a_{2n-1}, a_{2n}^{\text{rev}}, a_1, a_2^{\text{rev}},\dots, a_{2n-3}, a_{2n-2}^{\text{rev}}) .
\end{align}
Such a symmetry action is anomaly-free~\cite{gannon2019vanishing}.
For the symmetry fractionalization class $\mathfrak w$, because $H^2_{\tilde{\rho}}[\mathbb{Z}^{\bR}_n, \cA_{\cD}] = \mathbb{Z}_1$, the torsor $\mathfrak{w}$ is trivial.
Besides, the $\mathbb{Z}^{\bR}_n$ SET order can be modified by stacking a 2d $\mathbb{Z}^{\bR}_n$-SPT order corresponding to the torsor $\alpha \in H^3[\mathbb{Z}^{\bR}_n, \mathrm U(1)]$.
As a result, the topological spin of $\mathbb{Z}^{\bR}_n$ symmetry defect $x_{\bR}$ is changed
\begin{gather}
     \tilde{\theta}_{x_{\bR}}^{n} = \theta_{x_{\bR}}^n \alpha(\bR, {\bR}^{n-1}, \bR) 	,
\end{gather}
where the factor $\alpha(\bR, {\bR}^{n-1}, \bR) = e^{i\frac{2\pi p}{n}},\quad p \in  \{0, 1,\dots, n-1 \} $.
In order to correspond to the original single layer $C_n$ SET order, we study a $\mathbb{Z}^{\bR}_n$ SET order by fixing $\theta_{(\bm{1},\dots,\bm{1})_{\bR}}^n = 1$.

In the $2n$ layers geometry, the two boundaries are characterized by two algebras $A_1, A_2$ respectively
\begin{equation}
\begin{aligned}
	A_1 = \oplus_{a_1,\dots, a_n}[\boxtimes_i a_i]_1, \\
    A_2 = \oplus_{a_1,\dots, a_n}[\boxtimes_i a_i]_2,
\end{aligned}
\end{equation}
where we denote $(a_1, \bar{a}_1^{\text{rev}},\dots, a_n, \bar{a}_n^{\text{rev}})$ as $[\boxtimes_i a_i]_1$ and $(a_1, \bar{\rho}_{\br}(a_2)^{\text{rev}},\dots, a_n, \bar{\rho}_{\br}(a_1)^{\text{rev}})$ as $[\boxtimes_i a_i]_2$.
Then, we choose proper gauge to fix symmetry data of the bulk 
\begin{equation}
\begin{gathered}
	U_{\bg}(a_{\cD}, b_{\cD}; c_{\cD})_{\mu \nu} = \delta_{\mu \nu},\\
	\eta_{a_{\cD}}(\bg, \bh) = 1,
	\label{36}
\end{gathered}
\end{equation}
where $a_{\cD}, b_{\cD}$ and $c_{\cD}$ are anyons of $\cD$.
Apparently, algebras $A_1, A_2$ are both compatible with the symmetry $\mathbb{Z}^{\bR}_n$.
For the $A_1$-type boundary, \Eq{13} is written as
\begin{align}
	U^{(1)}_{\bg}([\boxtimes_i a_i]_1) U^{(1)}_{\bg}([\boxtimes_i b_i]_1) =U^{(1)}_{\bg}([\boxtimes_i c_i]_1) \label{38}.
\end{align}
The inequivalent solution class of $U^{(1)}_{\bg}([\boxtimes_i a_i]_1)$ is unique. 
According to \Eq{14}, we choose $A_1$-type boundary $U$-symbols
\begin{align}
	U^{(1)}_{\bg}([\boxtimes_i a_i]_1) = 1.
\end{align}
Moreover, according to the $\bR$-local condition, the symmetry defect $(\bm{1},\dots, \bm{1})_{\bR}$ is $\bR$-local.
The topological spin of the $\bR$-local defect satisfies $\theta_{(\bm{1},\dots,\bm{1})_{\bR}}^n = 1$, so the $\mathbb{Z}^{\bR}_n$-SPT phase is trivial on the other side.
It implies that the $H^3$ obstruction vanishes on the $\mathbb{Z}^{\bR}_n$-symmetric $A_1$-type boundary.
Besides, inequivalent additional VCLs $[\phi^{a_{\bg} b_{\bh}}_{c_{\bg \bh}, \alpha}]$ are unique due to the trivial classifying group $H^2[\mathbb{Z}^{\bR}_n, U(1)]$.
Then, we consider $\mathbb{Z}^{\bR}_n$-symmetric $A_2$-type boundary which is non-anomalous as well.
The $A_2$-type boundary $U$-symbols can be chosen 
\begin{gather}
	U^{(2)}_{\bg}([\boxtimes_i a_i]_2) = 1.
\end{gather}
Similarly, the inequivalent additional VCLs $[\varphi^{a_{\bg} b_{\bh}}_{c_{\bg \bh}, \alpha}]$ are unique.

\begin{figure}[H]
	\centering
	\includegraphics[width=0.2\textwidth, height=0.2\textwidth]{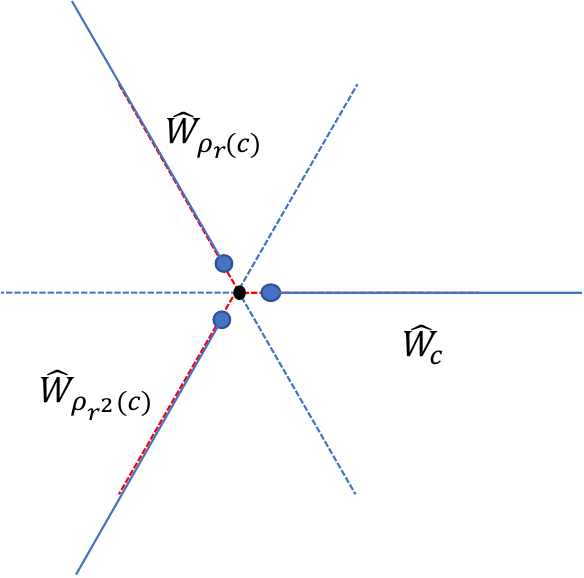}
	\caption{The symmetric string operators for $C_n$ SET orders}
	\label{fig:W(C_n)}
\end{figure}

Next, we investigate the $\mathbb{Z}^{\bR}_n$-symmetric junction between the two $\mathbb{Z}^{\bR}_n$-symmetric boundaries and denote common condensed anyon $(a, \bar{a}^{\text{rev}}, \rho_{{\br}^{n-1}}(a), \bar{\rho}_{{\br}^{n-1}}(a)^{\text{rev}},\dots, \rho_{\br}(a), \bar{\rho}_{\br}(a)^{\text{rev}})$ as $\ha$. 
In this case, \Eqs{19}{21} are written as
\begin{gather}
     \frac{\hL^{\hc}}{\hL^{\ha} \hL^{\hb}}=\frac{\varphi^{\ha \hb}_{\hc}}{\phi^{\ha \hb}_{\hc}} \label{41}, \\
	O_{\ha}({\bR}^k) = \frac{\hL^{\ha}}{\hL^{\hat{\rho}_{{\br}^{n-k}}(a)}} .
\end{gather}
Then ,at the junction, the $H^1$ obstruction is characterized by an element $\cO \in H^1_{\tilde{\rho}}[\mathbb{Z}^{\bR}_n, [\cA_{\cC}]_{\cA_{A_1} \vee \cA_{A_2}}]$.
We note that 
\begin{gather}
	H^1_{\tilde{\rho}}[\mathbb{Z}^{\bR}_n, [\cA_{\cC}]_{\cA_{A_1} \vee \cA_{A_2}}] = H^1_{\rho}[C_n, \cA_{\cC}].
\end{gather}
In non-anomalous cases, considering the \textit{topologically equivalent relations} of junctions, the special $L$-symbols are classified by
\begin{align}
	[\cA_{\cC}]^{\mathbb{Z}^{\bR}_n}_{\cA_{A_1} \vee \cA_{A_2}}/([[\cA_{\cC}]^{\mathbb{Z}^{\bR}_n}_{\cA_{A_1}}]_{\cA_{A_2}} \vee [[\cA_{\cC}]^{\mathbb{Z}^{\bR}_n}_{\cA_{A_2}}]_{\cA_{A_1}})
\end{align}
which is isomorphic to $[\cA^{C_n}_{\cC}]_{n\cA_{\cC}}$ as an abelian group.
Here, the set of abelian anyons $n\cA_{\cC}$ is defined as
\begin{gather}
    n\cA_{\cC} = \{\prod_{k=0}^{n-1} {^{{\br}^{k}} c}| c \in \cA_{\cC} \}.	
\end{gather}
Next, we interpret the construction of $\mathbb{Z}^{\bR}_n$-symmetric string operators, which can relate topologically equivalent junctions.
We firstly construct $C_n$-symmetric string operators within the original single layer, as depicted in Fig.~\ref{fig:W(C_n)}.
\noindent
Each red crease is assigned a string operator $\hW_{^{{\br}^i}c}$ that lies along it, which makes the $C_n$ symmetry preserved.
One endpoint of these string operators converges at the rotation center, while the other endpoints are localized at infinity.
In the multilayer geometry, the $C_n$-symmetric string operators transform into a $\mathbb{Z}^{\bR}_n$-symmetric string operator $\hW_{c_{\cD}}$ that lies along the $A_2$-type boundary.
Here, $c_{\cD} = (c, \bm{1}, \dots, c, \bm{1})$.
As a result, only the torsor $[\mathfrak{c}] \in [\cA^{C_n}_{\cC}]_{n\cA_{\cC}}$ can produce distinct topologically inequivalent special $L$-symbols.
Additionally, there is no $H^2$ obstruction at the junction due to trivial cohomology group $H^2[C_n, \mathrm U(1)]$~\footnote{Strictly speaking, the cohomology group should be $H^2[\mathbb{Z}^{\bR}_n, \mathrm U(1)]$. However, at the junction, for the cohomology group, we can disregard the distinction between $\mathbb{Z}^{\bR}_n$ and $C_n$}.
Then, the additional $L$-symbols $\hL^{\bg}$ can be modified by a torsor $\nu \in H^1[C_n, \mathrm U(1)]$. 
Totally, the $\mathbb{Z}^{\bR}_n$-symmetric junction can be modified by the pair $(\mathfrak{c}, \nu)$.

Back to the original single layer system, the non-trivial symmetry data of $C_n$ SET orders appear at the rotation center, corresponding to the $\mathbb{Z}^{\bR}_n$-symmetric junction.
More specifically, the rotation center can be modified by absorbing a $C_n$-invariant abelian anyon $\mathfrak{c}$, which can be detected by rotating a bulk $C_n$-invariant anyon $a$ around the center for one full circle.
Due to the braiding between anyon $a$ and $\mathfrak{c}$, this operation will result in an additional phase factor $M_{a\mathfrak{c}}$, which may alter the symmetry fractionalization of the rotation symmetry.
In addition, the torsor $\nu$ corresponds to a 0D invertible phase with symmetry $C_n$~\cite{kong2020classification}, which can be used to decorate the rotation center, thereby generating a distinct rotation center.
Consequently, the $C_n$ SET orders are classified by $[\cA^{C_n}_{\cC}]_{n\cA_{\cC}}$ and $H^1[C_n, \mathrm U(1)]$.
In particular, we derive
\begin{equation}
\begin{aligned}
	[\cA^{C_n}_{\cC}]_{n\cA_{\cC}} &= H^2_{\tilde{\rho}}[\mathbb{Z}_n, \cA_\cC], \\
	H^1[C_n, \mathrm U(1)] &= 	H^3[\mathbb{Z}_n, \mathrm U(1)],
\end{aligned}	
\end{equation}
where the generator $\ba$ of $\mathbb{Z}_n$ satisfies the relation $\tilde{\rho}_{\ba} = \rho_{\br}$.
In addition, we emphasize that the symbols $"="$ indicates that the left-hand side and the right-hand side are isomorphic as groups.
It implies that the classification group of $C_n$ SET orders is isomorphic to the classification group of onsite $\mathbb{Z}_n $ SET orders, which is in accordance with the \textit{crystalline equivalence principle}~\cite{thorngren2018gauging}.

\section{$D_{2n}$ symmetry enriched topological orders}
\label{sec:D2n}

In this section, we study $D_{2n}$ SET orders.
As discussed in Sec.~\ref{sec:overview}, given an intrinsic topological order $\cC$, the differences between different $D_{2n}$ SET orders appear at the rotation center and the $n$ mirror axes.
Considering the rotation symmetry, we note that there is only one independent mirror axis for odd integers $n$ and two independent mirror axes for even integers $n$.
For a mirror axis, the associated symmetry data can be encoded at a $\mathbb{Z}_2$-symmetric boundary of a bilayer system, produced by folding the original system along the mirror axis~\cite{qi2019folding, ding2024anomalies}.
At the $\mathbb{Z}_2$-symmetric boundary, there exists a potential $H^2$ obstruction and a potential $H^3$ obstruction.
When the obstructions vanish, the mirror axis is classified by $H^1_{\rho}[\mathbb{Z}^{\mb_i}_2, \cA_{\cC}]$.
Here, different from the layer exchange symmetry group $\mathbb{Z}_2$, $\mathbb{Z}^{\mb_i}_2$ is the mirror symmetry group associated with the mirror axis under consideration, and it is a subgroup of $D_{2n}$.
For the rotation center, similar to $C_n$ SET orders, we encode the associated symmetry data in a multilayer system, produced by folding the original system $2n$ times along the mirror axes.
In this case, the rotation center turns into a $\mathbb{D}_{2n}$-symmetric junction.
At the junction, there is a potential $H^1$ obstruction and a potential $H^2$ obstruction.
Specifically, for odd integers $n$, the $H^2$ obstruction always vanishes because $H^2[D_{2n}, \mathrm U(1)]$ is trivial.
When the obstruction vanishes, the rotation center is classified by $[\cA^{D_{2n}}_{\cC}]_{2n\cA_{\cC}}$ and $H^1[D_{2n}, \mathrm U(1)]/\mathbb{Z}_{(n+1, 2)}$. 
Totally, the classification groups of the mirror axes and the rotation center classify the $D_{2n}$ SET order. 
\begin{figure}[H]
	\centering
	\includegraphics[width=0.4\textwidth, height=3.25cm]{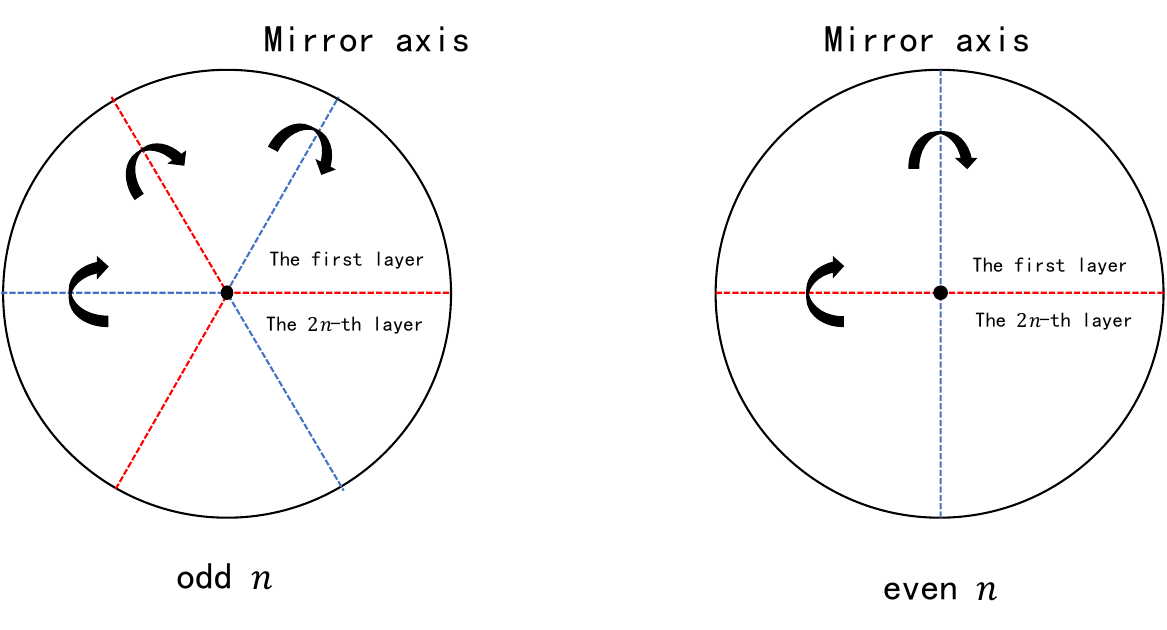}
	\caption{The mirror axes}
	\label{fig:axes}
\end{figure}
 
As depicted in Fig.~\ref{fig:axes}, there are $n$ mirror axes in the $D_{2n}$ SET order.
Focusing on a mirror axis between the first layer and $2n$-th layer, called $\mb_1$-mirror axis, we study it in associated bilayer system.
The mirror axis turns into a $\mathbb{Z}_2$-symmetric boundary of the bilayer system.
At the boundary, the potential $H^2$ obstruction is characterized by an element in~\cite{ding2024anomalies}
\begin{gather}
    H^2_{\rho}[\mathbb{Z}^{\mb_1}_2, \cA_{\cC}]	,
\end{gather}
where $\rho$ is a map which determines the anyon permutation under the associated $\mathbb{Z}^{\mb_1}_2$ mirror symmetry action
\begin{gather}
	\rho : \mathbb{Z}^{\mb_1}_2 \rightarrow \text{Aut}^*(\cC).
\end{gather}
When the $H^2$ obstruction vanishes, the boundary $U$-symbols are classified by 
\begin{gather}
    H^1_{\rho}[\mathbb{Z}^{\mb_1}_2, \cA_{\cC}].	
\end{gather}
Furthermore, the boundary $U$-symbols determine $\bg$-local symmetry defects on the $\mathbb{Z}_2$-symmetric boundary by the $\bg$-local condition, illustrated in Fig.~\ref{fig:local}.
The topological spins of the $\bg$-local symmetry defects can specify a 2D $\mathbb{Z}_2$-SPT phase on the other side.
The $\mathbb{Z}_2$-SPT phase is characterized by a cocycle $\alpha \in H^3[\mathbb{Z}_2, \mathrm U(1)]$, corresponding to the $H^3$ obstruction at the boundary.
Besides, the inequivalent additional VLCs is unique because of the trivial cohomology group $H^2[\mathbb{Z}_2, \mathrm U(1)] = \mathbb{Z}_1$.
Thus, the mirror axis is classified by $H^1_{\rho}[\mathbb{Z}^{\mb_1}_2, \cA_{\cC}]$.
Because the rotation symmetry actions can transform one mirror axis into another, the associated symmetry data of the mirror axes which can be connected by the rotation symmetry actions are identical.
Hence, for odd integers $n$, there is only one independent mirror axis, classified by $H^1_{\rho}[\mathbb{Z}^{\mb_1}_2, \cA_{\cC}]$. 
For even integers $n$, there are two distinct independent mirror axes, distinguished by red and blue dashed lines, as depicted in Fig.~\ref{fig:axes}, which are classified by two respective groups: $H^1_{\rho}[\mathbb{Z}^{\mb_1}_2, \cA_{\cC}]$ and $H^1_{\rho}[\mathbb{Z}^{\mb_2}_2, \cA_{\cC}]$.
Here, we apply the notation $\mathbb{Z}^{\mb_i}_2$ to distinguish different mirror symmetry groups.

In the following, we encode the symmetry data of the rotation center in a $2n$ layers system $\cD$.
As outlined in Sec.~\ref{sec:overview}, $\cD$ is produced by folding along the mirror axes, as shown in Fig.~\ref{fig:fold}.
In this case, $D_{2n} = \{ {\mb}^l {\br}^k|{\br}^n = {\mb}^2 = \be, \mb \br \mb = {\br}^{-1} \}$ nonlocal symmetry turns into $\mathbb{D}_{2n} = \{ {\bM}^l {\bR}^k|{\bR}^n = {\bM}^2 = \be, \bM \bR \bM = {\bR}^{-1} \}$ onsite symmetry.
Here, we denote the mirror symmetry along $\mb_1$-mirror axis as $\mb$.
The $\mathbb{D}_{2n}$-symmetry actions on anyons of $\cD$ are induced by the map
\begin{gather}
    \rho: D_{2n} \rightarrow \text{Aut}^*(\cC) .
\end{gather}
We note that $\rho_{\mb}$ is an anti-auto-equivalence of $\cC$, included in $\text{Aut}^*(\cC)$.
Additionally, the group $\text{Aut}^*(\cC)$ contains auto-equivalence $\rho_{\br}$ as well. 
Thus, we have the topological spins relation $\theta_{\ubr a} = \theta^*_{\ubm a} = \theta_a$.
As a result, the anyon $(a_1, {^{\br \mb} a}_2^{\text{rev}}, {\ubr a}_3, {^{{\br}^2 \mb}a}_4^{\text{rev}},\dots, {^{{\br}^{n-1}}a}_{2n-1}, {\ubm a}_{2n}^{\text{rev}})$ can be relabelled as $(a_1, a_2, a_3, a_4,\dots, a_{2n-1}, a_{2n})$ in the $2n$ layers system.
Benefiting from the new notation, the induced $\mathbb{D}_{2n}$-symmetry actions can be defined as 
\begin{equation}
\begin{aligned}
	\tilde{\rho}_{\bR} : &(a_1, a_2, a_3, a_4,
	\dots, a_{2n-1}, a_{2n}) \\ &\rightarrow (a_{2n-1}, a_{2n}, a_1, a_2,\dots, a_{2n-3}, a_{2n-2}) , \\
	\tilde{\rho}_{\bM}: &(a_1, a_2,\dots, a_n, a_{n+1},\dots, a_{2n-1}, a_{2n}) \\ &\rightarrow (a_{2n}, a_{2n-1},\dots, a_{n+1}, a_{n},\dots, a_2, a_1)
	\label{44}.
\end{aligned}
\end{equation}
Essentially, the symmetry actions of $\mathbb{D}_{2n}$ are layer exchanges which implies non-anomalous bulk~\cite{gannon2019vanishing} and trivial symmetry fractionalization because $H^2_{\tilde{\rho}}[\mathbb{D}_{2n}, \cA_{\cD}]$ is trivial.
Besides, in order to correspond to the original $D_{2n}$ SET order, we choose a bulk $\mathbb{D}_{2n}$ SET order by selecting topological spins of symmetry defects as 
\begin{gather}
    \theta_{(\bm{1},\dots,\bm{1})_{\bR}}^n = \theta_{(\bm{1},\dots,\bm{1})_{\bM \bR^k}}^2 = 1	.
\end{gather}

Next, we move on to two gapped boundaries of $\cD$, formed by the blue and red dashed lines respectively.
They are characterized by two algebras, $A_1$ and $A_2$
\begin{equation}
\begin{aligned}
	A_1 = \oplus_{a_1,\dots, a_n}[\boxtimes_i a_i]_1,\\
    A_2 = \oplus_{a_1,\dots, a_n}[\boxtimes_i a_i]_2.
\end{aligned}
\end{equation} 
Here, we denote condensed anyons $(a_1, \bar{\rho}_{\br \mb}(a_1),\dots, a_n, \bar{\rho}_{\br \mb}(a_n))$ as $[\boxtimes_i a_i]_1$, and $(a_1, \bar{\rho}_{\mb}(a_2), a_2,\dots, \bar{\rho}_{\mb}(a_n), a_n, \bar{\rho}_{\mb}(a_1))$ as $[\boxtimes_i a_i]_2$.
If the $H^2$ obstructions vanish on the all $n$ mirror axes, the two boundaries can be both compatible with the $\mathbb{D}_{2n}$ symmetry.
And then, we fix bulk $U, \eta$ symbols to be trivial by choosing proper gauge in bulk.
For the $A_2$-type boundary, the boundary $U$-symbols can be totally constructed by the symmetry data of the $\mb_1$-mirror axis between the first and $n$-th layer, shown as
\begin{equation}
\begin{aligned}
	 &U^{(2)}_{\bM}([\boxtimes_i a_i]_2) = \prod^n_{i=1} U^{(\mb_1)}_{\bM}(a_i, \bar{\rho}_{\mb}(a_i)), \\
	 &U^{(2)}_{\bR}([\boxtimes_i a_i]_2) = 1.
	 \label{65}
\end{aligned}	
\end{equation}
Here, the $\mb_1$-mirror axis can be regarded as a boundary in bilayer geometry, characterized by the Lagrangian algebra $A_{\mb_1} = \oplus_{a}(a, \bar{\rho}_{\mb}(a))$, and the $A_{\mb_1}$-type boundary $U$-symbols are denoted as $U^{(\mb_1)}_{\bM}(a, \bar{\rho}_{\mb}(a))$.
Verifying the classification groups of the $A_2$-type boundary and the $A_{\mb_1}$-type boundary, we derive
\begin{gather}
    H^1_{\tilde{\rho}}[\mathbb{D}_{2n}, [\cA_{\cD}]_{\cA_{A_2}}]	 = H^1_{\rho}[\mathbb{Z}^{\mb}_2, \cA_{\cC}].
\end{gather}
It implies that the \Eq{65} is a reasonable choice for the construction of $A_2$-type boundary $U$-symbols. 
For odd integers $n$, the $A_1$-type boundary $U$-symbols are also entirely constructed by the identical symmetry data of the $\mb_1$-mirror axis.
This is because the blue dashed lines, which constitute the $A_1$-type boundary, share the same mirror axes with the red dashed lines which form the $A_2$-type boundary.
On the other side of the two boundaries, the phase consists of $n$ copies of a $\mathbb{Z}_2$-SPT phase, corresponding to the $H^3$ obstruction of the $\mb_1$-mirror axis, denoted as $\alpha$.
Because the $\mathbb{Z}_2$-SPT phase invertible, we have the relation $n\alpha = \alpha$ which implies that the $H^3$ obstructions of the two boundaries is identical to the $H^3$ obstruction of the $\mb_1$-mirror axis.
For even integers $n$, the $A_1$-type boundary $U$-symbols are entirely determined by the symmetry data of $\mb_2$-mirror axis which lies between the first and the second layers.
Different from the case that $n$ is an odd integer, the $A_1$-type boundary $U$-symbols is independent from the $A_2$-type boundary $U$-symbols, which is classified by 
\begin{gather}
    H^1_{\tilde{\rho}}[\mathbb{D}_{2n}, [\cA_{\cD}]_{\cA_{A_1}}]	 = H^1_{\rho}[\mathbb{Z}^{\br \mb}_2, \cA_{\cC}],	
\end{gather}
where $\mathbb{Z}^{\br \mb}_2$ is the mirror symmetry group associated with the $\mb_2$-mirror axis.
On the other side of the two boundaries, the phase is trivial, since even copies of the $\mathbb{Z}_2$-SPT phase is always trivial.
It implies that the $H^3$ obstructions vanish at the two boundaries.
Furthermore, we obtain the additional VLCs by solving \Eqss{17}{24}.
For odd $n$, since $H^2[\mathbb{D}_{2n}, \mathrm U(1)] = \mathbb{Z}_1$, there is only one set of gauge-inequivalent solutions for the additional VLCs.
For even $n$, since $H^2[\mathbb{D}_{2n}, \mathrm U(1)] = \mathbb{Z}_2$, there are two set of gauge-inequivalent solutions.
As shown in \Eq{Twist BdryU}, the modification of additional VLCs by a torsor alters the boundary $U$-symbols of $\bg$-local defects. 
Thus, in order to correspond to the original $D_{2n}$-SET order, we determine a canonical choice of the additional VLCs by fixing boundary $U$-symbols of $\bg$-local defects as
\begin{equation}
\begin{aligned}
    	U^{(1)}_{\bh}(a^1_{\bg}) = 1, \\
    	U^{(2)}_{\bh}(a^2_{\bg}) = 1, \label{68}
\end{aligned}
\end{equation}
where $\bg$ is a nontrivial element and $a^{1(2)}_{\bg}$ is an $\bg$-local defect of $A_{1(2)}$-type boundary.
Consequently, combining with \Eqs{17}{24}, \Eq{68} can fully determine the additional VLCs of the two boundaries.
Hence, the nontrivial symmetry data of the two boundaries only involve the boundary $U$-symbols of condensed anyons, which can be totally determined by the symmetry data of mirror axes.
 
\begin{figure}[H]
	\centering
	\includegraphics[width=0.24\textwidth, height=0.2\textwidth]{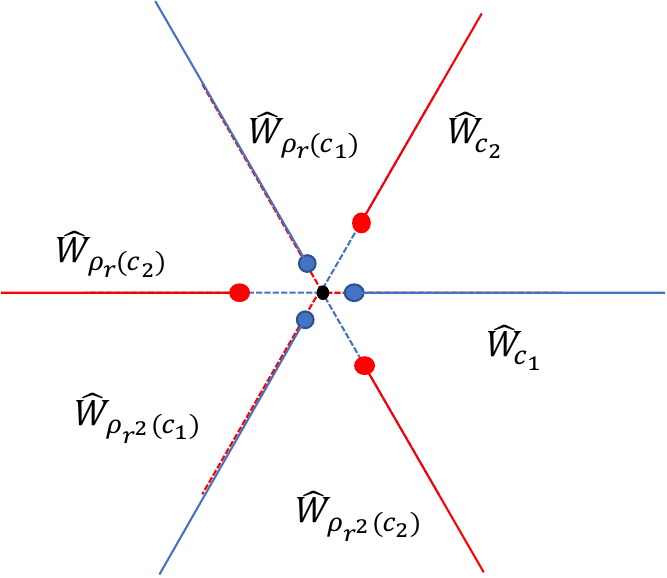}
	\caption{The $D_6$-symmetric operators for $D_6$-SET orders}
	\label{fig:W(D_2n)}
\end{figure}

Next, we study the $\mathbb{D}_{2n}$-symmetric junction between the two boundaries. 
For convenience, we denote common condensed anyon $(a, \bar{\rho}_{\br \mb}(a), \rho_{{\br}^{n-1}}(a), \bar{\rho}_{{\br}^2 \mb}(a),\dots, \rho_{\br}(a), \bar{\rho}_{\mb}(a))$ as $\ha$.
According to \Eqs{19}{21}, there is a potential $H^1$ obstruction, characterized by an element in
\begin{gather}
     H^1_{\tilde{\rho}}[\mathbb{D}_{2n}, [\cA_{\cD}]_{\cA_{A_1} \vee \cA_{A_2}}] = H^1_{\rho}[D_{2n}, \cA_{\cC}]	.
\end{gather}
In the absence of the $H^1$ obstruction, we can obtain special $L$-symbols of common condensed anyons $\ha$ by \Eqs{19}{20}.
The special $L$-symbols are classified by 
\begin{align}
	[\cA_{\cC}]^{\mathbb{D}_{2n}}_{\cA_{A_1} \vee \cA_{A_2}}/([[\cA_{\cC}]^{\mathbb{D}_{2n}}_{\cA_{A_1}}]_{\cA_{A_2}} \vee [[\cA_{\cC}]^{\mathbb{D}_{2n}}_{\cA_{A_2}}]_{\cA_{A_1}})
\end{align}
which is isomorphic to $[\cA^{D_{2n}}_{\cC}]_{2n\cA_{\cC}}$ as an abelian group.
Here, the set of abelian anyons $2n\cA_{\cC}$ is defined as
\begin{gather}
    2n\cA_{\cC} = \{\prod_{k_1, k_2=0}^{n-1} {^{{\br}^{k_1}} c_1}{^{{\br}^{k_2}} \bar{c}_2} | c_1, c_2 \in \cA_{\cC}      \}	.	
\end{gather}
In this case, the $\mathbb{D}_{2n}$-symmetric string operators, relating distinct topologically equivalent junctions, can also be transformed by constructed $D_{2n}$-symmetric operators within the original single layer.     
In the single layer geometry, the $D_{2n}$-symmetric operators consist of two independent groups of string operators, $\hW_{\rho_{\br^{k_1}}(c_1)}$ and $\hW_{\rho_{\br^{k_2}}(c_2)}$, which are situated along blue dashed lines and red dashed lines respectively, as depicted in Fig.~\ref{fig:W(D_2n)}.
\noindent
Here, the abelian anyons $c_1$ and $c_2$ satisfy
\begin{equation}
\begin{aligned}
	c_1 &= \rho_{\mb}(c_1) ,\\
	c_2 &= \rho_{\mb \br}(c_2).
\end{aligned}
\end{equation}
which makes the $D_{2n}$-symmetry preserved.
One endpoints of these string operators, $\hW_{\rho_{\br^{k_1}}(c_1)}$ and $\hW_{\rho_{\br^{k_2}}(c_2)}$, converges at the rotation center.
In $2n$ layers geometry, the two groups of string operators turn into two $\mathbb{D}_{2n}$-symmetric string operators $\hW_{c^{(1)}_{\cD}}$ and $\hW_{c^{(2)}_{\cD}}$, which lie on the $A_1$-type boundary and $A_2$-type boundary, respectively.
Here, $c^{(i)}_{\cD}$ denotes the anyon $(c_i, \bm{1},\dots, c_i, \bm{1})$.
The action of these two string operators, $\hW_{c^{(1)}_{\cD}}$ and $\hW_{c^{(2)}_{\cD}}$, is equivalent to decorating the junction with an abelian anyon, 
\begin{align}
	  \prod_{k_1, k_2=0}^{n-1} {^{{\br}^{k_1}} c_1}{^{{\br}^{k_2}} \bar{c}_2},
\end{align}
which may alter the special $L$-symbols.
Thus, only the torsor $[\mathfrak{c}] \in [\cA^{D_{2n}}_{\cC}]_{2n\cA_{\cC}}$ can modify the junction in a topologically inequivalent manner.
Furthermore, at the junction, when the special $L$-symbols are determined, there is the potential $H^2$ obstruction, characterized by an element in $H^2[D_{2n}, \mathrm U(1)]$~\footnote{Similar to the case of $C_n$-SETs, for the group cohomology, we disregard the distinction between $\mathbb{D}_{2n}$ and $D_{2n}$}.
Particularly, we note that this cohomology group is always trivial for odd integers $n$.
When the $H^2$ obstruction vanishes, the additional $L$-symbols $\hL^{\bg}$ can be modified by a torsor $\nu \in H^1[D_{2n}, \mathrm U(1)]$.
Totally, the $\mathbb{D}_{2n}$-symmetric junction can be modified by a pair $(\mathfrak{c}, \nu)$.

\begin{figure}[H]
	\centering
	\includegraphics[width=0.44\textwidth, height=0.2\textwidth]{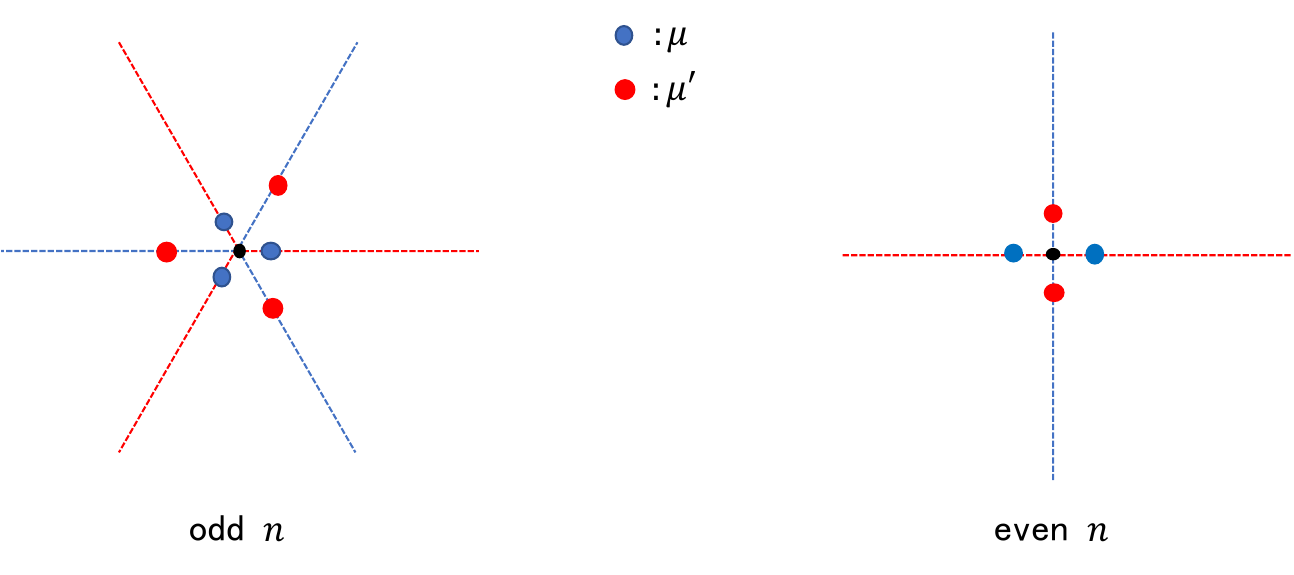}
	\caption{The decoration for $D_{2n}$ SET orders}
	\label{fig:D(D_2n)}
\end{figure}

Similar to the $C_n$-SET orders, upon returning to the single layer geometry, the torsor $\mathfrak{c}$ corresponds to an abelian anyon that is absorbed by the rotation center.
This abelian anyon $\mathfrak{c}$ modifies the symmetry fractionalization of the rotation symmetry due to the braiding between $\mathfrak{c}$ and the bulk anyons.
The other torsor, $\nu$, corresponds to a 0D $D_{2n}$-SPT phase, which is used to decorate the rotation center, leading to distinct rotation centers.
However, the decoration associated with the 0D $D_{2n}$-SPT phase $\nu$ may be topologically trivial.
For a torsor $\nu$, if it can be decomposed into several 0D $\mathbb{Z}_2$-SPT phases in a $D_{2n}$-symmetric manner, as illustrated in Fig.~\ref{fig:D(D_2n)}, then the new rotation centers, decorated by $\nu$, are considered to be topologically equivalent to the original ones.
This process is referred to as \textit{trivilization}~\cite{cheng2022rotation}. 
\noindent
Here, $\mu$ and $\mu'$ represent 0D $\mathbb{Z}_2$-SPT phases associated with mirror symmetries.
Since $H^1[\mathbb{Z}_2, \mathrm U(1)] = \mathbb{Z}_2$, $2\mu$ is a trivial phase, implying that all of the above decorations are trivial for even integers $n$ at the rotation center.
Conversely, for odd integers $n$, $n\mu$ is identical to $\mu$, potentially yielding a distinct rotation center, which is viewed as topologically equivalent to the original one.
As a result, only the decorations by the 0D $\mathbb{Z}_2$-SPT phases $\nu \in H^1[D_{2n}, \mathrm U(1)]/\mathbb{Z}_{(n+1, 2)}$ can produce topological inequivalent junctions.
Considering the topological inequivalent modifications by $\mathfrak{c}$, the rotation center is classified by $[\cA^{D_{2n}}_{\cC}]_{2n\cA_{\cC}}$ and $H^1[D_{2n}, \mathrm U(1)]/\mathbb{Z}_{(n+1, 2)}$.
Totally, the $D_{2n}$-SET orders are classified by two levels of groups.
The first level is
\begin{widetext}
\begin{equation}
\begin{aligned}
	&H^1_{\rho}[\mathbb{Z}^{\mb}_2, \cA_{\cC}] \oplus [\cA^{D_{2n}}_{\cC}]_{2n\cA_{\cC}} \quad &\text{for odd} \quad n ,\\
	&H^1_{\rho}[\mathbb{Z}^{\mb}_2, \cA_{\cC}] \oplus H^1_{\rho}[\mathbb{Z}^{\br \mb}_2, \cA_{\cC}] \oplus [\cA^{D_{2n}}_{\cC}]_{2n\cA_{\cC}} \quad &\text{for even} \quad n,
\end{aligned}
\end{equation} 
\end{widetext} 
which is isomorphic to $H^2_{\tilde{\rho}}[\mathbb{D}_{2n}, \cA_{\cC}]$, as shown in Appendix~\ref{app:H^2}.
Here, the map $\tilde{\rho}$ is defined as
\begin{equation}
\begin{aligned}
    \tilde{\rho} &: \bM \rightarrow \bar{\rho}_{\mb},  \\
    \tilde{\rho} &: \bR \rightarrow \rho_{\br}	.
    \label{ReAction}
\end{aligned}
\end{equation}
The second level is 
\begin{gather}
    H^1[D_{2n}, \mathrm U(1)]/\mathbb{Z}_{(n+1, 2)}	,
\end{gather}
which is isomorphic to $H^3[\mathbb{D}_{2n}, \mathrm U^{\bM}(1)]$.
Here, the superscript $\bM$ means that the symmetry operator corresponding to the element $\bM$ is anti-unitary.
Consequently, the classification of $D_{2n}$-SET orders is equivalent to the classification of onsite $\mathbb{D}_{2n}$-SET orders, which is in accordance with the \textit{crystalline equivalence principle}~\cite{thorngren2018gauging}.

\section{An anomalous example with $H^1$ obstruction at rotation center}
\label{sec:H1}
In this section, we investigate the $\mathbb{D}_{16}$ gauge theory enriched by the $C_2$ symmetry.
In the case that a special symmetry action $\rho_{\br}$ is chosen, we find that there is nontrivial $H^1$ obstruction at the rotation center.
The $H^1$ obstruction corresponds to $H^3$ obstruction of $\mathbb{D}_{16}$ gauge theory enriched by onsite $\mathbb{Z}_2$ symmetry~\cite{fidkowski2017realizing}.

We follow notations in Ref.~\cite{qi2019folding}.
The gauge group $\mathbb{D}_{16}$ is expressed as $ \{\bs^n \ba^m|\bs^2=\ba^8=\be, \bs \ba \bs=\ba^{-1}   \}$.
Anyons of $\mathbb{D}_{16}$ gauge theory are denoted as $([\bg], \phi)$ where $[\bg]$ is a conjugacy class of $\mathbb{D}_{16}$ and $\phi$ is an irreducible representation of centralizer group $Z_{\bg}$.
Particularly, the anyons $([\be], \phi_i)$ are referred to as pure charges.
Without ambiguity, the pure charges can be denoted as $\phi_i$ alone.
Then, we define the nontrivial symmetry action as
\begin{gather}
    \rho_{\br}: ([\bg], \phi) \rightarrow ([f(\bg)], \phi \circ f^{-1} ),
\end{gather}
where $f$ is an automorphism of $\mathbb{D}_{16}$ shown as 
\begin{gather}
    f(\ba) = \ba^5, \quad f(\bs) = \bs \ba.	
\end{gather}
After folding, $C_2$ symmetry transform into onsite $\mathbb{Z}^{\bR}_2$ symmetry.
We study the $\mathbb{Z}^{\bR}_2$ SET order in four layers geometry.
Particularly, the associated $A_1(A_2)$-type gapped boundary can be regarded as a stack comprising two identical gapped boundaries of a bilayer system, as shown
\begin{equation}
	\begin{aligned}
	A_1 &= \oplus_{a_1, a_2}(a_1, \bar{a}_1, a_2, \bar{a}_2) = \tilde{B} \boxtimes \tilde{B}, \\
	A_2 &= \oplus_{a_1, a_2}(a_1, \bar{\rho}_{\br}(a_2), a_2, \bar{\rho}_{\br}(a_1)) = B \boxtimes B,
\end{aligned}
\end{equation}
where $\tilde{B} = \sum_{a}(a, \bar{a})$ and $B = \sum_a (a, \bar{\rho}_{\br}(a)^{\text{rev}})$.
In this case, by selecting appropriate gauge and normalization factors, all VLCs of  the $A_1$-type boundary become trivial.
As a result, focusing on pure charges, \Eq{41} can be written as 
\begin{align}
	\frac{\hL(\phi_k)}{\hL(\phi_i)\hL(\phi_j)} m^{{\ubr \phi_i} {\ubr \phi_j}}_{\ubr \phi_k} = m^{\phi_i \phi_j}_{\phi_k},
\end{align}
where we denote gauge independent $L$-symbols of pure charges to $\hL(\phi_i)$ and $m^{\phi_i \phi_j}_{\phi_k}$ represent the VLCs of pure charges for the $B$-type gapped boundary of the bilayer system.
By employing \Eqs{2}{3} for the $B$-type boundary, referring to the derivation of $H^2$ obstruction in an anomalous mirror SET order~\cite{ding2024anomalies}, we can derive 
\begin{gather}
    O_{(\alpha_2, \alpha_2, \alpha_2, \alpha_2)}(\bR) = -1,	
\end{gather}
where $\alpha_2$ is a pure charge and corresponds to a 2D irreducible representation of $\mathbb{D}_{16}$ with $\alpha_2(\ba)=\text{diag}[ i, -i] $.
It implies that there is nontrivial $H^1$ obstruction.
Without loss of generality, we can select
\begin{gather}
    \cO(\bR)=(([{\ba}^4], \phi_1), \bm{1}, \bm{1}, \bm{1}),	
\end{gather}
where $\phi_1$ is trivial 1D irreducible representation of $\mathbb{D}_{16}$.
The above anyon $([{\ba}^4], \phi_1)$ is the exact value of $H^3$ obstruction for an anomalous example, $\mathbb{D}_{16}$ gauge theory enriched by onsite $\mathbb{Z}_2$ symmetry in Ref.~\cite{fidkowski2017realizing}.

\section{Three anomalous examples with $H^2$ obstruction at the symmetric junction}
\label{sec:H2}
In this section, we study three examples, with nontrivial relative $H^2$ obstructions at symmetric junctions. 
Based on the definition of relative $H^2$ obstructions, the relative $H^2$ obstructions in the three examples originate distinct torsors.
In the first example, the the nontrivial $H^2$ obstruction arises from torsors $\beta_i$.
In the later two example , the the nontrivial $H^2$ obstruction originates from the interplay between the torsors $\mathfrak{v}^i$ and $\mathfrak{c}$.   

\subsection{trivial bosonic state enriched by symmetry group $G$}
Considering trivial bosonic state $\cC = \{ \bm{1}\}$ enriched by symmetry group $G$, the non-trivial symmetry datas of bulk only involve $\alpha \in H^3[G, \mathrm U(1)]$ which corresponds to $F$-symbols $F^{\bm{1}_{\bg}\bm{1}_{\bh}\bm{1}_{\bk}}$.
Here, $\bm{1}_{\bg}$ is a symmetry defect.
Moreover, we investigate a $G$-symmetric junction between two $G$-symmetric boundary which are characterized by $(A_1, \phi^{\bm{1}_{\bg}\bm{1}_{\bh}}_{\bm{1}_{\bg\bh}})$ and $(A_2, \varphi^{\bm{1}_{\bg}\bm{1}_{\bh}}_{\bm{1}_{\bg\bh}})$ respectively.
Since the two algebras $A_1 = A_2 = \bm{1}$, we omit the trivial boundary $U$-symbols.
In addition, we assume that the bulk is a trivial $G$-SPT, which ensure that the $H^3$ obstruction vanishes on the two boundary.
According to \Eq{17}, we can derive 
\begin{align*}
	\phi^{\bm{1}_{\bg}\bm{1}_{\bh}}_{\bm{1}_{\bg\bh}} = \beta_1(\bg, \bh), \quad
	\varphi^{\bm{1}_{\bg}\bm{1}_{\bh}}_{\bm{1}_{\bg\bh}} = \beta_2(\bg, \bh),
\end{align*}
where $\beta_1, \beta_2 \in H^2[G, \mathrm U(1)]$. 
It follows that the $H^2$ obstruction can be derived 
\begin{gather}
   \gamma(\bg, \bh) = \frac{\beta_1(\bg, \bh)}{\beta_2(\bg, \bh)}.	
\end{gather}
As a result, if $\beta_1 \neq \beta_2$, there is nontrivial $H^2$ obstruction $\gamma$ at the $G$-symmetric junction. 
Physically, the junction can be regarded as a boundary of a nontrivial 1D $G$-SPT phase characterized by the $H^2$ obstruction $\gamma$.

\subsection{The semion theory enriched by $C_2 \times \mathbb{Z}_2$ symmetry}
\label{sec:semion-c2-z2}
\begin{table*}[ht]
\centering
\begin{tabular}{|c|c|c|c|c|}
\hline
    $\mathfrak{v}^{(1)}(\mathbbm{1}) $   & $ U^{(1)}_{\mathbbm{1}}(\bb, \bm{1})$ & $ U^{(1)}_{\mathbbm{1}}(\bm{1}, \bb)$  &  $\bg$-local symmetry defects \\ \hline
    $(\bm{1}, \bm{1}) $  &  1 &   1    &  $A_1 \otimes (\bm{1}, \bm{1})_{\mathbbm{1}}, A_1 \otimes (\bm{1}, \bm{1})_{\bR}, A_1 \otimes (\bm{1}, \bm{1})_{\mathbbm{R}}$ \\ \hline
    $(s, \bar{s}) $  &  -1 &   -1  &  $A_1 \otimes (s, s)_{\mathbbm{1}}, A_1 \otimes (\bm{1}, \bm{1})_{\bR}, A_1 \otimes (\bm{1}, \bm{1})_{\mathbbm{R}}$       \\ \hline
\end{tabular}
\caption{The symmetry data of $A_1$-type boundary(E.2)}
\label{tab:BA1}
\end{table*}

\begin{table*}[ht]
\centering
\begin{tabular}{|c|c|c|c|c|}
\hline
     $\mathfrak{v}^{(2)}(\mathbbm{1}) $   & $ U^{(2)}_{\mathbbm{1}}(s, \bar{s})$ & $ U^{(2)}_{\mathbbm{1}}(\bar{s}, s)$  &  $\bg$-local symmetry defects \\ \hline
    $(\bm{1}, \bm{1}) $  &  1 &   1    &  $A_2 \otimes (\bm{1}, \bm{1})_{\mathbbm{1}}, A_2 \otimes (\bm{1}, \bm{1})_{\bR}, A_2 \otimes (\bm{1}, \bm{1})_{\mathbbm{R}}$ \\ \hline
    $(\bm{1}, b) $  &  -1 &   -1  &  $A_2 \otimes (s, s)_{\mathbbm{1}}, A_2 \otimes (\bm{1}, \bm{1})_{\bR}, A_2 \otimes (\bm{1}, \bm{1})_{\mathbbm{R}}$       \\ \hline
\end{tabular}
\caption{The symmetry data of $A_2$-type boundary(E.2)}
\label{tab:BA2}
\end{table*}
The second example is the semion theory $\cC = \{ \bm{1}, s \}$ enriched by $C_2 \times \mathbb{Z}_2$ symmetry.
In this case, the symmetry action is trivial, with no other alternative options available. 
We study it in four layers geometry and regard the bulk as a bilayer double semion theory.
The double semion theory involves anyons $ \bm{1}, s, \bar{s}, \bb $ where $\bb$ is a boson and $\bar{s}$ is reverse counterpart of $s$.
Thus, anyons of the bulk are denoted as $(a_1, a_2)$ where $a_1$ and $a_2$ belong to the double semion theory.
Additionally, the original symmetry $C_2 \times \mathbb{Z}_2$ turns into onsite symmetry $\mathbb{Z}^{\bR}_2 \times \mathbb{Z}_2 = \{\be, \bR, \mathbbm{1},\mathbbm{R} \}$.
Here, $\bR$ and $\mathbbm{1}$ are generators of $\mathbb{Z}^{\bR}_2$ and $\mathbb{Z}_2$ respectively.
The induced symmetry action is 
\begin{equation}
\begin{aligned}
	\rho_{\mathbbm{1}} : (a_1, a_2) \rightarrow (a_1, a_2), \\
	\rho_{\bR} : (a_1, a_2) \rightarrow (a_2, a_1).
\end{aligned}
\end{equation}
Then, we choose trivial symmetry fractionalization class as
\begin{align}
	\eta_{(a_1, a_2)}(\bg, \bh) = 1.
\end{align}  
Besides, we fix topological spins of symmetry defects as
\begin{gather}
   	\theta^2_{(\bm{1}, \bm{1})_{\mathbbm{1}}}=1, \quad \theta^2_{(a, a)_{\bR}} = \theta^2_{(a, a)_{\mathbbm{R}}} = \theta_a.
\end{gather}
The two boundaries are characterized by 
\begin{equation}
\begin{aligned}
	A_1 &= (\bm{1}, \bm{1}) \oplus (\bm{1}, \bb) \oplus (\bb, \bm{1}) \oplus (\bb, \bb), \\
	A_2 &= (\bm{1}, \bm{1}) \oplus (s, \bar{s}) \oplus (\bar{s}, s) \oplus (\bb, \bb).
\end{aligned}
\end{equation}
There are two nonequivalent boundary $U$-symbols for the $A_{1(2)}$-type boundary which are shown in Table~\ref{tab:BA1}(\ref{tab:BA2}).
The $H^3$ obstructions on these boundaries all vanish.
In this case, we choose the original standard junction by fixing the boundary $U$-symbols:
\begin{align}
	U^{(1)}_{\bg}(a_1)= U^{(2)}_{\bg}(a_2) = 1
\end{align} 
where $a_i$ is an $A_i$-type condensed anyon.
Then, the original common $\bg$-local defects can be chosen as  $(\bm{1}, \bm{1})_{\mathbbm{1}}$,$(\bm{1}, \bm{1})_{\bR}$ and $(\bm{1}, \bm{1})_{\mathbbm{R}}$.
Considering a specific combination, $\mathfrak{v}^{(1)}(\mathbbm{1}) = (s, \bar{s})$ for $A_1$-type boundary and $\mathfrak{v}^{(2)}(\mathbbm{1}) = (\bm{1}, \bm{1})$ for $A_2$-type boundary, there is no $H^1$ obstruction at the junction.
For the relative $H^2$ obstruction, it is known that $H^2[\mathbb{Z}^{\bR}_2 \times \mathbb{Z}_2, \mathrm U(1)] = \mathbb{Z}_2$, so we define relative $H^2$ anomaly indicator for $\mathbb{Z}^{\bR}_2 \times \mathbb{Z}_2$ symmetry:
\begin{gather}
    \eta_r = \frac{\gamma(\mathbbm{1}, \bR)}{\gamma(\bR, \mathbbm{1})},
\end{gather}
which is gauge independent.
More specifically, $\eta_r = -1$ implies a nontrivial modification of the $H^2$ obstruction at the rotation center, while $\eta_r = 1$ leaves it unmodified.
In this case, referring to the definition~(\ref{Def.H^2}), we can derive
\begin{align}
	\eta_r = M^*_{(b, b)\mathfrak{c}}
\end{align}
Since the torsor $\mathfrak{c}$ takes values in the set $\{(\bm{1}, \bm{1}), (\bm{1}, s)\}$, the $H^2$ obstruction can be modified by the nontrivial relative $H^2$ anomaly indicator $M^*_{(b, b)\mathfrak{c}}$.
This implies that even though we do not know how to calculate the absolute $H^2$ obstruction, we can always select a proper torsor $\mathfrak{c}$ to modify the rotation center, thereby leading to a nontrivial $H^2$ obstruction at the rotation center.

\subsection{Toric code enriched by $D_4$ symmetry}
\begin{table*}[ht]
\centering
\begin{tabular}{|c|c|c|c|c|}
\hline
     $\mathfrak{v}^{(1)}(\bM_1) $   & $ U^{(1)}_{\bR}(a, b)_1$ & $U^{(1)}_{\bM_1}(a, b)_1$ & $U^{(1)}_{\bM_2}(a, b)_1$ & $\bg$-local symmetry defects \\ \hline
    $(\bm{1}, \bm{1}, \bm{1}, \bm{1}) $  &  1 &   1  & 1  &  $A_1 \otimes [\bm{1}, \bm{1}]_{\bR}, A_1 \otimes [\bm{1}, \bm{1}]_{\bM_2}, [\bm{1}, \bm{1}]_{\bM_1}$ \\ \hline
    $(\bm{1}, e, e, \bm{1})$    &  1   & $M^*_{ae}M^*_{be}$ & $M^*_{ae}M^*_{be}$ &  $A_1 \otimes [\bm{1}, \bm{1}]_{\bR}, A_1 \otimes [\bm{1}, \bm{1}]_{\bM_2}, [e, e]_{\bM_1}$    \\ \hline
    $(\bm{1}, m, m, \bm{1})$    &  1   & $M^*_{am}M^*_{bm}$ & $M^*_{am}M^*_{bm}$ &  $A_1 \otimes [\bm{1}, \bm{1}]_{\bR}, A_1 \otimes [\bm{1}, \bm{1}]_{\bM_2}, [m, m]_{\bM_1}$     \\ \hline
    $(\bm{1}, f, f, \bm{1})$    &  1   & $M^*_{af}M^*_{bf}$ & $M^*_{af}M^*_{bf}$ &  $A_1 \otimes [\bm{1}, \bm{1}]_{\bR}, A_1 \otimes [\bm{1}, \bm{1}]_{\bM_2}, [f, f]_{\bM_1}$       \\ \hline
\end{tabular}
\caption{The symmetry data of $A_1$-type boundary(E.3)}
\label{tab:CA1}
\end{table*}

\begin{table*}[ht]
\centering
\begin{tabular}{|c|c|c|c|c|}
\hline
     $\mathfrak{v}^{(2)}(\bM_2) $   & $ U^{(2)}_{\bR}(a, b)_2$ & $U^{(2)}_{\bM_1}(a, b)_2$ & $U^{(2)}_{\bM_2}(a, b)_2$ & $\bg$-local symmetry defects \\ \hline
    $(\bm{1}, \bm{1}, \bm{1}, \bm{1}) $  &  1 &   1  & 1  &  $A_2 \otimes [\bm{1}, \bm{1}]_{\bR}, A_2 \otimes [\bm{1}, \bm{1}]_{\bM_1}, [\bm{1}, \bm{1}]_{\bM_2}$ \\ \hline
    $(e, e, \bm{1}, \bm{1})$    &  1   & $M^*_{ae}M^*_{be}$ & $M^*_{ae}M^*_{be}$ &  $A_2 \otimes [\bm{1}, \bm{1}]_{\bR}, A_2 \otimes [\bm{1}, \bm{1}]_{\bM_1}, [e, e]_{\bM_2}$    \\ \hline
    $(m, m, \bm{1}, \bm{1})$    &  1   & $M^*_{am}M^*_{bm}$ & $M^*_{am}M^*_{bm}$ &  $A_2 \otimes [\bm{1}, \bm{1}]_{\bR}, A_2 \otimes [\bm{1}, \bm{1}]_{\bM_1}, [m, m]_{\bM_2}$     \\ \hline
    $(f, f, \bm{1}, \bm{1})$    &  1   & $M^*_{af}M^*_{bf}$ & $M^*_{af}M^*_{bf}$ &  $A_2 \otimes [\bm{1}, \bm{1}]_{\bR}, A_2 \otimes [\bm{1}, \bm{1}]_{\bM_1}, [f, f]_{\bM_2}$       \\ \hline
\end{tabular}
\caption{The symmetry data of $A_2$-type boundary(E.3)}
\label{tab:CA2}
\end{table*}
The third example is toric code enriched by $D_4$ symmetry.
Due to the non-locality of $D_4$ symmetry, we study the SET order in four layers geometry.
And then, symmetry $D_4$ turns into onsite symmetry, with the group $\mathbb{D}_4 = \{\be, \bM_1, \bM_2, \bR \}$.
Considering trivial $D_4$ symmetry action, the induced anyon permutation for $\mathbb{D}_4$ symmetry is
\begin{equation}
\begin{aligned}
	\rho_{\bM_1} : (a, b, c, d) \rightarrow (b, a, d, c),  \\
	\rho_{\bM_2} : (a, b, c, d) \rightarrow (d, c, b, a).  
\end{aligned}
\end{equation}
It implies non-anomalous bulk and trivial symmetry fractionalization.
We choose the topological spins of symmetry defects as
\begin{gather}
   \theta^2_{[a, b]_{\bM_1}} = \theta^2_{[a, b]_{\bM_2}}	 = \theta^2_{[a, b]_{\bR}} = \theta_{a}\theta_{b},
\end{gather}
where we denote the symmetry defects $(a, a, b, b)_{\bM_1}$, $(a, b, b, a)_{\bM_2}$ and $(a, b, a, b)_{\bR}$ as $[a, b]_{\bM_1}$, $[a, b]_{\bM_2}$ and $[a, b]_{\bR}$ respectively.
The two boundaries are characterized by
\begin{equation} 
\begin{gathered}
	A_1 = \oplus_{a, b} (a, b)_1 ,\\
	A_2 = \oplus_{a, b} (a, b)_2 ,
\end{gathered}
\end{equation}
where we denote $(a, a, b, b)$ as $(a, b)_1$ and $(a, b, b, a)$ to $(a, b)_2$.
There are four nonequivalent boundary $U$-symbols of the $A_1$-type boundary shown in Table~\ref{tab:CA1}.
Similarly, four nonequivalent boundary $U$-symbols of the $A_2$-type boundary are shown in Table~\ref{tab:CA2}.
Regardless of the chosen boundary $U$-symbols, there is no $H^3$ obstruction on the two boundaries which correspond to the result of Sec.\ref{sec:D2n}.
Additionally, the $H^1$ obstruction also vanishes at the symmetric junction.
Thus, there are 16 combinations to construct the symmetric junction which correspond to different common $\bg$-local defects.
Similar to the second example, we choose the original standard model by fixing the boundary $U$-symbols:
\begin{align}
	U^{(1)}_{\bg}(a_1)= U^{(2)}_{\bg}(a_2) = 1
\end{align} 
where $a_i$ is an $A_i$-type condensed anyon.
Then, the original common $\bg$-local defects can be selected as $[\bm{1}, \bm{1}]_{\bM_1}$, $[\bm{1}, \bm{1}]_{\bM_2}$ and $[\bm{1}, \bm{1}]_R$.

Considering the case where $\mathfrak{v}^{(1)}(\bM_1) = (\bm{1}, e, e, \bm{1})$ for $A_1$-type boundary and $\mathfrak{v}^{(2)}(\bM_2) = (m, m, \bm{1}, \bm{1})$ for $A_2$-type boundary.
Following the discussion, we also define the relative $H^2$ anomaly indicator as
\begin{align}
	\eta_r = \frac{\gamma(\bM_1, \bM_2)}{\gamma(\bM_2, \bM_1)}.
\end{align} 
Referring to \Eq{Def.H^2}, we can obtain
\begin{equation} 
\begin{aligned}
	\eta_r = M^*_{(f, f, f, f)\mathfrak{c}},
\end{aligned}
\end{equation}
Since the torsor $\mathfrak{c}$ takes values in the set $\{(\bm{1}, \bm{1}, \bm{1}, \bm{1}), (\bm{1}, \bm{1}, \bm{1}, e), (\bm{1}, \bm{1}, \bm{1}, m), (\bm{1}, \bm{1}, \bm{1}, f) \}$, there is nontrivial relative anomaly indicator $M^*_{(f, f, f, f)\mathfrak{c}}$.
Similar to the second example, regardless of whether the original system is anomaly-free, we can always choose an appropriate torsor to construct a modified rotation center with a nontrivial $H^2$ obstruction.

\section{Conclusion}
\label{sec:conclusion}

In this work, we study the classification of SET orders enriched by 2D point-group symmetries using the folding approach.
Similar to the case of mirror-SETs, we fold the two-dimensional topological order into a multilayer system occupying the fundamental domain of the point group, which generally has the shape of a sector and contains two boundaries joining at a junction located at the rotation center.
The point-group symmetries then becomes onsite layer-permuting symmetries in this multilayer system.
The classification of point-group SETs is then solved by classifying symmetric gapped boundaries and junctions between two boundaries.
The gapped boundaries encode mirror-SETs, and was studied in previous works~\cite{qi2019folding,ding2024anomalies}.
In this work, we develop a general theory of classifying symmetric gapped junctions between boundaries, and apply it to study the junction at the rotation center, which contains additional information of point-group SETs.
Using this approach, we show that the junction may have two potential obstruction: an $H^1$ obstruction class in $H^1[G, \mathcal A_{\mathcal C}]$ and an $H^2$ obstruction class in $H^2[G, \mathrm U(1)]$.
%When the obstructions vanish, the junction is classified by an element in $H^0[G, \mathcal A_{\mathcal C}]/n\mathcal A_{\mathcal C}$ and a cohomology class in $H^1[G, \mathrm U(1)]$
Our classification result is consistent with the classification of onsite-SETs~\cite{barkeshli2019symmetry} and the crystalline equivalence principle~\cite{thorngren2018gauging}, where the $H^1$ and $H^2$ obstructions corresponds to the $H^3$ and $H^4$ obstructions for onsite-SETs.

In this work, we only give explicit formulas for the relative anomalies in the study of symmetric junctions in Sec.~\ref{sec:symmetric} and correspondingly when applying it to Sec.~\ref{sec:Cn} and Sec.~\ref{sec:D2n}.
It may be possible to derive formulas for the absolute anomalies, at least for simple topological orders such as Abelian ones.
We will leave this to future works.
Our classification result may also be used to construct lattice models of point-group SETs, where the interior of the sector can be constructed using Levin-Wen models without symmetry enrichment, and the boundaries between sectors and the rotation center are constructed using the symmetric boundary and junction theories.
We also leave this to future works.

\begin{acknowledgments}
	The authors thank Chenjie Wang for insightful discussions.
	Y.Q. acknowledges support from National Key R\&D Program of China (Grant No. 2022YFA1403402), from the National Natural Science Foundation of China (Grant No. 12174068), and from the Science and Technology Commission of Shanghai Municipality (Grant Nos. 24LZ1400100 and 23JC1400600).
\end{acknowledgments}

\appendix
\section{Gauge equivalences of the $L$-symbols}
\label{App:gauge}
According to \Eq{9}, after fixing the special gauge-independent $L$-symbols $\hL^a$, it is known that there are different solutions of the other $L$-symbols $[L^{a}_{X}]_{(i, \mu_{i})(j, \mu_{j})}$.
We aim to prove that the different solutions are gauge equivalent.

Solving \Eq{9}, there are two sets of solutions $[L^a_{X}]$ and $[\tilde{L}^a_{X}]$, where we fix the special $L$-symbols as $\hL^a = \hat{\tilde{L}}^a$.
Then, we choose topological bases $\ket{a; i, \mu_{i}, n^{(1)}_{i}}_{X}$ and $\ket{a; j, \mu_{j}, n^{(2)}_{j}}_{X}$, which can be related by $L$-symbols $[L^a_{X}]$, as
\begin{align}
	\ket{a; i, \mu_{i}, n^{(1)}_{i}}_{X} &= \notag \\
	&\sum_{j, \mu_{j}, n^{(2)}_{j}}  [L^{a}_{X}]_{(i, \mu_{i})(j, \mu_{j})} \ket{a; j, \mu_{j}, n^{(2)}_{j}}_{X}.
\end{align}
We consider the gauge transformations as
\begin{equation}
\begin{aligned}
    &\widetilde{\ket{b; A_1, 1, 1}}_{X} = \ket{b; A_1, 1, 1}_{X},\\
	&\widetilde{\ket{c; i, \mu_{i}, n^{(1)}_{i}}}_{X} = \frac{[\tilde{L}^c_X]_{(i, \mu_i)(A_2, 1)}}{[L^c_X]_{(i, \mu_i)(A_2, 1)}} \ket{c; i, \mu_{i}, n^{(1)}_{i}}_{X},\\
	&\widetilde{\ket{c; A_2, 1, 1}}_{X} = \ket{c; A_2, 1, 1}_{X},\\
	&\widetilde{\ket{b; j, \mu_{j}, n^{(2)}_{j}}}_{X} = \frac{[L^b_X]_{(A_1, 1)(j, \mu_j)}}{[\tilde{L}^b_X]_{(A_1, 1)(j, \mu_j)}} \ket{b; j, \mu_{j}, n^{(2)}_{j}}_{X},\\
	\label{gt1}
\end{aligned}
\end{equation}
where $b$ and $c$ are condensed anyons on the $A_1$-type boundary and $A_2$-type boundary, respectively.
For the other bases, the factors of the gauge transformations can be determined by 
\begin{equation}
\begin{aligned}
	\zeta^a_i = \zeta^b_{A_1} \zeta^c_i \quad \text{for $A_1$-type boundary},\\
	\zeta^a_j = \zeta^b_j \zeta^c_{A_2} \quad \text{for $A_2$-type boundary},
	\label{gt2}
\end{aligned}
\end{equation}
which ensures that the VLCs remain invariant under the gauge transformations.
Here, the anyon $a$ can be always produced by the fusion of an $A_1$-type condensed anyon $b$ and an $A_2$-type condensed anyon $c$.
Then, combining \Eq{9}, we can verify 
\begin{align}
	\widetilde{\ket{a; i, \mu_{i}, n^{(1)}_{i}}}_{X} &= \notag \\
	&\sum_{j, \mu_{j}, n^{(2)}_{j}}  [\tilde{L}^{a}_{X}]_{(i, \mu_{i})(j, \mu_{j})} \widetilde{\ket{a; j, \mu_{j}, n^{(2)}_{j}}}_{X}.
\end{align}
It implies that the two sets of solutions, $[L^a_{X}]$ and $[\tilde{L}^a_{X}]$, can be related by the gauge transformations~(\ref{gt1}, \ref{gt2}).

\section{The existence of common $\bg$-local defects for abelian cases}
\label{app:com-def}
At a junction, when $H^1$ obstruction vanishes, we have \Eq{20}.
Then, we consider the following moves.
\begin{figure}[H]
	\centering
	\includegraphics[width=0.44\textwidth, height=0.35\textwidth]{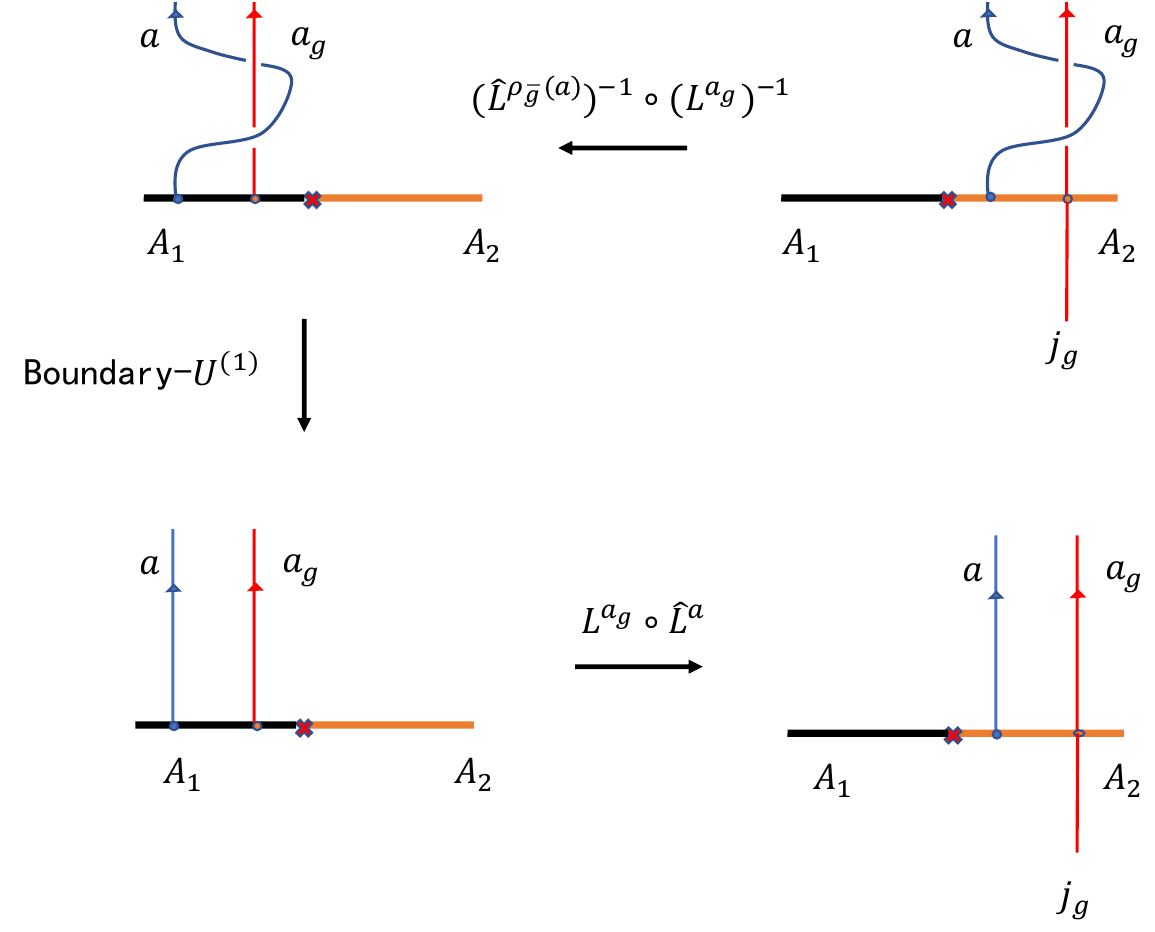}
\end{figure}
\noindent
Here, $a$ is a common condensed anyon.
The defect $a_{\bg}$ is an $\bg$-local symmetry defect of the $A_1$-type boundary and can become boundary defect $j_{\bg}$ on the $A_2$-type boundary.
Applying the \Eq{20}, for the common condensed anyon $a$, the defect $a_{\bg}$ satisfies associated $\bg$-local condition on the $A_2$-type boundary. 

Since anyons are all abelian, there are closed fusion rules for condensed anyons on the two boundary.
Thus, we define two fusion subcategories $\cE_1$ and $\cE_2$, whose simple objects consist of the condensed anyons on the respective boundaries.
Furthermore, we can define another two subcategories $\cE_1 \vee \cE_2$ and $\cE_1 \cap \cE_2$.
The former is the smallest fusion subcategory of $\cC$ that contains $\cE_1$ and $\cE_2$, while the later is the biggest common fusion subcategory of $\cE_1$ and $\cE_2$.
For the case that the bulk $\cC$ is non-degenerate, which is typically considered in physics, we have~\cite{etingof2015tensor}
\begin{equation}
\begin{aligned}
	& \text{Dim}(\cE_1 \cap \cE_2) \text{Dim}(\cE_1 \vee \cE_2) = \text{Dim}(\cE_{1})\text{Dim}(\cE_{2}) ,\\
	&\text{Dim}(\cE_1 \cap \cE_2)\text{Dim}((\cE_1 \cap \cE_2)') = \text{Dim}(\cC)
	\label{A1}
\end{aligned}
\end{equation}
where $(\cE_1 \cap \cE_2)'$ denotes the centralizer of $\cE_1 \cap \cE_2$.
Since both $A_1$ and $A_2$ are Lagrangian algebras of $\cC$, we have~\cite{kong2014anyon}
\begin{align}
	\text{Dim}^2(\cE_1) = \text{Dim}^2(\cE_2) = \text{Dim}(\cC)
	\label{A2}
\end{align}
Referring to the braidings between condensed anyons, we directly have $\cE_1 \vee \cE_2 \subseteq (\cE_1 \cap \cE_2)'$.
Furthermore, comparing \Eqs{A1}{A2}, we obtain 
\begin{align}
	(\cE_1 \cap \cE_2)' = \cE_1 \vee \cE_2.
	\label{A3}
\end{align}
 
Next, we aim to prove that there exists common $\bg$-local defects by reductio.
Assuming that there is no common $\bg$-local defect, the $\bg$-local defect $a_{\bg}$ of the $A_1$-type boundary should absorb an anyon $c$ to become a $\bg$-local defect $c_{\bg}$ of the $A_2$-type boundary.
In this case, the anyon $c$ can neither condense on the $A_1$-type boundary nor on the $A_2$-type boundary, meaning $c \notin \cE_1 \vee \cE_2$. 
Furthermore, considering common condensed anyons, both $a_{\bg}$ and $c_{\bg}$ should satisfy associated $\bg$-local conditions on the two boundaries.
It implies that the absorbed anyon $c$ must have trivial braiding with the common condensed anyons, meaning $c \in (\cE_1 \cap \cE_2)'$.
According to \Eq{A3}, the above conclusion leads to a contradiction.
Therefore, our initial assumption must be false, and we conclude that the common $\bg$-local defects indeed exists.

\section{The calculation of a $H^2$ cohomology group}
\label{app:H^2} 
We construct a proper resolution by the free resolution~\cite{wall1961resolutions} of group $\mathbb{D}_{2n}(n \geq 2 )$
\begin{equation}
\begin{aligned}
    (b^4_i)F_4 &: ~a_{0, 4}; ~ \bM a_{0, 4}+ a_{1, 3} + \bM a_{2, 2}; ~ a_{2, 2}; ~a_{3, 1}; ~a_{4, 0}, \\
	(b^3_i)F_3 &: ~a_{0, 3}; ~ \bM a_{0, 3}+ a_{1, 2} + \bM a_{2, 1}; ~ a_{2, 1}; ~ a_{3, 0}, \\
	(b^2_i)F_2 &: ~ a_{0, 2}; ~ \bM a_{0, 2} +  a_{1, 1} + \bM a_{2, 0}; ~ a_{2, 0}, \\
	(b^1_i)F_1 &: ~ a_{0, 1}; ~ \bM a_{0, 1}+ a_{1, 0}, \\
	(b^0_i)F_0 &: ~ a_{0, 0},
\end{aligned}
\end{equation}
where $a_{i, j}$ is the free resolution.
Considering a $\mathbb{D}_{2n}$-module $N$, we have 
\begin{gather}
     H^2_{\tilde{\rho}}[\mathbb{D}_{2n}, N] = H^2(\text{Hom}_{\mathbb{D}_{2n}}(F, N)),
\end{gather}
where the map $\tilde{\rho}$ is defined as \Eq{ReAction}.
Thus, we can obtain $H^2_{\tilde{\rho}}[\mathbb{D}_{2n}, N]$ by calculating $H^2(\text{Hom}_{\mathbb{D}_{2n}}(F, N))$.
For $\alpha \in \text{Hom}_{\mathbb{D}_{2n}}(F, N)$, we list the 3-rank coboundary maps
\begin{equation}
\begin{aligned}
	d\alpha(b^3_1)=& \frac{^{\bM}\alpha(b^2_1)}{\alpha(b^2_1)},\\
	d\alpha(b^3_2)=& \frac{{^{\bR \bM}\alpha}(b^2_2)}{\alpha(b^2_2)}\frac{\prod^{n-2}_{k=0}{^{{\bR}^k}\alpha}(b^2_3)}{\alpha^{n-2}(b^2_3) {^{\bR}\alpha}(b^2_3)} ,\\ 
	d\alpha(b^3_3)=& \frac{ \prod^{n-1}_{k=0} {^{{\bR}^k}\alpha}(b^2_2) }{ ^{\bR \bM}\alpha(b^2_3) \alpha(b^2_3) \prod^{n-1}_{k=0} {^{{\bR}^k \bM} \alpha}(b^2_1) } ,\\
	d\alpha(b^3_4)=&  \frac{{^{\bR}\alpha}(b^2_3)}{\alpha(b^2_3)}.
\end{aligned}	
\end{equation}
Similarly, the 2-rank coboundary maps are 
\begin{equation}
\begin{aligned}
  	d\alpha(b^2_1) =& \alpha(b^1_1) {^{\bM}\alpha(b^1_1) } ,\\
  	d\alpha(b^2_2) =& \alpha(b^1_2) {^{\bR \bM} \alpha(b^1_2) } ,\\
  	d\alpha(b^2_3) =& \frac{\prod^{n-1}_{k=0} {^{ {\bR}^k } \alpha}(b^1_2) }{\prod^{n-1}_{k=0} {^{{\bR}^k \bM} \alpha}(b^1_1)}.
\end{aligned}
\end{equation}
After taking the quotient by the coboundary $d\alpha(b^2_1)$, we can derive the basis $\alpha(b^2_1)$ by solving the cocycle equation $d\alpha(b^3_1) = \text{id}_N$ where the identity of $N$ is denoted as $\text{id}_N$.
The basis $\alpha(b^2_1)$ are classified by the cohomology group $H^1_{\rho}[\mathbb{Z}^{\mb}_2, N]$.
And then, we simplify the cocycle equation $d\alpha(b^3_2) = \text{id}_N$ by the cocycle equation $d\alpha(b^3_4)=\text{id}_N$. 
That is 
\begin{align}
	{^{\bR \bM}\alpha}(b^2_2) = \alpha(b^2_2).
\end{align}
When $n$ is an odd integer, the cocycle equation $d\alpha(b^3_3) = \text{id}_N$ implies that the basis $\alpha(b^2_2)$ can always be expressed as a coboundary $a {^{\bR \bM}a}$.
Here, $a$ is
\begin{gather}
     a = \frac{\prod_{k = 1}^{\frac{n-1}{2}} {^{{\bR}^k}}\alpha(b^2_2) }	{\alpha(b^2_3)}.
\end{gather}
In this case, the basis $\alpha(b^2_2)$ is trivial.
When $n$ is an even integer, the basis $\alpha(b^2_2)$ is classified by cohomology group $H^1_{\rho}[\mathbb{Z}^{\br \mb}_2, N]$.
Furthermore, when we fix the two bases $\alpha(b^2_1) = \alpha(b^2_2) = \text{id}_N$, the coboundary maps force
\begin{equation}
\begin{aligned}
	 &\alpha(b^1_1) {^{\bM}\alpha(b^1_1) } &= \text{id}_N ,\\
  	 &\alpha(b^1_2) {^{\bR \bM} \alpha(b^1_2) } &= \text{id}_N.
\end{aligned}
\end{equation}
Then, the cocycle equations $d\alpha(b^3_3) = \text{id}_N$ and $d\alpha(b^3_4) = \text{id}_N$ imply that the basis $\alpha(b^2_3)$ is $D_{2n}$-invariant.
As a result, the basis $\alpha(b^2_3)$ is classified by $[\cA^{D_{2n}}_{\cC}]_{2n\cA_{\cC}}$.
Totally, we have proved 
\begin{widetext}
\[ H^2_{\tilde{\rho}}[\mathbb{D}_{2n}, N] =
	\begin{cases}
		H^1_{\rho}[\mathbb{Z}^{\mb}_2, N] \oplus [\cA^{D_{2n}}_{\cC}]_{2n\cA_{\cC}} \quad  &\text{for odd} ~n,\\
	 H^1_{\rho}[\mathbb{Z}^{\mb}_2, N] \oplus H^1_{\rho}[\mathbb{Z}^{\br \mb}_2, N] \oplus [\cA^{D_{2n}}_{\cC}]_{2n\cA_{\cC}} \quad  &\text{for even} ~ n.
	\end{cases}
\]
\end{widetext}

\section{The $F$-symbols and additional VLCs for modified $\bg$-local defects}
\label{app:modified} 
We present the following two figures to illustrate the $F$-symbols and additional VLCs of the modified $\bg$-local defects.
\begin{figure}[H]
	\centering
	\includegraphics[width=0.44\textwidth, height=0.28\textwidth]{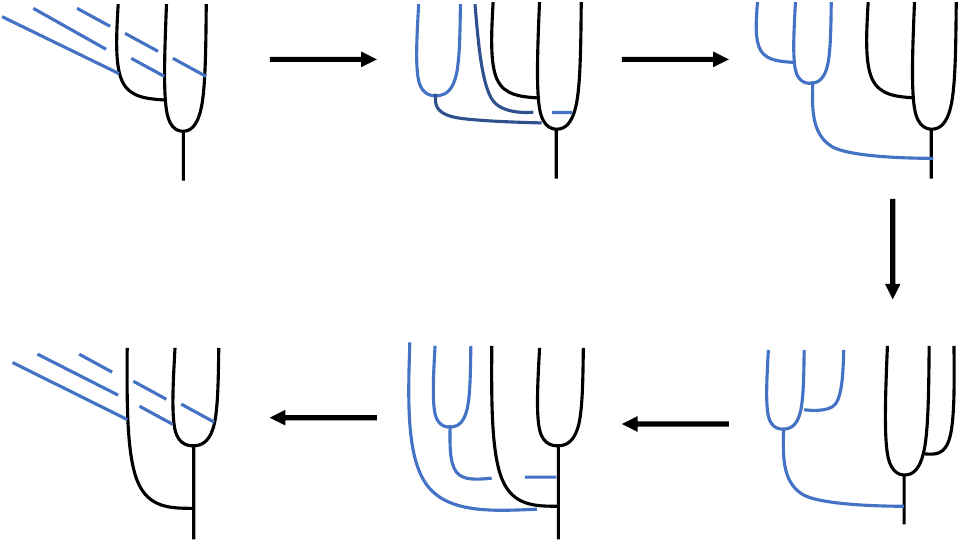}
\end{figure}
\noindent      
In this case, the $F$-move of the modified $\bg$-local defects can be decomposed into a series of moves that include the $F$-move of the original $\bg$-local defects.
\begin{figure}[H]
	\centering
	\includegraphics[width=0.44\textwidth, height=0.28\textwidth]{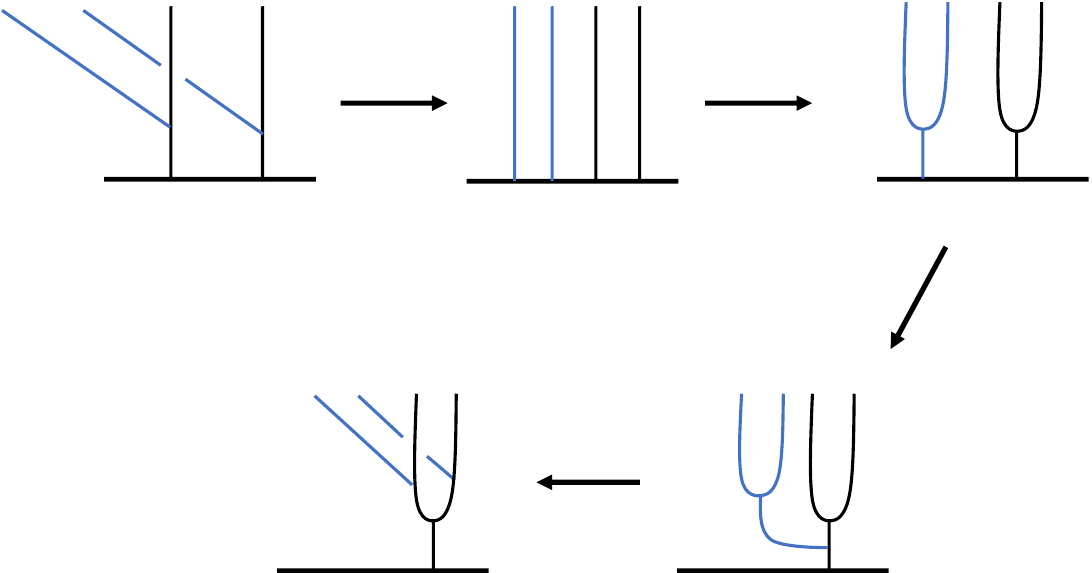}
\end{figure}
\noindent
As shown in the figure above, the additional VLCs of the modified $\bg$-local defects can be related to the additional VLCs of the original $\bg$-local defects.

\bibliography{reference}

\begin{thebibliography}{57}%
\makeatletter
\providecommand \@ifxundefined [1]{%
 \@ifx{#1\undefined}
}%
\providecommand \@ifnum [1]{%
 \ifnum #1\expandafter \@firstoftwo
 \else \expandafter \@secondoftwo
 \fi
}%
\providecommand \@ifx [1]{%
 \ifx #1\expandafter \@firstoftwo
 \else \expandafter \@secondoftwo
 \fi
}%
\providecommand \natexlab [1]{#1}%
\providecommand \enquote  [1]{``#1''}%
\providecommand \bibnamefont  [1]{#1}%
\providecommand \bibfnamefont [1]{#1}%
\providecommand \citenamefont [1]{#1}%
\providecommand \href@noop [0]{\@secondoftwo}%
\providecommand \href [0]{\begingroup \@sanitize@url \@href}%
\providecommand \@href[1]{\@@startlink{#1}\@@href}%
\providecommand \@@href[1]{\endgroup#1\@@endlink}%
\providecommand \@sanitize@url [0]{\catcode `\\12\catcode `\$12\catcode `\&12\catcode `\#12\catcode `\^12\catcode `\_12\catcode `\%12\relax}%
\providecommand \@@startlink[1]{}%
\providecommand \@@endlink[0]{}%
\providecommand \url  [0]{\begingroup\@sanitize@url \@url }%
\providecommand \@url [1]{\endgroup\@href {#1}{\urlprefix }}%
\providecommand \urlprefix  [0]{URL }%
\providecommand \Eprint [0]{\href }%
\providecommand \doibase [0]{https://doi.org/}%
\providecommand \selectlanguage [0]{\@gobble}%
\providecommand \bibinfo  [0]{\@secondoftwo}%
\providecommand \bibfield  [0]{\@secondoftwo}%
\providecommand \translation [1]{[#1]}%
\providecommand \BibitemOpen [0]{}%
\providecommand \bibitemStop [0]{}%
\providecommand \bibitemNoStop [0]{.\EOS\space}%
\providecommand \EOS [0]{\spacefactor3000\relax}%
\providecommand \BibitemShut  [1]{\csname bibitem#1\endcsname}%
\let\auto@bib@innerbib\@empty
%</preamble>
\bibitem [{\citenamefont {Wen}(1990)}]{Wen1990}%
  \BibitemOpen
  \bibfield  {author} {\bibinfo {author} {\bibfnamefont {X.~G.}\ \bibnamefont {Wen}},\ }\bibfield  {title} {\bibinfo {title} {Topological orders in rigid states},\ }\href {https://doi.org/10.1142/S0217979290000139} {\bibfield  {journal} {\bibinfo  {journal} {Int. J. Mod. Phys. B}\ }\textbf {\bibinfo {volume} {04}},\ \bibinfo {pages} {239} (\bibinfo {year} {1990})}\BibitemShut {NoStop}%
\bibitem [{\citenamefont {Wilczek}(1990)}]{wilczek1990fractional}%
  \BibitemOpen
  \bibfield  {author} {\bibinfo {author} {\bibfnamefont {F.}~\bibnamefont {Wilczek}},\ }\href {https://www.worldscientific.com/worldscibooks/10.1142/0961#t=aboutBook} {\emph {\bibinfo {title} {Fractional statistics and anyon superconductivity}}},\ Vol.~\bibinfo {volume} {5}\ (\bibinfo  {publisher} {World scientific},\ \bibinfo {year} {1990})\BibitemShut {NoStop}%
\bibitem [{\citenamefont {Wen}(2015)}]{Wen2015nsr}%
  \BibitemOpen
  \bibfield  {author} {\bibinfo {author} {\bibfnamefont {X.-G.}\ \bibnamefont {Wen}},\ }\bibfield  {title} {\bibinfo {title} {{A theory of 2+1D bosonic topological orders}},\ }\href {https://doi.org/10.1093/nsr/nwv077} {\bibfield  {journal} {\bibinfo  {journal} {Natl. Sci. Rev.}\ }\textbf {\bibinfo {volume} {3}},\ \bibinfo {pages} {68} (\bibinfo {year} {2015})}\BibitemShut {NoStop}%
\bibitem [{\citenamefont {Wen}(2002)}]{XGWenPSG2002}%
  \BibitemOpen
  \bibfield  {author} {\bibinfo {author} {\bibfnamefont {X.-G.}\ \bibnamefont {Wen}},\ }\bibfield  {title} {\bibinfo {title} {Quantum orders and symmetric spin liquids},\ }\href {https://doi.org/10.1103/PhysRevB.65.165113} {\bibfield  {journal} {\bibinfo  {journal} {Phys. Rev. B}\ }\textbf {\bibinfo {volume} {65}},\ \bibinfo {pages} {165113} (\bibinfo {year} {2002})}\BibitemShut {NoStop}%
\bibitem [{\citenamefont {Kitaev}(2006)}]{Kitaev2006}%
  \BibitemOpen
  \bibfield  {author} {\bibinfo {author} {\bibfnamefont {A.}~\bibnamefont {Kitaev}},\ }\bibfield  {title} {\bibinfo {title} {Anyons in an exactly solved model and beyond},\ }\href {https://doi.org/https://doi.org/10.1016/j.aop.2005.10.005} {\bibfield  {journal} {\bibinfo  {journal} {Ann. Phys.}\ }\textbf {\bibinfo {volume} {321}},\ \bibinfo {pages} {2} (\bibinfo {year} {2006})}\BibitemShut {NoStop}%
\bibitem [{\citenamefont {Chen}\ \emph {et~al.}(2010)\citenamefont {Chen}, \citenamefont {Gu},\ and\ \citenamefont {Wen}}]{XieChen2010LUT}%
  \BibitemOpen
  \bibfield  {author} {\bibinfo {author} {\bibfnamefont {X.}~\bibnamefont {Chen}}, \bibinfo {author} {\bibfnamefont {Z.-C.}\ \bibnamefont {Gu}},\ and\ \bibinfo {author} {\bibfnamefont {X.-G.}\ \bibnamefont {Wen}},\ }\bibfield  {title} {\bibinfo {title} {Local unitary transformation, long-range quantum entanglement, wave function renormalization, and topological order},\ }\href {https://doi.org/10.1103/PhysRevB.82.155138} {\bibfield  {journal} {\bibinfo  {journal} {Phys. Rev. B}\ }\textbf {\bibinfo {volume} {82}},\ \bibinfo {pages} {155138} (\bibinfo {year} {2010})}\BibitemShut {NoStop}%
\bibitem [{\citenamefont {Mesaros}\ and\ \citenamefont {Ran}(2013)}]{Mesaros2013}%
  \BibitemOpen
  \bibfield  {author} {\bibinfo {author} {\bibfnamefont {A.}~\bibnamefont {Mesaros}}\ and\ \bibinfo {author} {\bibfnamefont {Y.}~\bibnamefont {Ran}},\ }\bibfield  {title} {\bibinfo {title} {Classification of symmetry enriched topological phases with exactly solvable models},\ }\href {https://doi.org/10.1103/PhysRevB.87.155115} {\bibfield  {journal} {\bibinfo  {journal} {Phys. Rev. B}\ }\textbf {\bibinfo {volume} {87}},\ \bibinfo {pages} {155115} (\bibinfo {year} {2013})}\BibitemShut {NoStop}%
\bibitem [{\citenamefont {Essin}\ and\ \citenamefont {Hermele}(2013)}]{Essin2013}%
  \BibitemOpen
  \bibfield  {author} {\bibinfo {author} {\bibfnamefont {A.~M.}\ \bibnamefont {Essin}}\ and\ \bibinfo {author} {\bibfnamefont {M.}~\bibnamefont {Hermele}},\ }\bibfield  {title} {\bibinfo {title} {Classifying fractionalization: Symmetry classification of gapped ${\mathbb{z}}_{2}$ spin liquids in two dimensions},\ }\href {https://doi.org/10.1103/PhysRevB.87.104406} {\bibfield  {journal} {\bibinfo  {journal} {Phys. Rev. B}\ }\textbf {\bibinfo {volume} {87}},\ \bibinfo {pages} {104406} (\bibinfo {year} {2013})}\BibitemShut {NoStop}%
\bibitem [{\citenamefont {Qi}\ and\ \citenamefont {Fu}(2015{\natexlab{a}})}]{YQi2015}%
  \BibitemOpen
  \bibfield  {author} {\bibinfo {author} {\bibfnamefont {Y.}~\bibnamefont {Qi}}\ and\ \bibinfo {author} {\bibfnamefont {L.}~\bibnamefont {Fu}},\ }\bibfield  {title} {\bibinfo {title} {Detecting crystal symmetry fractionalization from the ground state: Application to ${\mathbb{z}}_{2}$ spin liquids on the kagome lattice},\ }\href {https://doi.org/10.1103/PhysRevB.91.100401} {\bibfield  {journal} {\bibinfo  {journal} {Phys. Rev. B}\ }\textbf {\bibinfo {volume} {91}},\ \bibinfo {pages} {100401(R)} (\bibinfo {year} {2015}{\natexlab{a}})}\BibitemShut {NoStop}%
\bibitem [{\citenamefont {Tarantino}\ \emph {et~al.}(2016)\citenamefont {Tarantino}, \citenamefont {Lindner},\ and\ \citenamefont {Fidkowski}}]{Tarantino_2016}%
  \BibitemOpen
  \bibfield  {author} {\bibinfo {author} {\bibfnamefont {N.}~\bibnamefont {Tarantino}}, \bibinfo {author} {\bibfnamefont {N.~H.}\ \bibnamefont {Lindner}},\ and\ \bibinfo {author} {\bibfnamefont {L.}~\bibnamefont {Fidkowski}},\ }\bibfield  {title} {\bibinfo {title} {Symmetry fractionalization and twist defects},\ }\href {https://doi.org/10.1088/1367-2630/18/3/035006} {\bibfield  {journal} {\bibinfo  {journal} {New J. Phys.}\ }\textbf {\bibinfo {volume} {18}},\ \bibinfo {pages} {035006} (\bibinfo {year} {2016})}\BibitemShut {NoStop}%
\bibitem [{\citenamefont {Heinrich}\ \emph {et~al.}(2016)\citenamefont {Heinrich}, \citenamefont {Burnell}, \citenamefont {Fidkowski},\ and\ \citenamefont {Levin}}]{Heinrich2016}%
  \BibitemOpen
  \bibfield  {author} {\bibinfo {author} {\bibfnamefont {C.}~\bibnamefont {Heinrich}}, \bibinfo {author} {\bibfnamefont {F.}~\bibnamefont {Burnell}}, \bibinfo {author} {\bibfnamefont {L.}~\bibnamefont {Fidkowski}},\ and\ \bibinfo {author} {\bibfnamefont {M.}~\bibnamefont {Levin}},\ }\bibfield  {title} {\bibinfo {title} {Symmetry-enriched string nets: Exactly solvable models for set phases},\ }\href {https://doi.org/10.1103/PhysRevB.94.235136} {\bibfield  {journal} {\bibinfo  {journal} {Phys. Rev. B}\ }\textbf {\bibinfo {volume} {94}},\ \bibinfo {pages} {235136} (\bibinfo {year} {2016})}\BibitemShut {NoStop}%
\bibitem [{\citenamefont {Cheng}\ \emph {et~al.}(2017)\citenamefont {Cheng}, \citenamefont {Gu}, \citenamefont {Jiang},\ and\ \citenamefont {Qi}}]{MengCheng2017}%
  \BibitemOpen
  \bibfield  {author} {\bibinfo {author} {\bibfnamefont {M.}~\bibnamefont {Cheng}}, \bibinfo {author} {\bibfnamefont {Z.-C.}\ \bibnamefont {Gu}}, \bibinfo {author} {\bibfnamefont {S.}~\bibnamefont {Jiang}},\ and\ \bibinfo {author} {\bibfnamefont {Y.}~\bibnamefont {Qi}},\ }\bibfield  {title} {\bibinfo {title} {Exactly solvable models for symmetry-enriched topological phases},\ }\href {https://doi.org/10.1103/PhysRevB.96.115107} {\bibfield  {journal} {\bibinfo  {journal} {Phys. Rev. B}\ }\textbf {\bibinfo {volume} {96}},\ \bibinfo {pages} {115107} (\bibinfo {year} {2017})}\BibitemShut {NoStop}%
\bibitem [{\citenamefont {Sun}\ \emph {et~al.}(2018)\citenamefont {Sun}, \citenamefont {Wang}, \citenamefont {Fang}, \citenamefont {Qi}, \citenamefont {Cheng},\ and\ \citenamefont {Meng}}]{sun2018dynamical}%
  \BibitemOpen
  \bibfield  {author} {\bibinfo {author} {\bibfnamefont {G.-Y.}\ \bibnamefont {Sun}}, \bibinfo {author} {\bibfnamefont {Y.-C.}\ \bibnamefont {Wang}}, \bibinfo {author} {\bibfnamefont {C.}~\bibnamefont {Fang}}, \bibinfo {author} {\bibfnamefont {Y.}~\bibnamefont {Qi}}, \bibinfo {author} {\bibfnamefont {M.}~\bibnamefont {Cheng}},\ and\ \bibinfo {author} {\bibfnamefont {Z.~Y.}\ \bibnamefont {Meng}},\ }\bibfield  {title} {\bibinfo {title} {Dynamical signature of symmetry fractionalization in frustrated magnets},\ }\href {https://doi.org/10.1103/PhysRevLett.121.077201} {\bibfield  {journal} {\bibinfo  {journal} {Phys. Rev. Lett.}\ }\textbf {\bibinfo {volume} {121}},\ \bibinfo {pages} {077201} (\bibinfo {year} {2018})}\BibitemShut {NoStop}%
\bibitem [{\citenamefont {Barkeshli}\ \emph {et~al.}(2014)\citenamefont {Barkeshli}, \citenamefont {Berg},\ and\ \citenamefont {Kivelson}}]{barkeshli_coherent_2014}%
  \BibitemOpen
  \bibfield  {author} {\bibinfo {author} {\bibfnamefont {M.}~\bibnamefont {Barkeshli}}, \bibinfo {author} {\bibfnamefont {E.}~\bibnamefont {Berg}},\ and\ \bibinfo {author} {\bibfnamefont {S.}~\bibnamefont {Kivelson}},\ }\bibfield  {title} {\bibinfo {title} {Coherent transmutation of electrons into fractionalized anyons},\ }\href {https://doi.org/10.1126/science.1253251} {\bibfield  {journal} {\bibinfo  {journal} {Science}\ }\textbf {\bibinfo {volume} {346}},\ \bibinfo {pages} {722} (\bibinfo {year} {2014})}\BibitemShut {NoStop}%
\bibitem [{\citenamefont {Barkeshli}\ \emph {et~al.}(2013)\citenamefont {Barkeshli}, \citenamefont {Jian},\ and\ \citenamefont {Qi}}]{barkeshli2013twist}%
  \BibitemOpen
  \bibfield  {author} {\bibinfo {author} {\bibfnamefont {M.}~\bibnamefont {Barkeshli}}, \bibinfo {author} {\bibfnamefont {C.-M.}\ \bibnamefont {Jian}},\ and\ \bibinfo {author} {\bibfnamefont {X.-L.}\ \bibnamefont {Qi}},\ }\bibfield  {title} {\bibinfo {title} {Twist defects and projective non-abelian braiding statistics},\ }\href {https://doi.org/10.1103/PhysRevB.87.045130} {\bibfield  {journal} {\bibinfo  {journal} {Phys. Rev. B}\ }\textbf {\bibinfo {volume} {87}},\ \bibinfo {pages} {045130} (\bibinfo {year} {2013})}\BibitemShut {NoStop}%
\bibitem [{\citenamefont {Barkeshli}\ \emph {et~al.}(2019)\citenamefont {Barkeshli}, \citenamefont {Bonderson}, \citenamefont {Cheng},\ and\ \citenamefont {Wang}}]{barkeshli2019symmetry}%
  \BibitemOpen
  \bibfield  {author} {\bibinfo {author} {\bibfnamefont {M.}~\bibnamefont {Barkeshli}}, \bibinfo {author} {\bibfnamefont {P.}~\bibnamefont {Bonderson}}, \bibinfo {author} {\bibfnamefont {M.}~\bibnamefont {Cheng}},\ and\ \bibinfo {author} {\bibfnamefont {Z.}~\bibnamefont {Wang}},\ }\bibfield  {title} {\bibinfo {title} {Symmetry fractionalization, defects, and gauging of topological phases},\ }\href {https://doi.org/10.1103/PhysRevB.100.115147} {\bibfield  {journal} {\bibinfo  {journal} {Phys. Rev. B}\ }\textbf {\bibinfo {volume} {100}},\ \bibinfo {pages} {115147} (\bibinfo {year} {2019})}\BibitemShut {NoStop}%
\bibitem [{\citenamefont {Thorngren}\ and\ \citenamefont {Else}(2018)}]{thorngren2018gauging}%
  \BibitemOpen
  \bibfield  {author} {\bibinfo {author} {\bibfnamefont {R.}~\bibnamefont {Thorngren}}\ and\ \bibinfo {author} {\bibfnamefont {D.~V.}\ \bibnamefont {Else}},\ }\bibfield  {title} {\bibinfo {title} {Gauging spatial symmetries and the classification of topological crystalline phases},\ }\href {https://doi.org/10.1103/PhysRevX.8.011040} {\bibfield  {journal} {\bibinfo  {journal} {Phys. Rev. X}\ }\textbf {\bibinfo {volume} {8}},\ \bibinfo {pages} {011040} (\bibinfo {year} {2018})}\BibitemShut {NoStop}%
\bibitem [{\citenamefont {Song}\ \emph {et~al.}(2020)\citenamefont {Song}, \citenamefont {Fang},\ and\ \citenamefont {Qi}}]{Song2020}%
  \BibitemOpen
  \bibfield  {author} {\bibinfo {author} {\bibfnamefont {Z.}~\bibnamefont {Song}}, \bibinfo {author} {\bibfnamefont {C.}~\bibnamefont {Fang}},\ and\ \bibinfo {author} {\bibfnamefont {Y.}~\bibnamefont {Qi}},\ }\bibfield  {title} {\bibinfo {title} {Real-space recipes for general topological crystalline states},\ }\href {https://doi.org/10.1038/s41467-020-17685-5} {\bibfield  {journal} {\bibinfo  {journal} {Nat. Commun.}\ }\textbf {\bibinfo {volume} {11}},\ \bibinfo {pages} {4197} (\bibinfo {year} {2020})}\BibitemShut {NoStop}%
\bibitem [{\citenamefont {Else}\ and\ \citenamefont {Thorngren}(2019)}]{Else2019}%
  \BibitemOpen
  \bibfield  {author} {\bibinfo {author} {\bibfnamefont {D.~V.}\ \bibnamefont {Else}}\ and\ \bibinfo {author} {\bibfnamefont {R.}~\bibnamefont {Thorngren}},\ }\bibfield  {title} {\bibinfo {title} {Crystalline topological phases as defect networks},\ }\href {https://doi.org/10.1103/PhysRevB.99.115116} {\bibfield  {journal} {\bibinfo  {journal} {Phys. Rev. B}\ }\textbf {\bibinfo {volume} {99}},\ \bibinfo {pages} {115116} (\bibinfo {year} {2019})}\BibitemShut {NoStop}%
\bibitem [{\citenamefont {Isobe}\ and\ \citenamefont {Fu}(2015)}]{isobe2015theory}%
  \BibitemOpen
  \bibfield  {author} {\bibinfo {author} {\bibfnamefont {H.}~\bibnamefont {Isobe}}\ and\ \bibinfo {author} {\bibfnamefont {L.}~\bibnamefont {Fu}},\ }\bibfield  {title} {\bibinfo {title} {Theory of interacting topological crystalline insulators},\ }\href {https://doi.org/10.1103/PhysRevB.92.081304} {\bibfield  {journal} {\bibinfo  {journal} {Phys. Rev. B}\ }\textbf {\bibinfo {volume} {92}},\ \bibinfo {pages} {081304} (\bibinfo {year} {2015})}\BibitemShut {NoStop}%
\bibitem [{\citenamefont {Song}\ \emph {et~al.}(2017)\citenamefont {Song}, \citenamefont {Huang}, \citenamefont {Fu},\ and\ \citenamefont {Hermele}}]{song2017topological}%
  \BibitemOpen
  \bibfield  {author} {\bibinfo {author} {\bibfnamefont {H.}~\bibnamefont {Song}}, \bibinfo {author} {\bibfnamefont {S.-J.}\ \bibnamefont {Huang}}, \bibinfo {author} {\bibfnamefont {L.}~\bibnamefont {Fu}},\ and\ \bibinfo {author} {\bibfnamefont {M.}~\bibnamefont {Hermele}},\ }\bibfield  {title} {\bibinfo {title} {Topological phases protected by point group symmetry},\ }\href {https://doi.org/10.1103/PhysRevX.7.011020} {\bibfield  {journal} {\bibinfo  {journal} {Phys. Rev. X}\ }\textbf {\bibinfo {volume} {7}},\ \bibinfo {pages} {011020} (\bibinfo {year} {2017})}\BibitemShut {NoStop}%
\bibitem [{\citenamefont {Huang}\ \emph {et~al.}(2017)\citenamefont {Huang}, \citenamefont {Song}, \citenamefont {Huang},\ and\ \citenamefont {Hermele}}]{huang2017building}%
  \BibitemOpen
  \bibfield  {author} {\bibinfo {author} {\bibfnamefont {S.-J.}\ \bibnamefont {Huang}}, \bibinfo {author} {\bibfnamefont {H.}~\bibnamefont {Song}}, \bibinfo {author} {\bibfnamefont {Y.-P.}\ \bibnamefont {Huang}},\ and\ \bibinfo {author} {\bibfnamefont {M.}~\bibnamefont {Hermele}},\ }\bibfield  {title} {\bibinfo {title} {Building crystalline topological phases from lower-dimensional states},\ }\href {https://doi.org/10.1103/PhysRevB.96.205106} {\bibfield  {journal} {\bibinfo  {journal} {Phys. Rev. B}\ }\textbf {\bibinfo {volume} {96}},\ \bibinfo {pages} {205106} (\bibinfo {year} {2017})}\BibitemShut {NoStop}%
\bibitem [{\citenamefont {Qi}\ \emph {et~al.}(2019)\citenamefont {Qi}, \citenamefont {Jian},\ and\ \citenamefont {Wang}}]{qi2019folding}%
  \BibitemOpen
  \bibfield  {author} {\bibinfo {author} {\bibfnamefont {Y.}~\bibnamefont {Qi}}, \bibinfo {author} {\bibfnamefont {C.-M.}\ \bibnamefont {Jian}},\ and\ \bibinfo {author} {\bibfnamefont {C.}~\bibnamefont {Wang}},\ }\bibfield  {title} {\bibinfo {title} {Folding approach to topological order enriched by mirror symmetry},\ }\href {https://doi.org/10.1103/PhysRevB.99.085128} {\bibfield  {journal} {\bibinfo  {journal} {Phys. Rev. B}\ }\textbf {\bibinfo {volume} {99}},\ \bibinfo {pages} {085128} (\bibinfo {year} {2019})}\BibitemShut {NoStop}%
\bibitem [{Note1()}]{Note1}%
  \BibitemOpen
  \bibinfo {note} {In this paper, we use blackboard bold letters to denote the onsite symmetry groups in order to distinguish them from point-groups. For example, an interlayer exchange symmetry group in multilayer systems is denoted as $\protect \mathbb {Z}_n$, which originates from the rotation symmetry group $C_n$.}\BibitemShut {Stop}%
\bibitem [{\citenamefont {Cheng}\ and\ \citenamefont {Williamson}(2020)}]{cheng2020relative}%
  \BibitemOpen
  \bibfield  {author} {\bibinfo {author} {\bibfnamefont {M.}~\bibnamefont {Cheng}}\ and\ \bibinfo {author} {\bibfnamefont {D.~J.}\ \bibnamefont {Williamson}},\ }\bibfield  {title} {\bibinfo {title} {Relative anomaly in ($1+1$)d rational conformal field theory},\ }\href {https://doi.org/10.1103/PhysRevResearch.2.043044} {\bibfield  {journal} {\bibinfo  {journal} {Phys. Rev. Res.}\ }\textbf {\bibinfo {volume} {2}},\ \bibinfo {pages} {043044} (\bibinfo {year} {2020})}\BibitemShut {NoStop}%
\bibitem [{\citenamefont {Ding}\ and\ \citenamefont {Qi}(2024)}]{ding2024anomalies}%
  \BibitemOpen
  \bibfield  {author} {\bibinfo {author} {\bibfnamefont {Z.}~\bibnamefont {Ding}}\ and\ \bibinfo {author} {\bibfnamefont {Y.}~\bibnamefont {Qi}},\ }\bibfield  {title} {\bibinfo {title} {Anomalies in mirror symmetry enriched topological orders},\ }\href {https://doi.org/10.1103/PhysRevB.110.155151} {\bibfield  {journal} {\bibinfo  {journal} {Phys. Rev. B}\ }\textbf {\bibinfo {volume} {110}},\ \bibinfo {pages} {155151} (\bibinfo {year} {2024})}\BibitemShut {NoStop}%
\bibitem [{Note2()}]{Note2}%
  \BibitemOpen
  \bibinfo {note} {In this work, we adapt the notation that $D_{2n}$ is the order-$2n$ dihedral group. This notation is often used in mathematical literatures, while the dihedral groups $D_{2n}$ is denoted as $D_n$ in crystallography.}\BibitemShut {Stop}%
\bibitem [{Note3()}]{Note3}%
  \BibitemOpen
  \bibinfo {note} {$C_1$ is the trivial point group containing no nontrivial symmetry operations; $D_2$ is the mirror symmetry group.}\BibitemShut {Stop}%
\bibitem [{\citenamefont {Chen}\ \emph {et~al.}(2013)\citenamefont {Chen}, \citenamefont {Gu}, \citenamefont {Liu},\ and\ \citenamefont {Wen}}]{chen2013symmetry}%
  \BibitemOpen
  \bibfield  {author} {\bibinfo {author} {\bibfnamefont {X.}~\bibnamefont {Chen}}, \bibinfo {author} {\bibfnamefont {Z.-C.}\ \bibnamefont {Gu}}, \bibinfo {author} {\bibfnamefont {Z.-X.}\ \bibnamefont {Liu}},\ and\ \bibinfo {author} {\bibfnamefont {X.-G.}\ \bibnamefont {Wen}},\ }\bibfield  {title} {\bibinfo {title} {Symmetry protected topological orders and the group cohomology of their symmetry group},\ }\href {https://doi.org/10.1103/PhysRevB.87.155114} {\bibfield  {journal} {\bibinfo  {journal} {Phys. Rev. B}\ }\textbf {\bibinfo {volume} {87}},\ \bibinfo {pages} {155114} (\bibinfo {year} {2013})}\BibitemShut {NoStop}%
\bibitem [{Note4()}]{Note4}%
  \BibitemOpen
  \bibinfo {note} {Generally, there might be another $H^2$ obstruction class in $H^2[G, \protect \mathrm U(1)]$, but the cohomology group $H^2[C_n, \protect \mathrm U(1)]$ is trivial.}\BibitemShut {Stop}%
\bibitem [{Note5()}]{Note5}%
  \BibitemOpen
  \bibinfo {note} {(m, n) is the greatest divisor of m and n.}\BibitemShut {Stop}%
\bibitem [{\citenamefont {Qi}\ and\ \citenamefont {Fu}(2015{\natexlab{b}})}]{qi2015anomalous}%
  \BibitemOpen
  \bibfield  {author} {\bibinfo {author} {\bibfnamefont {Y.}~\bibnamefont {Qi}}\ and\ \bibinfo {author} {\bibfnamefont {L.}~\bibnamefont {Fu}},\ }\bibfield  {title} {\bibinfo {title} {Anomalous crystal symmetry fractionalization on the surface of topological crystalline insulators},\ }\href {https://doi.org/10.1103/PhysRevLett.115.236801} {\bibfield  {journal} {\bibinfo  {journal} {Phys. Rev. Lett.}\ }\textbf {\bibinfo {volume} {115}},\ \bibinfo {pages} {236801} (\bibinfo {year} {2015}{\natexlab{b}})}\BibitemShut {NoStop}%
\bibitem [{\citenamefont {Barkeshli}\ \emph {et~al.}(2020)\citenamefont {Barkeshli}, \citenamefont {Bonderson}, \citenamefont {Cheng}, \citenamefont {Jian},\ and\ \citenamefont {Walker}}]{barkeshli_reflection_2020}%
  \BibitemOpen
  \bibfield  {author} {\bibinfo {author} {\bibfnamefont {M.}~\bibnamefont {Barkeshli}}, \bibinfo {author} {\bibfnamefont {P.}~\bibnamefont {Bonderson}}, \bibinfo {author} {\bibfnamefont {M.}~\bibnamefont {Cheng}}, \bibinfo {author} {\bibfnamefont {C.-M.}\ \bibnamefont {Jian}},\ and\ \bibinfo {author} {\bibfnamefont {K.}~\bibnamefont {Walker}},\ }\bibfield  {title} {\bibinfo {title} {Reflection and {Time} {Reversal} {Symmetry} {Enriched} {Topological} {Phases} of {Matter}: {Path} {Integrals}, {Non}-orientable {Manifolds}, and {Anomalies}},\ }\href {https://doi.org/10.1007/s00220-019-03475-8} {\bibfield  {journal} {\bibinfo  {journal} {Commun Math Phys}\ }\textbf {\bibinfo {volume} {374}},\ \bibinfo {pages} {1021} (\bibinfo {year} {2020})}\BibitemShut {NoStop}%
\bibitem [{\citenamefont {Barkeshli}\ and\ \citenamefont {Cheng}(2018)}]{barkeshli2018time}%
  \BibitemOpen
  \bibfield  {author} {\bibinfo {author} {\bibfnamefont {M.}~\bibnamefont {Barkeshli}}\ and\ \bibinfo {author} {\bibfnamefont {M.}~\bibnamefont {Cheng}},\ }\bibfield  {title} {\bibinfo {title} {Time-reversal and spatial-reflection symmetry localization anomalies in (2+1)-dimensional topological phases of matter},\ }\href {https://doi.org/10.1103/PhysRevB.98.115129} {\bibfield  {journal} {\bibinfo  {journal} {Phys. Rev. B}\ }\textbf {\bibinfo {volume} {98}},\ \bibinfo {pages} {115129} (\bibinfo {year} {2018})}\BibitemShut {NoStop}%
\bibitem [{\citenamefont {Cheng}\ and\ \citenamefont {Wang}(2022)}]{cheng2022rotation}%
  \BibitemOpen
  \bibfield  {author} {\bibinfo {author} {\bibfnamefont {M.}~\bibnamefont {Cheng}}\ and\ \bibinfo {author} {\bibfnamefont {C.}~\bibnamefont {Wang}},\ }\bibfield  {title} {\bibinfo {title} {Rotation symmetry-protected topological phases of fermions},\ }\href {https://doi.org/10.1103/PhysRevB.105.195154} {\bibfield  {journal} {\bibinfo  {journal} {Phys. Rev. B}\ }\textbf {\bibinfo {volume} {105}},\ \bibinfo {pages} {195154} (\bibinfo {year} {2022})}\BibitemShut {NoStop}%
\bibitem [{\citenamefont {Chen}\ \emph {et~al.}(2015)\citenamefont {Chen}, \citenamefont {Burnell}, \citenamefont {Vishwanath},\ and\ \citenamefont {Fidkowski}}]{chen2015anomalous}%
  \BibitemOpen
  \bibfield  {author} {\bibinfo {author} {\bibfnamefont {X.}~\bibnamefont {Chen}}, \bibinfo {author} {\bibfnamefont {F.~J.}\ \bibnamefont {Burnell}}, \bibinfo {author} {\bibfnamefont {A.}~\bibnamefont {Vishwanath}},\ and\ \bibinfo {author} {\bibfnamefont {L.}~\bibnamefont {Fidkowski}},\ }\bibfield  {title} {\bibinfo {title} {Anomalous symmetry fractionalization and surface topological order},\ }\href {https://doi.org/10.1103/PhysRevX.5.041013} {\bibfield  {journal} {\bibinfo  {journal} {Phys. Rev. X}\ }\textbf {\bibinfo {volume} {5}},\ \bibinfo {pages} {041013} (\bibinfo {year} {2015})}\BibitemShut {NoStop}%
\bibitem [{\citenamefont {Fidkowski}\ and\ \citenamefont {Vishwanath}(2017)}]{fidkowski2017realizing}%
  \BibitemOpen
  \bibfield  {author} {\bibinfo {author} {\bibfnamefont {L.}~\bibnamefont {Fidkowski}}\ and\ \bibinfo {author} {\bibfnamefont {A.}~\bibnamefont {Vishwanath}},\ }\bibfield  {title} {\bibinfo {title} {Realizing anomalous anyonic symmetries at the surfaces of three-dimensional gauge theories},\ }\href {https://doi.org/10.1103/PhysRevB.96.045131} {\bibfield  {journal} {\bibinfo  {journal} {Phys. Rev. B}\ }\textbf {\bibinfo {volume} {96}},\ \bibinfo {pages} {045131} (\bibinfo {year} {2017})}\BibitemShut {NoStop}%
\bibitem [{\citenamefont {Eli\"ens}\ \emph {et~al.}(2014)\citenamefont {Eli\"ens}, \citenamefont {Romers},\ and\ \citenamefont {Bais}}]{eliens2014diagrammatics}%
  \BibitemOpen
  \bibfield  {author} {\bibinfo {author} {\bibfnamefont {I.~S.}\ \bibnamefont {Eli\"ens}}, \bibinfo {author} {\bibfnamefont {J.~C.}\ \bibnamefont {Romers}},\ and\ \bibinfo {author} {\bibfnamefont {F.~A.}\ \bibnamefont {Bais}},\ }\bibfield  {title} {\bibinfo {title} {Diagrammatics for bose condensation in anyon theories},\ }\href {https://doi.org/10.1103/PhysRevB.90.195130} {\bibfield  {journal} {\bibinfo  {journal} {Phys. Rev. B}\ }\textbf {\bibinfo {volume} {90}},\ \bibinfo {pages} {195130} (\bibinfo {year} {2014})}\BibitemShut {NoStop}%
\bibitem [{\citenamefont {Kong}(2014)}]{kong2014anyon}%
  \BibitemOpen
  \bibfield  {author} {\bibinfo {author} {\bibfnamefont {L.}~\bibnamefont {Kong}},\ }\bibfield  {title} {\bibinfo {title} {Anyon condensation and tensor categories},\ }\href {https://doi.org/https://doi.org/10.1016/j.nuclphysb.2014.07.003} {\bibfield  {journal} {\bibinfo  {journal} {Nucl. Phys. B.}\ }\textbf {\bibinfo {volume} {886}},\ \bibinfo {pages} {436} (\bibinfo {year} {2014})}\BibitemShut {NoStop}%
\bibitem [{\citenamefont {Cong}\ \emph {et~al.}(2016)\citenamefont {Cong}, \citenamefont {Cheng},\ and\ \citenamefont {Wang}}]{cong2016topological}%
  \BibitemOpen
  \bibfield  {author} {\bibinfo {author} {\bibfnamefont {I.}~\bibnamefont {Cong}}, \bibinfo {author} {\bibfnamefont {M.}~\bibnamefont {Cheng}},\ and\ \bibinfo {author} {\bibfnamefont {Z.}~\bibnamefont {Wang}},\ }\href {https://arxiv.org/abs/1609.02037} {\bibinfo {title} {Topological quantum computation with gapped boundaries}} (\bibinfo {year} {2016}),\ \Eprint {https://arxiv.org/abs/1609.02037} {arXiv:1609.02037 [quant-ph]} \BibitemShut {NoStop}%
\bibitem [{\citenamefont {Lou}\ \emph {et~al.}(2021)\citenamefont {Lou}, \citenamefont {Shen}, \citenamefont {Chen},\ and\ \citenamefont {Hung}}]{lou2021dummy}%
  \BibitemOpen
  \bibfield  {author} {\bibinfo {author} {\bibfnamefont {J.}~\bibnamefont {Lou}}, \bibinfo {author} {\bibfnamefont {C.}~\bibnamefont {Shen}}, \bibinfo {author} {\bibfnamefont {C.}~\bibnamefont {Chen}},\ and\ \bibinfo {author} {\bibfnamefont {L.-Y.}\ \bibnamefont {Hung}},\ }\bibfield  {title} {\bibinfo {title} {A (dummy’s) guide to working with gapped boundaries via (fermion) condensation},\ }\href {https://doi.org/10.1007/JHEP02(2021)171} {\bibfield  {journal} {\bibinfo  {journal} {J. High Energy Phys.}\ }\textbf {\bibinfo {volume} {2021}}\bibinfo  {number} { (2)},\ \bibinfo {pages} {171}}\BibitemShut {NoStop}%
\bibitem [{\citenamefont {Wang}\ \emph {et~al.}(2022)\citenamefont {Wang}, \citenamefont {Hu},\ and\ \citenamefont {Wan}}]{wang_extend_2022}%
  \BibitemOpen
\bibfield  {number} {  }\bibfield  {author} {\bibinfo {author} {\bibfnamefont {H.}~\bibnamefont {Wang}}, \bibinfo {author} {\bibfnamefont {Y.}~\bibnamefont {Hu}},\ and\ \bibinfo {author} {\bibfnamefont {Y.}~\bibnamefont {Wan}},\ }\bibfield  {title} {\bibinfo {title} {Extend the {Levin}-{Wen} model to two-dimensional topological orders with gapped boundary junctions},\ }\href {https://doi.org/10.1007/JHEP07(2022)088} {\bibfield  {journal} {\bibinfo  {journal} {J. High Energy Phys.}\ }\textbf {\bibinfo {volume} {2022}}\bibinfo  {number} { (7)},\ \bibinfo {pages} {88}}\BibitemShut {NoStop}%
\bibitem [{Note6()}]{Note6}%
  \BibitemOpen
\bibfield  {number} {  }\bibinfo {note} {In the subsequent sections of this paper, we focus exclusively on boundaries where the multiplicity of anyon condensing is single.}\BibitemShut {Stop}%
\bibitem [{\citenamefont {Lan}\ \emph {et~al.}(2015)\citenamefont {Lan}, \citenamefont {Wang},\ and\ \citenamefont {Wen}}]{lan2015gapped}%
  \BibitemOpen
  \bibfield  {author} {\bibinfo {author} {\bibfnamefont {T.}~\bibnamefont {Lan}}, \bibinfo {author} {\bibfnamefont {J.~C.}\ \bibnamefont {Wang}},\ and\ \bibinfo {author} {\bibfnamefont {X.-G.}\ \bibnamefont {Wen}},\ }\bibfield  {title} {\bibinfo {title} {Gapped domain walls, gapped boundaries, and topological degeneracy},\ }\href {https://doi.org/10.1103/PhysRevLett.114.076402} {\bibfield  {journal} {\bibinfo  {journal} {Phys. Rev. Lett.}\ }\textbf {\bibinfo {volume} {114}},\ \bibinfo {pages} {076402} (\bibinfo {year} {2015})}\BibitemShut {NoStop}%
\bibitem [{\citenamefont {Fuchs}\ \emph {et~al.}(2002)\citenamefont {Fuchs}, \citenamefont {Runkel},\ and\ \citenamefont {Schweigert}}]{fuchs2002tft}%
  \BibitemOpen
  \bibfield  {author} {\bibinfo {author} {\bibfnamefont {J.}~\bibnamefont {Fuchs}}, \bibinfo {author} {\bibfnamefont {I.}~\bibnamefont {Runkel}},\ and\ \bibinfo {author} {\bibfnamefont {C.}~\bibnamefont {Schweigert}},\ }\bibfield  {title} {\bibinfo {title} {Tft construction of rcft correlators i: partition functions},\ }\href {https://doi.org/https://doi.org/10.1016/S0550-3213(02)00744-7} {\bibfield  {journal} {\bibinfo  {journal} {Nucl. Phys. B.}\ }\textbf {\bibinfo {volume} {646}},\ \bibinfo {pages} {353} (\bibinfo {year} {2002})}\BibitemShut {NoStop}%
\bibitem [{\citenamefont {Shen}\ and\ \citenamefont {Hung}(2019)}]{shen2019defect}%
  \BibitemOpen
  \bibfield  {author} {\bibinfo {author} {\bibfnamefont {C.}~\bibnamefont {Shen}}\ and\ \bibinfo {author} {\bibfnamefont {L.-Y.}\ \bibnamefont {Hung}},\ }\bibfield  {title} {\bibinfo {title} {Defect verlinde formula for edge excitations in topological order},\ }\href {https://doi.org/10.1103/PhysRevLett.123.051602} {\bibfield  {journal} {\bibinfo  {journal} {Phys. Rev. Lett.}\ }\textbf {\bibinfo {volume} {123}},\ \bibinfo {pages} {051602} (\bibinfo {year} {2019})}\BibitemShut {NoStop}%
\bibitem [{\citenamefont {Bischoff}\ \emph {et~al.}(2019)\citenamefont {Bischoff}, \citenamefont {Jones}, \citenamefont {Lu},\ and\ \citenamefont {Penneys}}]{bischoff2019spontaneous}%
  \BibitemOpen
  \bibfield  {author} {\bibinfo {author} {\bibfnamefont {M.}~\bibnamefont {Bischoff}}, \bibinfo {author} {\bibfnamefont {C.}~\bibnamefont {Jones}}, \bibinfo {author} {\bibfnamefont {Y.-M.}\ \bibnamefont {Lu}},\ and\ \bibinfo {author} {\bibfnamefont {D.}~\bibnamefont {Penneys}},\ }\bibfield  {title} {\bibinfo {title} {Spontaneous symmetry breaking from anyon condensation},\ }\href {https://doi.org/10.1007/JHEP02(2019)062} {\bibfield  {journal} {\bibinfo  {journal} {J. High Energy Phys.}\ }\textbf {\bibinfo {volume} {2019}}\bibinfo  {number} { (2)},\ \bibinfo {pages} {62}}\BibitemShut {NoStop}%
\bibitem [{Note7()}]{Note7}%
  \BibitemOpen
\bibfield  {number} {  }\bibinfo {note} {In this paper, for the convenience of exposition, we only present the consistency equations for unitary cases. For anti-unitary cases, the extension is straightforward.}\BibitemShut {Stop}%
\bibitem [{\citenamefont {Meir}\ and\ \citenamefont {Musicantov}(2012)}]{meir2012module}%
  \BibitemOpen
  \bibfield  {author} {\bibinfo {author} {\bibfnamefont {E.}~\bibnamefont {Meir}}\ and\ \bibinfo {author} {\bibfnamefont {E.}~\bibnamefont {Musicantov}},\ }\bibfield  {title} {\bibinfo {title} {Module categories over graded fusion categories},\ }\href {https://doi.org/https://doi.org/10.1016/j.jpaa.2012.03.014} {\bibfield  {journal} {\bibinfo  {journal} {J Pure Appl Algebra}\ }\textbf {\bibinfo {volume} {216}},\ \bibinfo {pages} {2449} (\bibinfo {year} {2012})}\BibitemShut {NoStop}%
\bibitem [{\citenamefont {Barkeshli}\ and\ \citenamefont {Cheng}(2020)}]{barkeshli2020relative}%
  \BibitemOpen
  \bibfield  {author} {\bibinfo {author} {\bibfnamefont {M.}~\bibnamefont {Barkeshli}}\ and\ \bibinfo {author} {\bibfnamefont {M.}~\bibnamefont {Cheng}},\ }\bibfield  {title} {\bibinfo {title} {{Relative anomalies in (2+1)D symmetry enriched topological states}},\ }\href {https://doi.org/10.21468/SciPostPhys.8.2.028} {\bibfield  {journal} {\bibinfo  {journal} {SciPost Phys.}\ }\textbf {\bibinfo {volume} {8}},\ \bibinfo {pages} {028} (\bibinfo {year} {2020})}\BibitemShut {NoStop}%
\bibitem [{Note8()}]{Note8}%
  \BibitemOpen
  \bibinfo {note} {In the rest of paper, without ambiguity, we use additional VLCs to specifically refer to as the VLCs $[\phi ^{a_{\protect \ensuremath {\protect \mathbf g}} b_{\protect \ensuremath {\protect \mathbf h}}}_{c_{\protect \ensuremath {\protect \mathbf g}\protect \ensuremath {\protect \mathbf h}}, \alpha }]$, associated with $\protect \ensuremath {\protect \mathbf g}$-local defects.}\BibitemShut {Stop}%
\bibitem [{\citenamefont {Etingof}\ \emph {et~al.}(2015)\citenamefont {Etingof}, \citenamefont {Gelaki}, \citenamefont {Nikshych},\ and\ \citenamefont {Ostrik}}]{etingof2015tensor}%
  \BibitemOpen
  \bibfield  {author} {\bibinfo {author} {\bibfnamefont {P.}~\bibnamefont {Etingof}}, \bibinfo {author} {\bibfnamefont {S.}~\bibnamefont {Gelaki}}, \bibinfo {author} {\bibfnamefont {D.}~\bibnamefont {Nikshych}},\ and\ \bibinfo {author} {\bibfnamefont {V.}~\bibnamefont {Ostrik}},\ }\href {https://bookstore.ams.org/surv-205} {\emph {\bibinfo {title} {Tensor categories}}},\ Vol.\ \bibinfo {volume} {205}\ (\bibinfo  {publisher} {American Mathematical Soc.},\ \bibinfo {year} {2015})\BibitemShut {NoStop}%
\bibitem [{\citenamefont {Gannon}\ and\ \citenamefont {Jones}(2019)}]{gannon2019vanishing}%
  \BibitemOpen
  \bibfield  {author} {\bibinfo {author} {\bibfnamefont {T.}~\bibnamefont {Gannon}}\ and\ \bibinfo {author} {\bibfnamefont {C.}~\bibnamefont {Jones}},\ }\bibfield  {title} {\bibinfo {title} {Vanishing of {Categorical} {Obstructions} for {Permutation} {Orbifolds}},\ }\href {https://doi.org/10.1007/s00220-019-03288-9} {\bibfield  {journal} {\bibinfo  {journal} {Commun Math Phys}\ }\textbf {\bibinfo {volume} {369}},\ \bibinfo {pages} {245} (\bibinfo {year} {2019})}\BibitemShut {NoStop}%
\bibitem [{Note9()}]{Note9}%
  \BibitemOpen
  \bibinfo {note} {Strictly speaking, the cohomology group should be $H^2[\protect \mathbb {Z}^{\protect \ensuremath {\protect \mathbf R}}_n, \protect \mathrm U(1)]$. However, at the junction, for the cohomology group, we can disregard the distinction between $\protect \mathbb {Z}^{\protect \ensuremath {\protect \mathbf R}}_n$ and $C_n$}\BibitemShut {NoStop}%
\bibitem [{\citenamefont {Kong}\ \emph {et~al.}(2020)\citenamefont {Kong}, \citenamefont {Lan}, \citenamefont {Wen}, \citenamefont {Zhang},\ and\ \citenamefont {Zheng}}]{kong2020classification}%
  \BibitemOpen
  \bibfield  {author} {\bibinfo {author} {\bibfnamefont {L.}~\bibnamefont {Kong}}, \bibinfo {author} {\bibfnamefont {T.}~\bibnamefont {Lan}}, \bibinfo {author} {\bibfnamefont {X.-G.}\ \bibnamefont {Wen}}, \bibinfo {author} {\bibfnamefont {Z.-H.}\ \bibnamefont {Zhang}},\ and\ \bibinfo {author} {\bibfnamefont {H.}~\bibnamefont {Zheng}},\ }\bibfield  {title} {\bibinfo {title} {Classification of topological phases with finite internal symmetries in all dimensions},\ }\href {https://doi.org/10.1007/JHEP09(2020)093} {\bibfield  {journal} {\bibinfo  {journal} {J. High Energy Phys.}\ }\textbf {\bibinfo {volume} {2020}}\bibinfo  {number} { (9)},\ \bibinfo {pages} {93}}\BibitemShut {NoStop}%
\bibitem [{Note10()}]{Note10}%
  \BibitemOpen
\bibfield  {number} {  }\bibinfo {note} {Similar to the case of $C_n$-SETs, for the group cohomology, we disregard the distinction between $\protect \mathbb {D}_{2n}$ and $D_{2n}$}\BibitemShut {NoStop}%
\bibitem [{\citenamefont {Wall}(1961)}]{wall1961resolutions}%
  \BibitemOpen
  \bibfield  {author} {\bibinfo {author} {\bibfnamefont {C.~T.~C.}\ \bibnamefont {Wall}},\ }\bibfield  {title} {\bibinfo {title} {Resolutions for extensions of groups},\ }\href {https://doi.org/10.1017/S0305004100035155} {\bibfield  {journal} {\bibinfo  {journal} {Math. Proc. Cambridge}\ }\textbf {\bibinfo {volume} {57}},\ \bibinfo {pages} {251–255} (\bibinfo {year} {1961})}\BibitemShut {NoStop}%
\end{thebibliography}%
\end{document}